\pdfoutput=1
\documentclass[%
prd,amsmath,aps,floats,amssymb, floatfix, superscriptaddress,nofootinbib,twocolumn,preprintnumbers
]{revtex4-2}

\usepackage{graphicx}
\usepackage{dcolumn}
\usepackage{bm}
\usepackage{cprotect}
\usepackage{orcidlink}

\def\apj{{Astrophys.~J.}}       
\def\apjs{{ApJS}}               

\def\mnras{Mon.~Not.~Roy.~Astron.~Soc.}             
\def\prd{{Phys.~Rev.~D}}        
\def\pasj{{Publ. Astron. Soc. Jap.}} 
\newcommand{\beq}{\begin{equation}}
\newcommand{\beqa}{\begin{eqnarray}}
\newcommand{\eeq}{\end{equation}}
\newcommand{\eeqa}{\end{eqnarray}}

\newcommand{\simgt}{\lower.5ex\hbox{$\; \buildrel > \over \sim \;$}}
\newcommand{\simlt}{\lower.5ex\hbox{$\; \buildrel < \over \sim \;$}}

\newcommand{\bd}[1]{\mbox{\boldmath $#1$}}

\usepackage{xcolor}
\newcommand{\rev}[1]{{\color{black}{#1}}}

\usepackage{amsmath,amssymb,natbib,latexsym,times}

\begin{document}


\title{Masses of Sunyaev-Zel'dovich Galaxy Clusters Detected by The Atacama Cosmology Telescope: Stacked Lensing Measurements with Subaru HSC Year 3 data} 

\author{Masato~Shirasaki\orcidlink{0000-0002-1706-5797}}
  \email{masato.shirasaki@nao.ac.jp}
\affiliation{%
National Astronomical Observatory of Japan (NAOJ), National Institutes of Natural Science, Mitaka, Tokyo 181-8588, Japan
}%
\affiliation{
The Institute of Statistical Mathematics,
Tachikawa, Tokyo 190-8562, Japan
}

\author{Crist\'{o}bal~Sif\'{o}n\orcidlink{0000-0002-8149-1352}}
\affiliation{%
Instituto de F\'{i}sica, Pontificia Universidad Cat\'{o}lica de Valpara\'{i}so, Casilla 4059, Valpara\'{i}so, Chile
}%

\author{Hironao~Miyatake\orcidlink{0000-0001-7964-9766}}
\affiliation{Kobayashi-Maskawa Institute for the Origin of Particles and the
Universe (KMI), Nagoya University, Nagoya, 464-8602, Japan}
\affiliation{Institute for Advanced Research, Nagoya University, Nagoya
464-8601, Japan}
\affiliation{Kavli Institute for the Physics and Mathematics of the Universe (WPI), The University of Tokyo Institutes for Advanced Study (UTIAS), The University of Tokyo, Chiba 277-8583, Japan}

\author{Erwin~Lau\orcidlink{0000-0001-8914-8885}}
\affiliation{%
Center for Astrophysics | Harvard \& Smithsonian, 60 Garden St, Cambridge, MA 02138, USA
}%

\author{Zhuowen~Zhang}
\affiliation{%
Department of Astronomy and Astrophysics, University of Chicago, 5640 S. Ellis Ave, Chicago, IL, USA, 60637
}%
\affiliation{%
Fermilab, Kirk \& Pine Road, Batavia, IL, USA, 60510
}%

\author{Neta~Bahcall}
\affiliation{Department of Astrophysical Sciences, Peyton Hall, Princeton University, Princeton, NJ USA 08544}

\author{Nicholas~Battaglia\orcidlink{0000-0001-5846-0411}}
\affiliation{%
Department of Astronomy, Cornell University, Ithaca, NY 14853, USA
}%

\author{Mark~Devlin\orcidlink{0000-0002-3169-9761}}
\affiliation{%
Department of Physics and Astronomy, University of Pennsylvania, 209 South 33rd Street, Philadelphia, PA, USA
}%
\affiliation{%
Fermilab, Kirk \& Pine Road, Batavia, IL, USA, 60510
}%

\author{Jo~Dunkley\orcidlink{0000-0002-7450-2586}}
\affiliation{Joseph Henry Laboratories of Physics, Jadwin Hall, Princeton University, Princeton, NJ, USA 08544}\affiliation{Department of Astrophysical Sciences, Peyton Hall, Princeton University, Princeton, NJ USA 08544}

\author{Arya~Farahi\orcidlink{0000-0003-0777-4618}}
\affiliation{%
Department of Statistics and Data Science, The University of Texas at Austin, TX 78712, USA
}

\author{Matt~Hilton\orcidlink{0000-0002-8490-8117}}
\affiliation{%
Wits Centre for Astrophysics, School of Physics, University of the Witwatersrand, Private Bag 3, 2050, Johannesburg, South Africa
}%
\affiliation{%
Astrophysics Research Centre, School of Mathematics, Statistics, and Computer Science, University of KwaZulu-Natal, Westville Campus, Durban 4041, South Africa
}%

\author{Yen-Ting~Lin}
\affiliation{Institute of Astronomy and Astrophysics, Academia Sinica, Taipei 10617, Taiwan}

\author{Daisuke~Nagai\orcidlink{0000-0002-6766-5942}}
\affiliation{%
Department of Physics, Yale University, New Haven, CT 06520, USA
}%

\author{Suzanne~T.~Staggs\orcidlink{0000-0002-7020-7301}}
\affiliation{%
Joseph Henry Laboratories of Physics, Jadwin Hall, Princeton University, Princeton, NJ, USA 08544
}%

\author{Tomomi~Sunayama\orcidlink{0009-0004-6387-5784}}
\affiliation{Steward Observatory, University of Arizona, Tucson, AZ 85704, USA}
\affiliation{Kobayashi-Maskawa Institute for the Origin of Particles and the Universe (KMI), Nagoya University, Nagoya, 464-8602, Japan}

\author{David~Spergel}
\affiliation{Simons Foundation, 160 Fifth Avenue NY, NY 10010, USA}

\author{Edward~J.~Wollack\orcidlink{0000-0002-7567-4451}}
\affiliation{NASA/Goddard Space Flight Center, Greenbelt, MD 20771, USA}



\date{\today}

\begin{abstract}
We present a stacked lensing analysis of 96 galaxy clusters selected  
by the thermal Sunyaev-Zel'dovich (SZ) effect in maps of the cosmic microwave background (CMB).
We select foreground galaxy clusters with a $5\sigma$-level SZ threshold in CMB observations from the Atacama Cosmology Telescope, while we define background source galaxies for the lensing analysis with secure photometric redshift cuts in Year 3 data of the Subaru Hyper Suprime Cam survey.
We detect the stacked lensing signal in the range of $0.1 < R\, [h^{-1}\mathrm{Mpc}] < 100$ in each of three cluster redshift bins, $0.092<z\le0.445$, $0.445<z\le0.695$, and $0.695<z\le1.180$, with 32 galaxy clusters in each bin. The cumulative signal-to-noise ratios of the lensing signal are $14.6$, $12.0$, and $6.6$, respectively.
Using a halo-based forward model, we then constrain statistical relationships
between the mass inferred from the SZ observation (i.e.~SZ mass) and the total mass 
derived from our stacked lensing measurements. 
At the average SZ mass in the cluster sample ($2.1-2.4\times10^{14}\, h^{-1}M_\odot$), our likelihood analysis shows that the average total mass differs from the SZ counterpart by a factor of $1.3 \pm 0.2$, $1.6 \pm 0.2$, and $1.6 \pm 0.3$ (68\%) in the aforementioned redshift ranges, respectively. 
Our limits are consistent with previous lensing measurements, and we find that the cluster modeling choices can introduce a $1\sigma$-level difference in our parameter inferences.
\end{abstract}


\maketitle

\section{Introduction}

Galaxy clusters are the largest gravitationally bound objects in the universe.
Analytic approaches (e.g.~\cite{Press:1973iz, Sheth:1998ew}) 
and numerical simulations (e.g.~\cite{Tinker:2008ff, 2011ApJ...732..122B, Watson:2012mt, Despali:2015yla, Shirasaki:2021orc}) predict that 
the number density of galaxy clusters as a function of their masses is 
primarily determined by the variance in linear fluctuations in cosmic mass density within a Lagrangian comoving scale of $R = [3M/(4\pi \bar{\rho}_\mathrm{m0})]^{1/3}$, 
where $M$ is the cluster mass and $\bar{\rho}_\mathrm{m0}$ is 
the average cosmic mass density in comoving coordinates.
Hence,  number counts of galaxy clusters offer a unique means of measuring the formation rate of cosmological structures at their most massive end, $\sim 10^{14}\, h^{-1}M_\odot$,
giving physical insight into the mass variance at the scale of $R=8\, h^{-1}\mathrm{Mpc}$, the quantity referred to as $\sigma_8$ in the literature.
This enables one to constrain physics behind fundamental components in our universe such as dark energy by counting the number of galaxy clusters (see Ref.~\cite{Allen:2011zs} for a review).

Cosmological inference with galaxy clusters requires careful assessment of both selection effects and biases in mass estimates.
Among several cluster finders, observations of thermal Sunyaev Zel'dovich (SZ) effects on the cosmic microwave background (CMB) \cite{Sunyaev:1972eq} have received great attention because 
the thermal SZ effect in a single cluster does not depend on the cluster redshift.
Recent observations of the CMB with multiple frequency channels allow 
searching for galaxy clusters based on their thermal SZ effects alone.
Several large catalogs of SZ-selected clusters have been constructed \cite{Planck:2015koh, SPT:2018njh, ACT:2020lcv}, some containing more than $1000$ clusters. In this work, we employ a cluster catalog from the Atacama Cosmology Telescope (ACT) \cite{ACT:2020lcv}.
Future CMB observations can further increase the number of the SZ-selected clusters, enhancing the constraining power of galaxy clusters in fundamental physics \cite{Madhavacheril:2017onh}.
Several cosmological analyses with the SZ-selected clusters have been performed \cite{Hasselfield:2013wf, Planck:2015lwi, SPT:2018njh, Salvati:2021gkt}; however, their cosmological constraints hinge on systematic uncertainties in the estimate of cluster masses. 

Masses of individual clusters are commonly estimated from X-ray observations under the assumption of hydrostatic equilibrium (see Ref.~\cite{Rosati:2002ts} for a review).
For several reasons, the assumption behind the X-ray mass estimates can be invalid in practice.
Those include asphericity of the galaxy clusters, 
clumpiness in cluster gas density, gas accelerations at cluster outskirts,
and non-thermal pressure components induced by turbulent gas motions 
(see Ref~\cite{Pratt:2019cnf} and references therein for more details).
Recent progress in galaxy imaging surveys has opened an additional means of estimating cluster masses -- measuring the gravitational lensing effect on background galaxies.
The mass distributed around individual clusters can act as a lens for background galaxies,
introducing a coherent pattern of distortion in the galaxy shapes.
Tangential distortions in the shapes of background galaxies with respect to the foreground cluster position provide a direct measure of the cluster mass 
including invisible dark matter \cite{Umetsu:2020wlf}.

In this paper, we measure a gravitational-lensing signature induced by the SZ-selected clusters obtained from the Atacama Cosmology Telescope  \cite{ACT:2020lcv}. 
We employ a stacked lensing technique, which is suitable for studying the average mass density distribution around the clusters of interest \cite{Okabe:2009pf}.
The stacked lensing measurement allows us to investigate mass density profiles around the clusters because it combines the lensing effects of 
individual clusters into a single profile and thus increases the detection significance by reducing shot noise in the galaxy-shape measurements. 
Moreover, spherically-symmetric modeling of the mass density profile provides a good approximation to the stacked lensing measurements, because combining several lensing measurements suppresses any preferences in the orientation of real galaxy clusters given the statistical isotropy of the universe.
For the stacked lensing measurements, we use the galaxy shape catalog constructed from Year 3 data of the Subaru Hyper Suprime Cam (HSC) Survey \cite{Li:2021mvq}.
The HSC has the largest source number density among its competitors in stage-III lensing surveys, allowing us to study the stacked lensing measurements 
of the SZ-selected clusters over a wide range of cluster redshifts.

We improve upon our previous analysis in Ref.~\cite{Miyatake:2018lpb} in several ways.
The main improvements on the previous analysis are summarized below.

\begin{enumerate}
    \item[(i)] We increase the number of the SZ-selected clusters by a factor of 12. Our cluster sample now consists of 96 objects with a redshift range $0.096<z\le1.180$. 
    \item[(ii)] We estimate statistical errors in our stacked lensing measurements with a set of realistic mock catalogs of HSC source galaxies and SZ-selected clusters from ACT observation. Our mock catalogs are constructed from full-sky lensing simulations and their associated halo catalogs in Ref.~\cite{Takahashi:2017hjr}, including all sample-variance effects caused by non-linear structure formation.
    \item[(iii)] We adopt halo-based forward modeling for our stacked lensing measurements. As an input, the model requires an estimate of scatter\footnote{We mean the scatter is between clusters in the sample.} in the scaling relation between an SZ observable and the cluster mass; this is the dominant uncertainty in our modeling. We use the Illustris-TNG simulations \cite{Springel:2017tpz} to set an informative prior on the scatter caused by the variation of electron thermal pressure inside clusters. Also, we include the information on cluster number counts when inferring our model parameters with the stacked lensing measurements, helping to break inherent parameter degeneracy in our model.
    \item[(iv)] We incorporate possible baryonic effects on cluster mass density profiles with an efficient emulator of dark-halo statistics developed in Ref.~\cite{Nishimichi:2018etk}. Hence, our model includes fully non-linear gravitational effects in the mass distribution around cluster-sized halos as well as back-reaction effects on the cluster mass by the presence of stellar and gas components.
    We marginalize the modeling uncertainty due to baryonic effects to place constraints on the cluster masses.
\end{enumerate}

The goal of this paper is to constrain statistical relationships between an SZ observable 
and the total cluster mass with our stacked lensing measurements.
To be specific, we focus attention on the mass inferred from the SZ observation, referred to as SZ mass, and on a possible difference between the SZ and total cluster mass.
Given the completeness of our SZ-selected clusters as a function of SZ masses, 
we can infer the total cluster mass from our stacked lensing measurements 
by accounting for selection effects in the ACT observations.
Because the estimate of SZ masses relies on a model of cluster electron pressure calibrated with X-ray observations under the assumption of hydrostatic equilibrium, 
our inference of the cluster mass can validate the conventional hydrostatic-mass assumption in a direct manner. 

\begin{figure*}[!t]
\includegraphics[clip, width=2.0\columnwidth]{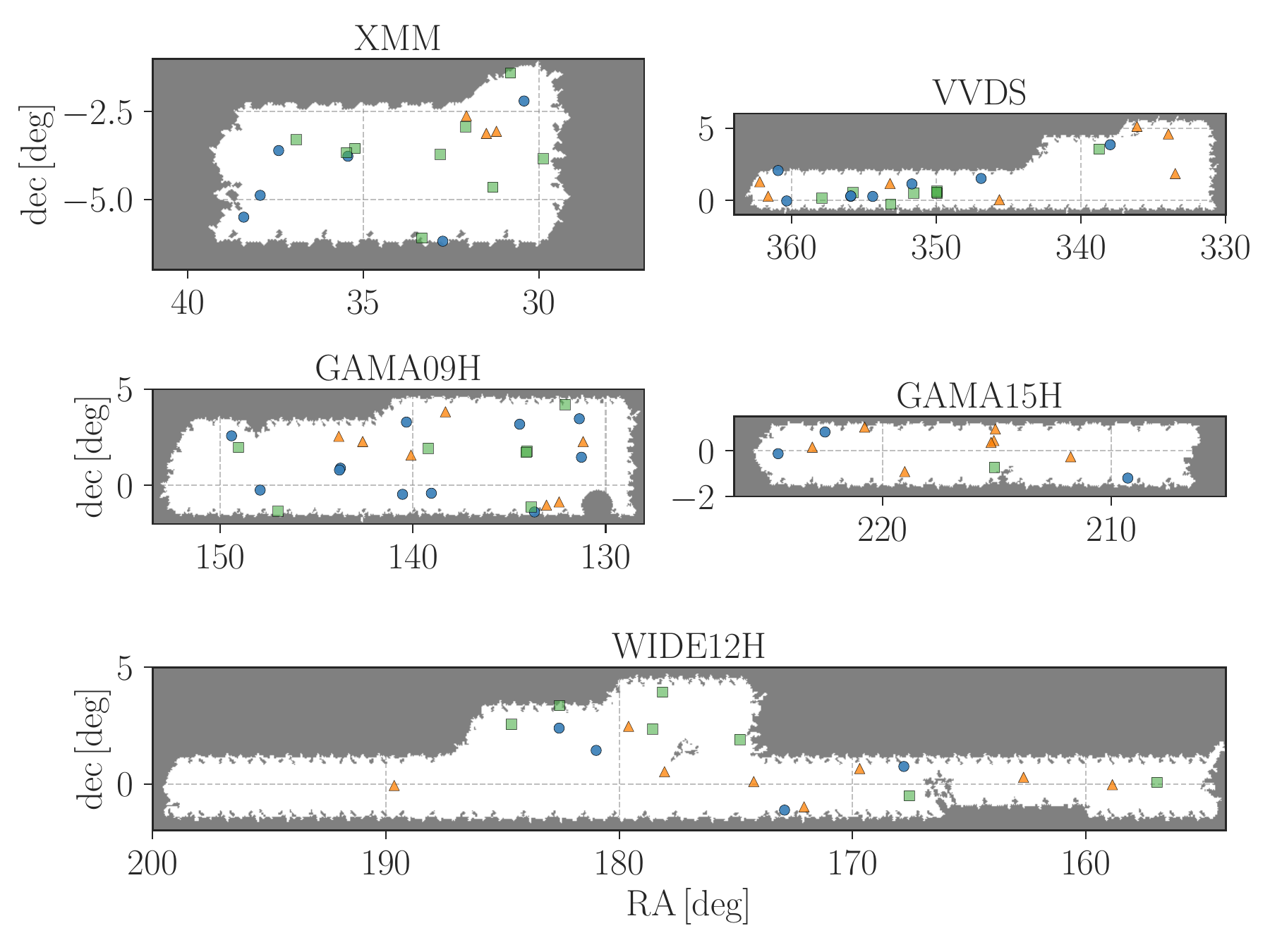}
\cprotect\caption{\label{fig:survey_overlap} 
ACT SZ-selected clusters in the HSC S19A fields. 
The colored points show 96 ACT clusters with the selection threshold of ${\tt fixed \verb|_| SNR} > 5$. 
The blue circles represent the clusters in the redshift range $0.092<z\le0.445$,
while orange triangles and the green squares show  clusters at 
$0.445<z\le0.695$ and $0.695<z\le1.180$, respectively.
The white regions in each panel highlight the full-depth, full-color (FDFC) regions in the HSC S19A observations.
}
\end{figure*}

Recently, Ref.~\cite{Robertson:2023gxe} presented a stacked lensing analysis of the ACT-selected galaxy clusters using the KiDS-1000 weak lensing catalog. 
Their analysis mainly focuses on the calibration of cluster masses at the average cluster redshift of $z\sim0.3-0.4$, whereas we aim at studying a possible redshift evolution of
stacked lensing effects around the clusters employing a secure selection of background galaxies based on HSC photometric redshift measurements.
Note that the authors in Ref.~\cite{Robertson:2023gxe} adopted a simple halo model based on the universal dark-matter density profile \cite{Navarro:1996gj} to infer the cluster masses. In this paper, we apply a more flexible model, accounting for baryonic effects and possible deviations from the universal density profile around boundaries of galaxy clusters.

The rest of this paper is organized as follows. 
We describe our data in Section.~\ref{sec:data}. 
Next, we summarize details of our stacked lensing measurements in Section.~\ref{sec:lensing}.
We explain our halo-based model of the stacked lensing measurements and
the setup of our parameter inference in Section~\ref{sec:analysis_method}.
Section~\ref{sec:results} presents the key results, 
whereas we discuss the limitations of our analysis in Section~\ref{sec:limit}. 
Finally, concluding remarks are provided in Section~\ref{sec:conclusion}.
In the following, $\ln$ represents the natural logarithm and 
$\log$ is the logarithm with base 10. 
Throughout this paper, we adopt $\Lambda$CDM cosmological parameters below; 
the average cosmic mass density $\Omega_\mathrm{m}=0.279$, 
the cosmological constant $\Omega_\Lambda = 1-\Omega_\mathrm{m} = 0.721$, 
the average baryon density $\Omega_\mathrm{b} = 0.046$,
the present-day Hubble parameter $ H_0 = 100h = 70.0 \, \mathrm{km}/\mathrm{s}/\mathrm{Mpc}$,
the spectral index of the power spectrum of primordial curvature
perturbations $n_s = 0.97$, 
and the square root of the linear mass variance within $8\, h^{-1}\mathrm{Mpc}$ 
being $\sigma_8 = 0.82$. Those parameters are consistent with statistical
analyses of the CMB anisotropies in Ref.~\cite{WMAP:2012nax}.
Our analysis requires mock catalogs of the HSC lensing data and the ACT cluster samples to precisely estimate statistical errors in our lensing measurements. The mock catalogs in this paper have been constructed from N-body simulations with the above cosmological model, which is slightly different from the latest version inferred by the Planck satellite \cite{Planck:2018vyg}.

\section{Data}\label{sec:data}

\subsection{ACT DR5 clusters}

In this paper, we use the cluster sample 
constructed from ACT observations from the 2008-2018 observing seasons \cite{ACT:2020lcv}.
The sample was extracted from 98 and 150 GHz observations of
a 13,211 $\mathrm{deg}^2$ sky, and form part of the ACT fifth data release (DR5; \cite{Naess:2020wgi}).
A multi-frequency matched filter (e.g.~\cite{Melin:2006qq, Williamson:2011jz})
was applied to the DR5 maps for cluster searches.
When searching for cluster candidates, the authors in Ref.~\cite{ACT:2020lcv} 
constructed a signal template from the Universal Pressure Profile (UPP; \cite{Arnaud:2009tt}) 
and its associated SZ signal-to-mass scaling relation.
We provide the explicit functional form of the UPP model in Appendix~\ref{apdx:UPP}.
Candidates were then confirmed as galaxy clusters by cross-matching 
with external optical and infrared (IR) data sets, including several spectroscopic surveys 
\cite{Blake:2016fcm, DES:2017ndx, SDSS-IV:2019txh, Scodeggio:2016jaf},
as well as optically-selected cluster samples 
in Sloan Digital Sky Survey (SDSS), Dark Energy Survey (DES), and HSC \cite{SDSS:2013jmz, DES:2016dkd, Oguri:2017khw}.
In addition, some cluster redshifts were estimated by the zCluster algorithm \cite{ACT:2017dgj}.
The detection significance of each cluster candidate is characterized 
by the signal-to-noise ratio at the reference 2.4-arcmin filter scale, referred to as ${\tt fixed\verb|_|SNR}$.
In this paper, we impose a threshold ${\tt fixed\verb|_|SNR} > 5$ 
to have a secure cluster sample, as explained in Ref.~\cite{ACT:2020lcv}.
Further details of the cluster search and optical/IR follow-up processes are 
summarized in Ref.~\cite{ACT:2020lcv}.
In the following, we refer to our cluster sample as ACT clusters.

\subsection{HSC-ACT survey overlap}

\begin{figure}[!t]
\includegraphics[clip, width=1.\columnwidth]{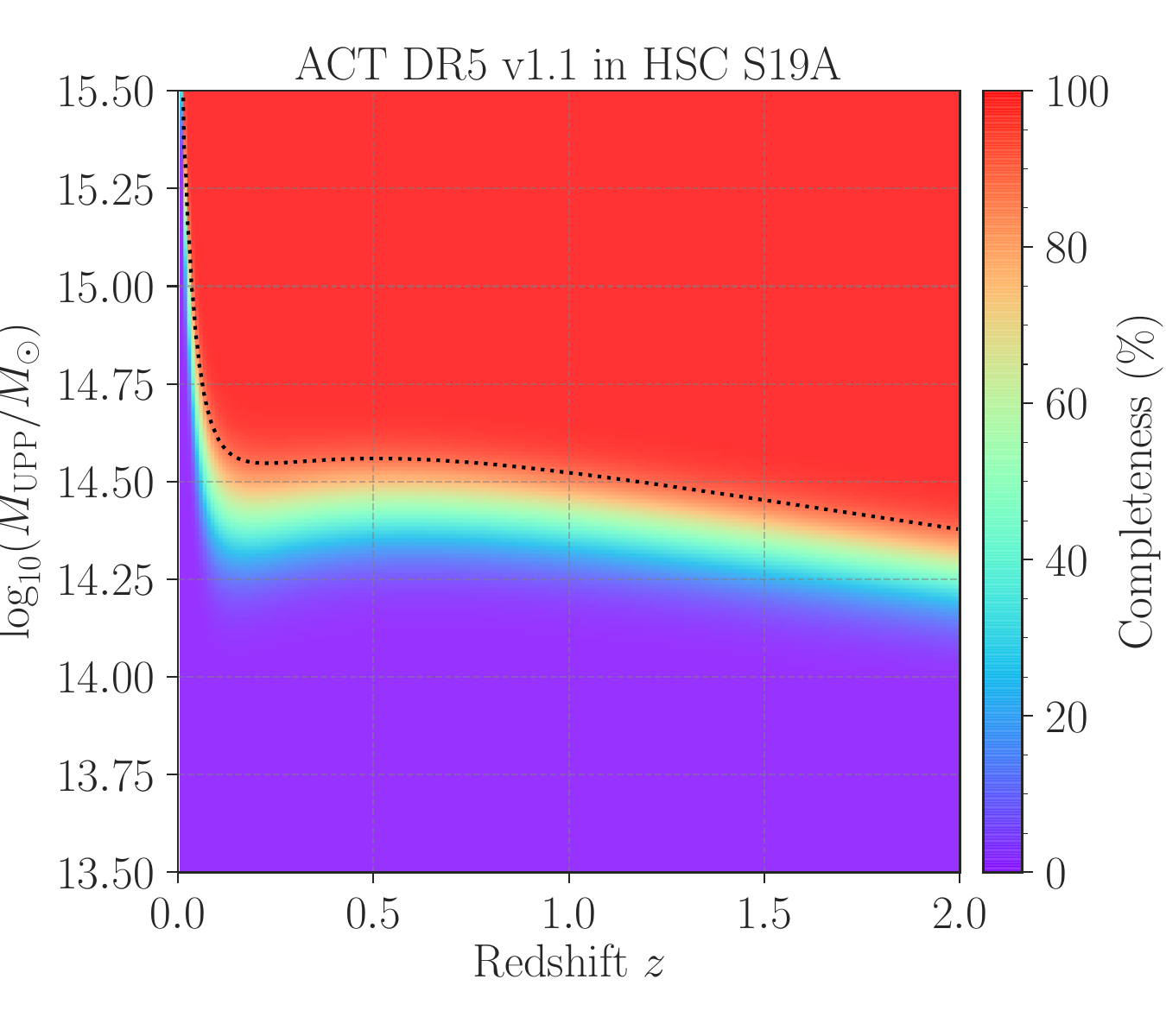}
\cprotect\caption{\label{fig:ACT_selecfunc} 
The completeness for ${\tt fixed \verb|_| SNR} > 5$ as a function
of redshift, in terms of the mass parameter in 
the UPP model $M_\mathrm{UPP}$.  This is computed for sources in the HSC S19A footprint overlapping the ACT DR5 map. 
Note that the UPP model in Ref.~\cite{Arnaud:2009tt} is adopted to estimate the completeness. The dotted line shows the 90\% completeness limit.
}
\end{figure}

\begin{figure}[!t]
\includegraphics[clip, width=1.\columnwidth]{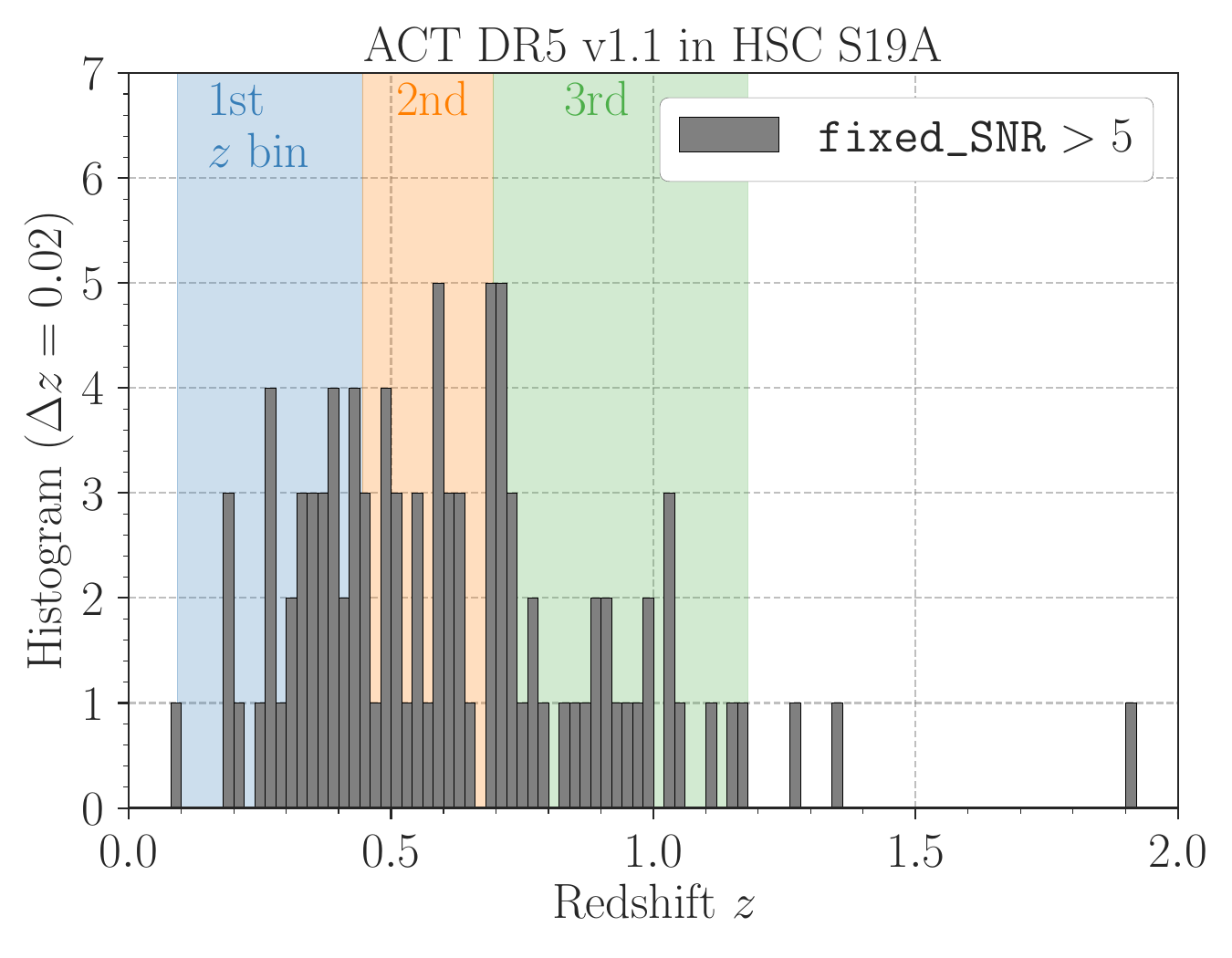}
\cprotect\caption{\label{fig:ACT_zdist} 
The redshift distribution of ACT clusters in the HSC S19A fields.
The gray histogram shows the distribution for objects selected by ${\tt fixed \verb|_| SNR} > 5$.
The colored shaded regions mark the boundaries in cluster redshifts used to define the three subsamples for our stacked lensing analysis. We set the boundary so that the number of each subsample is equal.
}
\end{figure}

In the HSC observations acquired from March 2014 to April 2019 (referred to as S19A),
the full-depth full-color (FDFC) footprint has $474.43\, \mathrm{deg}^2$ of overlap 
with the ACT DR5 cluster search area.
Fig.~\ref{fig:survey_overlap} represents the angular distribution of our cluster sample 
in five HSC S19A regions of interest, named XMM, VVDS, GAMA09H, GAMA15H, and WIDE12H.
We then estimate the completeness of the ACT cluster search in terms of the mass parameter in the UPP model $M_\mathrm{UPP}$ (denoted as the SZ mass hereafter) using Monte Carlo simulations as described in Ref.~\cite{ACT:2020lcv}.
It is beneficial to note that $M_\mathrm{UPP}$ is a fiducial mass estimate
and we use it for ease of comparison with previous works.
Fig.~\ref{fig:ACT_selecfunc} shows the completeness as a function of redshift and $M_\mathrm{UPP}$
across the HSC S19A regions\footnote{The completeness in this paper is computed with the Monte Carlo code as of August 2021. Since then, the code has been updated so that it improves beam convolutions and in-painting objects, introducing small differences in the completeness. We have confirmed that the latest completeness changes our model predictions of lensing signals at a $<2\%$ level, but this does not affect our final results significantly.}. 
In the figure, the dotted line highlights the 90\% mass completeness limit, 
which is similar to the equivalent limit when averaging over the whole 13,211 $\mathrm{deg}^2$ ACT cluster search area.
\rev{It is also worth noting that about 95\% of ACT clusters in the HSC S19A regions has been optically confirmed when our selection is adopted.}
Note that we confirmed that every cluster in our sample in the HSC S19A regions has complete redshift follow-up.
For our stacked lensing analysis, we divide our cluster sample into three subsamples by the cluster redshifts 
so that the number of objects in each subsample is equal (32). 
The gray histogram in Fig.~\ref{fig:ACT_zdist} shows the redshift distribution of ACT clusters 
in the HSC S19A regions with ${\tt fixed\verb|_|SNR} > 5$.
We then construct three equal subsamples by imposing limits $0.092<z\le0.445$, $0.445<z\le0.695$ and $0.695<z\le1.180$, as highlighted in colored shaded regions in Fig.~\ref{fig:ACT_zdist}. The average redshifts of the three samples are given by 0.324, 0.567, and 0.878, respectively.
For reference, the three samples have the average SZ masses in units of $10^{14}\, h^{-1}M_\odot$ of 2.41, 2.39, and 2.09 from lower to higher redshift bins. 

\subsection{HSC Y3 source galaxies}

HSC is a wide-field optical imager installed at the Subaru Telescope 
\cite{2018PASJ...70S...3F, 2018PASJ...70S...1M}.
Under the Subaru Strategic Survey Program (SSP; \cite{Aihara:2017paw}), 
HSC conducted a deep multi-band imaging survey with 
a target sky coverage of 1,400 $\mathrm{deg}^2$, starting in 2014.
The 330-night observations in the HSC SSP program have been completed.
In this paper, we use the HSC third-year galaxy shape catalog (HSC Y3; \cite{Li:2021mvq}) 
based on the HSC-SSP imaging 
data taken from March 2014 through April 2019. 
The HSC Y3 data consist of six distinct fields (HECTOMAP, GAMA09H, WIDE12H, GAMA15H, XMM, VVDS) and
cover $433.48 \, \mathrm{deg}^2$ in total. 
We ignore the
HECTOMAP field since it does not overlap with the ACT DR5 map.
The effective number density of source galaxies marginally depends on 
the full width at half maximum of $i$-band Point Spread Function (PSF)
reaching $19.9\, \mathrm{arcmin}^{-2}$ on average.
More details of the Y3 shape catalog are found in Ref.~\cite{Li:2021mvq}, 
but we summarize the basic information on the HSC Y3 source galaxies below.

The galaxy shapes were estimated on coadded $i$-band images
by the re-Gaussianization PSF correction method \cite{Hirata:2003cv}.
This moment-based method has been applied to SDSS imaging data 
and systematic errors have been extensively characterized \cite{Mandelbaum:2005wv, Mandelbaum:2012ay}. 
In the HSC Y3 shape catalog, the shapes (ellipticities to be more precise) of individual galaxies are defined by 
\beqa
(e_1, e_2) = \frac{1-(b/a)^2}{1+(b/a)^2}(\cos 2\phi, \sin 2\phi), \label{eq:ellipticity}
\eeqa
where $\phi$ is position angle and $b/a$ is the minor-to-major axes ratio.
Eq.~(\ref{eq:ellipticity}) provides an estimator of lensing distortion \cite{1991ApJ...380....1M}
and one needs to introduce a multiplicative correction of the shape to estimate lensing shear $\gamma$ as \cite{Bernstein:2001nz}
\beqa
\hat{\gamma}_\alpha = \frac{1}{2{\cal R}}\langle e_\alpha \rangle, \label{eq:shape2shear}
\eeqa
where $\alpha=1,2$ are the indices for the galaxy shapes and lensing shear, 
$\langle \cdots \rangle$ indicates a weighted average and $\cal R$ is called the shear responsivity.
An inverse variance weight is commonly used for the average in Eq.~(\ref{eq:shape2shear}) and it is given by
\beqa
w_i = \frac{1}{e^2_{\mathrm{RMS},i} + \sigma^2_{\mathrm{e},i}}, \label{eq:weight_hsc_shape}
\eeqa
where $e_{\mathrm{RMS},i}$ and $\sigma_{\mathrm{e},i}$ are 
the intrinsic shape noise and the error in the shape measurement for the $i$-th galaxy, respectively.
The measurement error $\sigma_\mathrm{e}$ is statistically estimated from the shape measurements performed on
simulated images and the intrinsic shapes are derived
by subtracting the measurement error 
from the ellipticity dispersion measured from the real data.
The shear responsivity for a given source population is then computed as
\beqa
{\cal R} = 1 - \frac{\sum_i w_i e^2_{\mathrm{RMS},i}}{\sum_i w_i}.
\eeqa
To eliminate possible biases in the shear estimation in the re-Gaussianization method, 
the galaxy shapes were further calibrated against image simulations 
generated with the open source software {\tt GalSim} \cite{2015A&C....10..121R}. 
We refer the reader to Ref.~\cite{Mandelbaum:2017ctf} for details of the production of image simulations.
The calibration assumes that Eq.~(\ref{eq:shape2shear}) can provide a biased estimator of the true shear expressed as 
\beqa
\hat{\gamma}_{\alpha,\mathrm{obs}} = (1 + m) \gamma_{\alpha, \mathrm{true}} + c_\alpha, 
\eeqa
where $\gamma_{\alpha, \mathrm{true}}$ represents the true shear, $m$ and $c_\alpha$ are correction factors,
referred to as the multiplicative and additive biases, respectively.
Extensive use of HSC-like image simulations allows us to calibrate $m$ and $c_\alpha$ 
as a function of galaxy properties.
For our stacked lensing analysis, we properly include the correction for both multiplicative and additive biases to produce an unbiased estimate of lensing shear as described in Section 3.1.2 and 3.1.3 in Ref.~\cite{Li:2021mvq}.

Five-band photometry in the HSC SSP allows us to estimate redshifts of the HSC Y3 source galaxies with photometric redshift (photo-z) measurements (see Ref.~\cite{Tanaka:2017lit} for the photo-z statistics in the first-year HSC data).  Among several photo-z methods, we use a neural network-based photo-z conditional
density estimation code ${\tt dnnz}$, as our fiducial choice. 
The ${\tt dnnz}$ architecture consists of multilayer perceptrons with 5 hidden layers. 
The training uses ${\tt cmodel}$ fluxes, unblended convolved fluxes, PSF fluxes, and galaxy shape information. The construction of the conditional density uses 100 nodes in the output layer, each representing a redshift histogram bin spanning the range $z = 0$ to $7$ (see Nishizawa et al. in prep. for details of ${\tt dnnz}$).
To define background source galaxies behind a given ACT cluster, we apply the P-cut method \cite{Oguri:2014eba} by using the ${\tt dnnz}$ estimates of the posterior redshift distribution of each HSC Y3 galaxy.
More detail about the P-cut method is provided in Section~\ref{subsec:background_selection}. 

\section{Stacked lensing measurements}\label{sec:lensing}

\subsection{Estimator}\label{subsec:dSigma_estimator}

Gravitational lensing effects by foreground clusters introduce a coherent distortion of observed shapes of background source galaxies. In the weak lensing regime, lensing effects by the cluster's gravitational potential 
leave a tangential distortion in galaxy shapes after one takes the azimuthal average of the source shapes 
around the cluster center.
For a source galaxy at a comoving transverse separation $R$ from the lens center,
the azimuthal averaged profile of tangential distortion $\gamma_t$ can be expressed as (see Ref.~\cite{Umetsu:2020wlf} for a review)
\beqa
\gamma_t(R) = \frac{\bar{\Sigma}(<R)-\Sigma(R)}{\Sigma_\mathrm{cr}(z_l, z_s)} \equiv \frac{\Delta \Sigma(R)}{\Sigma_\mathrm{cr}(z_l, z_s)}, \label{eq:gamma_t}
\eeqa
where $\Sigma(R)$ is the surface mass density of the lens cluster, 
$\bar{\Sigma}(< R)$ is the average projected mass density inside $R$,
and $\Sigma_\mathrm{cr} (z_l, z_s)$ represents the critical surface mass density which depends on
the redshifts of the source galaxy, $z_s$, and the lens, $z_l$.
For a single lens-source pair, 
$\Sigma_\mathrm{cr}(z_l, z_s)$ is defined as
\beqa
\Sigma_\mathrm{cr}(z_l, z_s) = \frac{c^2}{4\pi G}\frac{D_A(z_s)}{(1+z_l)^2 D_A(z_l) D_A(z_l, z_s)}, \label{eq:crit_dens_lens}
\eeqa
where 
$D_A(z_s)$, $D_A(z_l)$, and $D_A(z_l, z_s)$ represent 
the angular diameter distances of the source and lens from us, and between the two, respectively.
Note that the factor $(1+z_l)^{-2}$ in Eq.~(\ref{eq:crit_dens_lens}) arises from 
our use of comoving coordinates \cite{Mandelbaum:2006pw}.

Eq.~(\ref{eq:gamma_t}) motivates us to measure the excess surface mass density $\Delta \Sigma$
from real data on the background galaxies' shapes.
In this work, we aim at measuring the average of $\Delta \Sigma$ over a set of foreground clusters 
by stacking tangential components of the shapes of background galaxies, $e_t$.
In practice, the estimator of $\Delta \Sigma$ for our cluster sample is given by 
(e.g. see Ref.~\cite{Miyatake:2021sdd} for a similar analysis)
\beqa
\widehat{\Delta \Sigma}(R_i) &=& 
\frac{\sum_{\mathrm{ls}\in R_i} w_\mathrm{ls} \langle \Sigma^{-1}_\mathrm{cr}\rangle_\mathrm{ls} \left[e_{t,\mathrm{s}}/\left(2{\cal R}(R_i)\right)-c_{t,\mathrm{s}}\right]}{\left[1+K_\mathrm{sel}(R_i)\right]\left[1+K(R_i)\right]\sum_{\mathrm{ls}\in R_i}w_\mathrm{ls}}\Bigg{|}_{R_i} \nonumber \\
&&
\qquad
- (\mathrm{signal\, around\, random\, points})|_{R_i}, \label{eq:estimator_dSigma}
\eeqa
where the summation ``ls" run over all lens-source pairs that lie in the i-th radial bin 
$R_i \equiv D_A(z_l) \Delta \theta_\mathrm{ls}$; 
$D_A(z_l)$ is the comoving angular diameter distance to the l-th ACT lens cluster at the redshift $z_l$, 
and $\Delta \theta_\mathrm{ls}$ is the angular separation between lens 
and source in each pair; 
$e_{t,\mathrm{s}}$ is the tangential component of ellipticity of
the s-th HSC source galaxy; 
$c_{t,\mathrm{s}}$ is the additive
shear calibration factor as given in the HSC shape catalog;
$\langle \Sigma^{-1}_\mathrm{cr}\rangle_\mathrm{ls}$ 
is the average of the inverse critical surface mass density-weighted with
the photo-z posterior distribution function of each source
galaxy, $P_s(z_s)$, behind the l-th lens cluster:
\beqa
\langle \Sigma^{-1}_\mathrm{cr}\rangle_\mathrm{ls}
=\frac{\int_0^{z_\mathrm{max}}\mathrm{d}z_s\, P_s(z_s)\, \Sigma^{-1}_\mathrm{cr}(z_l, z_s)}{\int_0^{z_\mathrm{max}}\mathrm{d}z_s\, P_s(z_s)}, \label{eq:avg_inv_crit_dens_lens}
\eeqa
where $z_\mathrm{max}=7$ as given by ${\tt dnnz}$.
Note that we set $\Sigma^{-1}_\mathrm{cr} = 0$  in Eq.~(\ref{eq:avg_inv_crit_dens_lens}) when $z_s < z_l$. 
The factor $w_\mathrm{ls}$ in Eq.~(\ref{eq:estimator_dSigma}) represents 
the weight, for which we employ an inverse-variance weighting that is nearly optimal
in the shape-noise-dominated regime \cite{SDSS:2003slg, Mandelbaum:2005wv, Mandelbaum:2012ay, Shirasaki:2018cgl}:
\beqa
w_\mathrm{ls} = w_\mathrm{l} w_\mathrm{s} \langle \Sigma^{-1}_\mathrm{cr}\rangle_\mathrm{ls}^2
\eeqa
where $w_\mathrm{s}$ is the weight in the HSC shape catalog as in Eq.~(\ref{eq:weight_hsc_shape})
and we adopt $w_\mathrm{l} = 1$ for the weight of the ACT clusters.

In Eq.~(\ref{eq:estimator_dSigma}), we introduce the shear responsivity for the stacked lensing signal at a given radial bin as
\beqa
{\cal R}(R_i) = 1-\frac{\sum_{\mathrm{ls}\in R_i} w_\mathrm{ls} e^2_\mathrm{RMS,s}}{\sum_{\mathrm{ls}\in R_i} w_\mathrm{ls}}.
\eeqa
Also, we adopt the correction of the multiplicative bias in the shape measurements as
\beqa
1+K(R_i) = 1+\frac{\sum_{\mathrm{ls}\in R_i} w_\mathrm{ls}\, m_\mathrm{s}}{\sum_{\mathrm{ls}\in R_i} w_\mathrm{ls}}, \label{eq:mbias_correction}
\eeqa
where $m_\mathrm{s}$ is the multiplicative bias for the s-th source galaxy.  The factor of $1+K_\mathrm{sel}(R_i)$ accounts for the effect of selection bias.
We set $1+K_\mathrm{sel}(R_i)$ by following the procedure as in Section 3.1.4 in Ref.~\cite{Li:2021mvq}.
Note that Eq.~(\ref{eq:mbias_correction}) corrects a 1\%-level bias in our measurements.

Finally, the second term on the right-hand side of Eq.~(\ref{eq:estimator_dSigma}) 
denotes the signal around a set of random positions,
which is measured by replacing lens clusters with random points. 
We need to subtract this random signal to correct for observational systematic effects such as 
residual systematics in shape measurements due to an imperfect correction 
of optical distortions across the field of view. The number of random points is 100 times as large as that of lens clusters, where the random catalogs are generated assuming the hydrostatic equilibrium and analytic halo mass function in Ref.~\cite{Tinker:2008ff}.
We should point out here that this is done with the purpose of reproducing the redshift distribution because it impacts source selection, but the mass distribution is irrelevant.
In the end, the redshift distribution of the random points is slightly different from the counterpart of the actual ACT clusters.
To correct the difference in the redshift distribution in our lensing analysis, we set the weight for the random points (denoted as $w_\mathrm{r}$) as
\beqa
w_r(z) = \frac{{\cal H}_\mathrm{cl}(z;z_\mathrm{b},\Delta)}{{\cal H}_\mathrm{r}(z; z_\mathrm{b}, \Delta)},
\eeqa
where ${\cal H}_\mathrm{cl}(z;z_\mathrm{b},\Delta)$ is the histogram for the number of the clusters at their redshift range of $z_\mathrm{b}-\Delta/2 \le z < z_\mathrm{b} + \Delta/2$,
${\cal H}_\mathrm{r}(z;z_\mathrm{b},\Delta)$ is the histogram for the random points,
and each histogram is normalized to unity.
To compute the histogram, we employ 200 linear-spaced bins in the range of $0\le z<2$.
We found that the signal around random points starts to deviate from zero at $R \simgt 20\, h^{-1}\mathrm{Mpc}$,
which is caused by a finite area coverage in our lensing analysis.

\begin{figure*}[!t]
\includegraphics[clip, width=2.1\columnwidth]{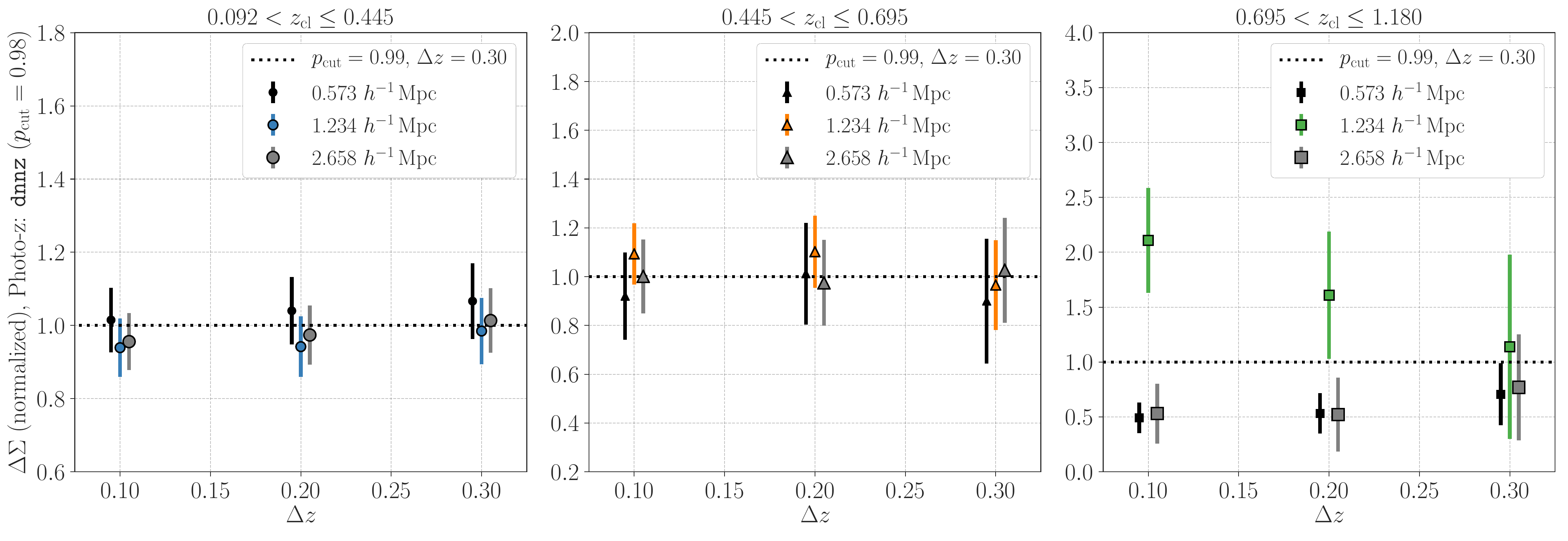}
\caption{\label{fig:Dsigma_varypzcut} 
Stacked lensing signal, $\Delta \Sigma$ (normalized by the signal with $p_\mathrm{cut}=0.99$ and $\Delta z=0.3$), as a function of lens redshift threshold $\Delta z$ in the P-cut selection (see Eq.~[\ref{eq:Pcut}] for the definition). The three panels summarize the results at different cluster redshift bins of 
$0.092<z\le0.445$, $0.445<z\le0.695$ and $0.695<z\le1.180$ from left to right.
In each panel, different symbols represent the measurements at different radial bins centered at $R=0.573\, h^{-1}\mathrm{Mpc}$, $1.234\, h^{-1}\mathrm{Mpc}$ and $2.658\, h^{-1}\mathrm{Mpc}$. 
This range of $R$ approximates the virial radii of a typical cluster-sized halo with $M \sim 10^{14-15}\, h^{-1}M_\odot$.
The horizontal dashed line indicates the most stringent selection criteria ($p_\mathrm{cut}=0.99$ and $\Delta z=0.3$). The error bars in each panel are set by the shot noise due to intrinsic scatters in galaxy shapes.
}
\end{figure*}

Throughout this paper, we employ a logarithmic binning with 18 radial bins ranging from $R=0.1\, h^{-1}\mathrm{Mpc}$ to $R=100\, h^{-1}\mathrm{Mpc}$ for our stacked lensing analysis.
For our fiducial choice, we define the center of each cluster as the position of the brightest cluster galaxy (BCG).
In Appendix~\ref{apdx:test_dSigma_measurement}, we perform several systematic tests on our stacked lensing analyses, including the divergence-free B-mode signal around the lens clusters, the B-mode signal around random points, 
and the measurement of the boost factor 
that arises from contamination of the source galaxy sample by galaxies physically associated with lens clusters themselves, since the source photo-z measurements may be imperfect. 
We did not find any evidence for such residual systematic effects in our stacked lensing measurement.

Apart from the null tests, we also examine how sensitive our stacked lensing analysis is to (i) the definition of cluster centers and (ii) the choice of photo-z methods (see Appendix~\ref{apdx:test_dSigma_measurement}).
We found that the magnitude of stacked lensing signal with the center inferred from the thermal SZ analysis (the SZ center) tends to become smaller than that using the BCG positions at $R\simlt 0.3\, h^{-1}\mathrm{Mpc}$ when stacking clusters in the lowest redshift bin (the difference is at a $\sim1\sigma$ level).  In contrast, the choice of the cluster centers does not introduce a significant effect in our stacked lensing analysis for the other two cluster redshift bins.
To investigate the photo-z impact, we measure $\Delta \Sigma$ by using photo-z information estimated with a different photo-z code, ${\tt mizuki}$, which uses template fitting and physical priors about galaxy properties \cite{2015ApJ...801...20T}.
Note that our default photo-z code ${\tt dnnz}$ takes a data-driven approach using a neural network, while ${\tt mizuki}$ falls into the category of intuitive photo-z methods with informative priors about galaxy physics. 
In this sense, the comparison of $\Delta \Sigma$ with the ${\tt dnnz}$ photo-z 
and the ${\tt mizuki}$ counterpart
enables us to bracket a typical uncertainty due to the choice of photo-z methods. 
Imposing the background selection as described in Section~\ref{subsec:background_selection},
we did not find any significant differences in our stacked lensing signals when 
using the photo-z from ${\tt dnnz}$ or ${\tt mizuki}$.

\subsection{Selecting background sources}\label{subsec:background_selection}

If member galaxies in a cluster of interest are misidentified as background galaxies,
they will introduce a systematic dilution in the $\Delta \Sigma$ measurement for that cluster.
To make a secure selection of background sources, we adopt the P-cut method as introduced in Ref.~\cite{Oguri:2017khw}. 
The method utilizes the photo-z posterior distributions of individual source galaxies and selects the source galaxy when it satisfies the condition below:
\beqa
p_\mathrm{cut} < \int_{z_{l}+\Delta z}^{7.0}\, \mathrm{d}z_s \, P_s(z_s), \label{eq:Pcut}
\eeqa
where $z_l$ is the redshift of the l-th cluster, 
$P_s(z_s)$ is the photo-z posterior distribution of the s-th source galaxy, 
$p_\mathrm{cut}$ and $\Delta z$ are parameters in the P-cut selection.
Ref.~\cite{Medezinski:2017eex} have extensively examined the P-cut selection by varying the parameters 
for optically-selected clusters and source galaxies in previous HSC data, 
finding that $p_\mathrm{max} = 0.98$ and $\Delta z =0.2$ 
are a reasonable choice to mitigate the dilution effect. 

We here investigate the choice of 
$p_\mathrm{cut} = 0.98$ and $\Delta z =0.2$ with the HSC Y3 source galaxies and the ACT clusters.
We perform the stacked lensing measurements by using the ${\tt dnnz}$ photo-z information, while
varying the parameters $p_\mathrm{cut}$ and $\Delta z$ used in the background selection.
We compare our fiducial $\Delta \Sigma$ signals with those obtained by assuming $p_\mathrm{cut} = 0.99$ and $\Delta z =0.3$ to examine
possible systematic effects in the P-cut background selection.
Fig.~\ref{fig:Dsigma_varypzcut} shows $\Delta \Sigma$ in 
three different radial bins, varying the parameter $\Delta z$ but fixing $p_\mathrm{cut}=0.98$.  These are compared to $\Delta \Sigma$ calculated with $p_\mathrm{cut} = 0.99$ and $\Delta z =0.3$.
As seen in the figure, we find that the selection with $p_\mathrm{cut} = 0.98$ and $\Delta z =0.2$ does not introduce any significant dilution effects compared to the more stringent selection criteria ($p_\mathrm{cut} = 0.99$ and $\Delta z =0.3$) at the cluster redshifts of $0.092<z\le 0.445$ and $0.445<z\le 0.695$.
In the highest redshift bin of $0.695<z\le 1.180$, we see a trend showing possible dilution in our stacked lensing measurement when adopting $p_\mathrm{cut} = 0.98$ and $\Delta z =0.2$, while  the choice of $p_\mathrm{cut} = 0.98$ and $\Delta z =0.3$ provides more converged results.
Given the results of this experiment, we decide to use $\Delta z=0.2$ for the first and second cluster redshift bins, but $\Delta z=0.3$ is applied to the $\Delta \Sigma$ measurement at the highest redshift bin. 

Because the P-cut method is applied to every single lens-source pair, it is not suitable to define a single population of background galaxies.
Nevertheless, it is worth computing a representative number density of background source galaxies 
by using a single value of the lens redshift.
Following the definition of the effective source number density $\bar{n}_\mathrm{gal}$ in Ref.~\cite{Chang:2013xja}, we find $\bar{n}_\mathrm{gal}=10.8$, $6.48$ and $0.40\, \mathrm{arcmin}^{-2}$
when applying the P-cut selection for the lens redshifts of $0.324$, $0.567$, and $0.878$, respectively.
Note that the small source number density at the highest cluster redshift bin prevents us from measuring $\Delta \Sigma$ at $R \simlt 0.3\, h^{-1}\mathrm{Mpc}$ (there are no pairs of random points and source galaxies inside this radius after we impose the P-cut selection).


\subsection{Statistical uncertainty}\label{subsec:covariance}

\begin{figure*}[!t]
\includegraphics[clip, width=2.1\columnwidth]{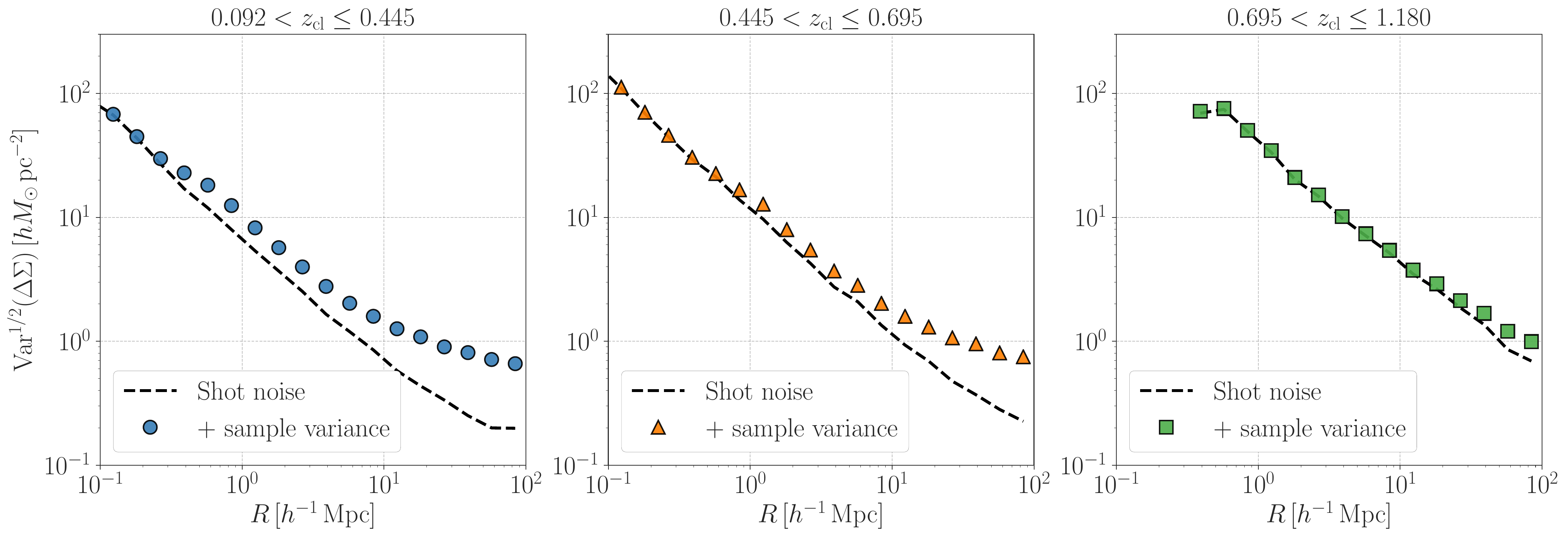}
\caption{\label{fig:Dsigma_variance} 
The diagonal components of the covariance matrix for measurements of the stacked lensing profiles for
the ACT clusters. We used 1404 mock catalogs of the ACT clusters and HSC-Y3 source galaxies 
to estimate the sample variance (see text for details). Three panels summarise the results at different cluster redshift bins ($0.092<z\le0.445$, $0.445<z\le0.695$, and $0.695<z\le 1.180$ from right to left).
The colored symbols in each panel show the full covariance for the lensing profile, while the dashed lines represent the shape noise contribution alone for comparison.
}
\end{figure*}


Precise estimations of the statistical error in $\Delta \Sigma$ measurements are essential
for a robust characterization of ACT clusters. 
In this paper, we follow the method given in Refs.~\cite{Murata:2017zdo, Miyatake:2020uhg}
to determine the covariance matrix of statistical errors for our stacked lensing analysis.
The covariance matrix in $\Delta \Sigma$ measurements can be decomposed into two parts:
\beqa
\mathrm{Cov}_{ij} = \mathrm{Cov}^\mathrm{int}_{ij} + \mathrm{Cov}^\mathrm{sample}_{ij}, \label{eq:cov_dSigma}
\eeqa
where the indices of $i$ and $j$ denote the i- and j-th radius bins,
$\mathrm{Cov}^\mathrm{int}_{ij}$ represents the statistical uncertainty due to intrinsic scatter in galaxy
shapes (so-called shape noise), and $\mathrm{Cov}^\mathrm{sample}_{ij}$ stands for the sample variance arising from uncorrelated large-scale structures along a line of sight, inherent variations in mass density distributions inside clusters, 
and modulation effects in underlying density fields inside our survey window caused by density fluctuations at scales longer than the survey window \cite{Takada:2013wfa, Takahashi:2018pze}.

To estimate the first term of the right-hand side in Eq.~(\ref{eq:cov_dSigma}), 
we use a random rotation of the HSC Y3 source galaxy shapes and real ACT clusters.
It is worth noting that we can remove the real lensing signal 
by rotating the ellipticity orientation of each source galaxy.
We perform the stacked lensing measurements with randomly-rotated shapes of HSC Y3 galaxies and the ACT clusters in the same manner as in the actual measurement, 
including the corrections of multiplicative bias discussed in Section~\ref{subsec:dSigma_estimator}.
We repeat the process of random rotation and stacked lensing analysis to produce 1000 realizations of $\Delta \Sigma$, allowing us to estimate the shape noise contribution to the covariance matrix.

To find the sample covariance of $\mathrm{Cov}^\mathrm{sample}_{ij}$ in Eq.~(\ref{eq:cov_dSigma}),
we use a set of mock catalogs of the HSC Y3 source galaxy shapes and ACT clusters.
To construct the mock catalogs in a light-cone volume,
we use 108 full-sky, light-cone simulations generated in Ref.~\cite{Takahashi:2017hjr}. 
As described in Appendix~\ref{apdx:mock} in detail, we populate
HSC-like galaxies and ACT-like clusters in the full-sky, light-cone
simulation. 
Since the HSC-Y3 data still has a small area coverage
(about 433 square degrees), we identify 13 footprints of the HSC-Y3 data in each of the all-sky, light-cone simulation realizations. We thus generate $108\times 13=1404$ mock catalogs of the HSC data and the ACT clusters in total.

We then perform measurements of $\Delta \Sigma$ from each realization
using the same analysis pipelines that are used in the actual measurements of the real HSC and ACT data.
Because the intrinsic scatter and measurement error 
in galaxy shapes does not affect the sample covariance of $\mathrm{Cov}^\mathrm{sample}_{ij}$, 
we use a modified version of $\Delta \Sigma$ estimator instead of Eq.~(\ref{eq:estimator_dSigma}):
\beqa
\widehat{\Delta \Sigma}^\prime(R_i) &=& 
\frac{\sum_{\mathrm{ls}\in R_i} w_\mathrm{ls} \langle \Sigma^{-1}_\mathrm{cr}\rangle_\mathrm{ls} \gamma^\mathrm{mock}_{t,\mathrm{s}}}{\sum_{\mathrm{ls}\in R_i}w_\mathrm{ls}}\Bigg{|}_{R_i} \nonumber \\
&&
\quad
- (\mathrm{signal\, around\, random\, points})|_{R_i}, \label{eq:modified_estimator_dSigma}
\eeqa
where $\gamma^\mathrm{mock}_{t,\mathrm{s}}$ represents the tangential component of lensing shear for 
the s-th HSC mock galaxies.
Finally, we estimate the sample covariance for the stacked lensing signal $\Delta \Sigma$ at three cluster redshift bins, from the scatters among the 1404 measurements of $\widehat{\Delta \Sigma}^\prime$.

Fig.~\ref{fig:Dsigma_variance} shows the diagonal components of the covariance matrix.
In each panel, dashed lines highlight the shape-noise covariance decreasing with radius as $1/R$.
The colored symbols in each panel show the full variance including the sample variance term, making the variance larger at $R\simgt 10\, h^{-1}\mathrm{Mpc}$ for the first and second cluster redshift bins.
Note that the shape-noise variance is dominant in the statistical error of our $\Delta \Sigma$ measurement at the highest cluster redshift bin because of a secure background selection of the HSC source galaxies.
We also find that the off-diagonal components in the $\Delta \Sigma$ covariance become prominent at $R\simgt 3\, h^{-1}\mathrm{Mpc}$ for the first and second cluster redshift bins, while they are only important at $R\simgt 10\, h^{-1}\mathrm{Mpc}$ for the highest cluster redshift bin. 



\section{model and inference}\label{sec:analysis_method}

In this section, we describe a theoretical model to predict our lensing measurements. Subsection~\ref{subsec:halo_model} provides key ingredients in our halo-based model, including the definition of halo masses, the scaling relation between the SZ and halo masses, the halo-model expression of our observables, and the choice of model parameters. Compared with our previous work \cite{Miyatake:2018lpb}, we adopt an informative prior of the scatter in the scaling relation extracted from a hydrodynamical simulation suite of Illustris TNG \cite{Springel:2017tpz}. In Subsection~\ref{subsec:baryonic_effects}, we explain how to incoorporate possible baryonic effects with our model by using a semi-analytic method \cite{Schneider:2015wta, Schneider:2018pfw, Giri:2021qin}. Subsection~\ref{subsec:param_dependence} gives some physical intuition about the model-parameter dependence of stacked lensing profiles $\Delta \Sigma$ based on our model. We then summarize the setup of our likelihood analysis in Subsection~\ref{subsec:likelihood}.

\subsection{Halo-based forward model}\label{subsec:halo_model}

In this paper, we adopt a forward modeling approach to predict cluster observables.
Our modeling pipeline relies heavily on the Dark Emulator, an efficient and 
accurate emulation tool for statistics of dark matter halos \cite{Nishimichi:2018etk}.
The Dark Emulator enables us to predict the halo mass function, halo autocorrelations, 
and halo-mass cross-correlations as functions of redshift, halo masses, and 
cosmological parameters in modest CPU time.
As in the Dark Emulator, we adopt $M \equiv M_\mathrm{200b} = 4\pi (R_\mathrm{200b})^3 \bar{\rho}_\mathrm{m0} \times 200/3$ for our fiducial halo mass definition, where $R_\mathrm{200b}$ is 
the spherical halo boundary radius within which the mean mass density is 
200 times the present-day mean mass density.
Note that the mass $M$ is not necessarily equal to the spherical overdensity mass, $M_\mathrm{200b}$, 
when accounting for baryonic effects as described in Section~\ref{subsec:baryonic_effects}.
Another spherical overdensity mass of $M_\mathrm{500c}$ is widely adopted for characterizations of cluster masses in the literature. Within a halo-model framework provided by the Dark Emulator, we define $M_\mathrm{500c}$ as
\beqa
M_\mathrm{500c} &\equiv& \frac{4\pi}{3} \, 500\, \rho_\mathrm{crit}(z)\, R^3_\mathrm{500c}\, (1+z)^{-3} \label{eq:def_M500c_base} \\
&=&
\int_0^{R_\mathrm{500c}}\mathrm{d}r\, 4\pi r^2 \, \rho_\mathrm{hm}(r, M, z, \mathbf{c}), \label{eq:def_M500c}
\eeqa
where 
$\rho_\mathrm{crit}(z)$ is the critical density in the universe at a redshift $z$,
$R_\mathrm{500c}$ is the halo boundary radius in comoving coordinates
and $\rho_\mathrm{hm}(r, M, z, \mathbf{c})$ represents the average halo mass density profile at the halo-centric radius $r$, as a function of the redshift $z$, halo mass $M$ and a set of cosmological parameters $\mathbf{c}$, given by the Dark Emulator. 
In the following, we omit the cosmological parameter dependence of halo statistics 
for simplicity.

Given inputs from the Dark Emulator, a model prediction of our stacked lensing measurements contains the integral of 
\beqa
&&\int \mathrm{d}\log M_\mathrm{UPP} \int \mathrm{d}M\, S(\log M_\mathrm{UPP}, z)\, {\cal P}(\log M_\mathrm{UPP} | M, z) \nonumber \\
&&
\qquad \qquad \qquad \qquad 
\times 
n_h(M, z)\, \Delta \Sigma_h(R, M, z), \label{eq:Dsigma_model_keys}
\eeqa
where $S(\log M_\mathrm{UPP}, z)$ represents the selection function of our cluster sample,
${\cal P}(\log M_\mathrm{UPP} | M, z)$ sets
the probability distribution of the SZ mass, 
$M_\mathrm{UPP}$, for halos with a given mass $M$ at redshift $z$,
$n_h(M,z)$ and $\Delta \Sigma_h(R, M, z)$
are the halo mass function and the excess surface mass density profile for a single halo,
respectively.
The Dark Emulator can predict two of $n_h(M,z)$ and 
$\Delta \Sigma_h(R, M, z)$ in an efficient way, while
the functional form of ${\cal P}(\log M_\mathrm{UPP} | M, z)$ must be given in advance.
A detailed description of two key ingredients ${\cal P}(\log M_\mathrm{UPP} | M, z)$ and $\Delta \Sigma_h(R, M, z)$ is found below.

\subsubsection*{Scaling relation between the SZ and halo masses}

Following our previous work \cite{Miyatake:2018lpb}, we assume a log-normal distribution 
for ${\cal P}(\log M_\mathrm{UPP}|M)$;
\beqa \label{eq:scatter}
&&
{\cal P}(\log M_\mathrm{UPP}|M) \, \mathrm{d}\log M_\mathrm{UPP} = 
\frac{\mathrm{d}\log M_\mathrm{UPP}}{\sqrt{2\pi}\sigma_{\log M_\mathrm{UPP}}} \nonumber \\
&&
\qquad \qquad 
\times
\exp\left[-\frac{[\log M_\mathrm{UPP} - \log M^\prime_\mathrm{UPP}(M)]^2}{2\sigma^2_{\log M_\mathrm{UPP}}}\right], \label{eq:prob_MUPP_given_M}
\eeqa
where $\log M^\prime_\mathrm{UPP}(M)$ sets the mean relation of $\log M_\mathrm{UPP}$ given halo masses 
and $\sigma_{\log M_\mathrm{UPP}}$ represents the rms in the log-normal distribution.
Because the SZ mass $M_\mathrm{UPP}$ is introduced in terms of the spherical overdensity mass $M_\mathrm{500c}$,
we parameterize the mean relation of $\log M^\prime_\mathrm{UPP}(M)$ as guided 
by the $M_\mathrm{500c}-M$ relation in Eq.~(\ref{eq:def_M500c}).
To be specific, we adopt the functional form below;
\beqa
\log M^\prime_\mathrm{UPP}(M) &=& 
\log a_0 + \log M \nonumber \\
&&
\qquad \qquad
+a_1 \left(\log M - 14.5\right), \label{eq:mean}
\eeqa
where $M_\mathrm{UPP}$ and $M$ are set in units of $M_\odot$ in Eq.~(\ref{eq:mean}).
For the standard $\Lambda$CDM cosmology, we find that the right-hand side of Eq.~(\ref{eq:mean}) with $0.45<a_0<0.6$ and $a_1 \sim -0.01$ provides a reasonable fit to the Dark-Emulator prediction of 
$M_\mathrm{500c}-M$ relations at $0 < z < 1$. 

Apart from the mean relation, the scatter of $\sigma_{\log M_\mathrm{UPP}}$ has a critical role in comparisons with the model and stacked lensing measurements. Because of a strong degeneracy between the mean and scatter, it is difficult to place a meaningful constraint on the scatter in cluster observable-mass relations with stacked lensing measurements alone \cite{Murata:2017zdo, Wu:2020xxy}. Hence, we introduce an informative prior in the scatter in our model by using the Illustris TNG simulations \cite{Springel:2017tpz}.
We estimate the covariance matrix of gas pressure profiles in cluster-sized halos in the simulations by using the Kernel Localized Linear Regression model \cite{kllr1, kllr2}, which allows us to infer mass-conditioned estimates of the gas pressure profile.
Given the covariance matrix, we can generate random realizations of the gas pressure profile at a fixed halo mass and redshift assuming the log-normality of the pressure profile.
More details about our calibration of the covariance model are given in Appendix~\ref{apdx:scatter_Pgas}.

To estimate the scatter, $\sigma_{\log M_\mathrm{UPP}}$, we take a five-step procedure as follows;
\begin{itemize}
\item[(i)] set a cluster redshift $z$,
\item[(ii)] set a cluster mass $\Upsilon = \log (M_\mathrm{500c} [h^{-1} M_\odot])$ by drawing a uniform random number between $\Upsilon=13.5$ and 16.0,
\item[(iii)] produce a random realization of the gas pressure profile $P_\mathrm{gas}(r)$ with the mean profile being the UPP at the redshift $z$ and logarithmic mass $\Upsilon$ (where $r$ stands for the halo-centric radius and we here assume $M_\mathrm{500c}=M_\mathrm{UPP}$ for simplicity),
\item[(iv)] perform a least chi-square fitting of $P_\mathrm{gas}(r)$ with the UPP profile varying a single mass parameter $M_\mathrm{UPP}$ and, 
\item[(v)] repeat the steps (ii)-(iv) 100,000 times and then estimate the standard deviation of $\log M_\mathrm{UPP, bestfit} - \log M_\mathrm{UPP, input}$ where $M_\mathrm{UPP, input}$ and $M_\mathrm{UPP, bestfit}$
are the input SZ mass parameter given in step (ii) and the best-fit estimate inferred in step (iv), respectively. We then identify the standard deviation with the scatter, $\sigma_{\log M_\mathrm{UPP}}$. 
\end{itemize}

We also examine a possible redshift dependence of $\sigma_{\log M_\mathrm{UPP}}$ with the above procedure by varying the cluster redshift between $z=0.05$ and $z=1.50$. In the end, we find that the estimated scatter from our random realizations of the gas pressure profiles can be described as
\beqa
&&\sigma^2_\mathrm{ref}(\log M_\mathrm{UPP}, z) = 
\alpha(z) \left\{\log(M_\mathrm{UPP}\, [M_\odot])-15\right\} \nonumber \\
&&
\qquad \qquad \qquad \qquad \qquad \qquad \qquad \qquad \qquad
+\beta(z), \label{eq:sigma_ref}\\
&&\alpha(z) = -1.13\times10^{-3} \, z + 5.73\times10^{-3}, \\
&&\beta(z) = -8.79 \times10^{-5} \, z + 1.54\times10^{-2},
\eeqa
where Eq.~(\ref{eq:sigma_ref}) is valid for 
$M_\mathrm{UPP} \simgt 10^{14}\, M_\odot$ and $0.05\le z \le 1.50$.
Although we expect that Eq.~(\ref{eq:sigma_ref}) will provide a reasonable estimate of $\sigma_{\log M_\mathrm{UPP}}$, we introduce a single nuisance parameter, $C_0$, to account for any differences with the simulation-based model and real observables;
\beqa
\sigma_{\log M_\mathrm{UPP}}(\log M_\mathrm{UPP}, z) = C_0 \sigma_\mathrm{ref}(\log M_\mathrm{UPP}, z),
\label{eq:log_MUPP_scatter}
\eeqa
where $C_0$ is our model parameter. In practice, we adopt $M_\mathrm{UPP} = 0.5 M$ in Eq.~(\ref{eq:log_MUPP_scatter}) to set $\sigma_{\log M_\mathrm{UPP}}$, allowing us to maintain the log-normal distribution of Eq.~(\ref{eq:prob_MUPP_given_M}). 

Hence we model the mass scaling relation in Eq.~(\ref{eq:prob_MUPP_given_M}) 
by three model parameters of $a_0, a_1$ and $C_0$. 
We will explore the best-fitting model parameters that 
can reproduce both the abundance and lensing profile measurements simultaneously.
Information on the cluster abundance is required to break 
a degeneracy between $a_0$ and $C_0$.
Because we perform a likelihood analysis in each of 
the three cluster redshift bins independently, 
we ignore possible redshift dependence of the mass scaling relation for simplicity.
It is worth noting that our lensing measurement is not 
designed to extract the mass dependence of the excess surface mass density $\Delta \Sigma$.
Therefore, the posterior distribution of $a_1$ is mostly determined by the prior information.
Nevertheless, there are no strong motivations 
to fix $a_1$ from an observational point of view at present.
To make our analysis conservative, we decide to vary $a_1$ over a wide range. 

\subsubsection*{Cluster number count}

To break a parameter degeneracy between $a_0$ and $C_0$, we include information on the cluster abundance for our parameter inference.
Given the completeness of our cluster search, the cluster abundance in the redshift range of $z_\mathrm{min}\le z \le z_\mathrm{max}$ can be written as
\beqa
&&
N_\mathrm{cl}(z_\mathrm{min}, z_\mathrm{max}) = 4\pi f_\mathrm{sky}\, \int_{z_\mathrm{min}}^{z_\mathrm{max}} \mathrm{d}z\, \frac{c\, D^2_A(z)}{H(z)} \nonumber \\
&&
\times \int \mathrm{d}\log M_\mathrm{UPP} \int \mathrm{d}M\, S(\log M_\mathrm{UPP}, z)\, {\cal P}(\log M_\mathrm{UPP} | M, z) \nonumber \\
&&
\qquad \qquad \qquad \qquad \qquad \qquad \qquad \qquad \qquad
\times
n_h(M, z), \label{eq:Ncl}
\eeqa
where $f_\mathrm{sky} = 0.0115$ is the area fraction of our cluster search region,
$H(z)$ is the Hubble parameter at a redshift $z$,
$D_A(z)$ is the comoving angular diameter distance to the redshift $z$,
$S(\log M_\mathrm{UPP},z)$ is the completeness of our ACT cluster catalog as shown in Fig.~\ref{fig:ACT_selecfunc},
${\cal P}(\log M_\mathrm{UPP}|M,z)$ represents the probability distribution given in Eq.~(\ref{eq:prob_MUPP_given_M}),
and $n_h(M, z)$ is the halo mass function at the redshift $z$ and halo mass $M$ computed by the Dark Emulator.
In Eq.~(\ref{eq:Ncl}), we set lower and upper limits in the double mass integrals to be 
$10^{13.5}\, [h^{-1}M_\odot]$ and $10^{16}\, [h^{-1}M_\odot]$, respectively.

\subsubsection*{Stacked lensing signal}

The stacked lensing profile for halos with mass $M$ and at redshift $z$ 
probes the average radial profile of matter distribution around the halos,
$\rho_\mathrm{hm}(r, M, z)$. 
Because a direct output from the Dark Emulator is 
the cross-correlation function between the halo distribution and the matter density fluctuation field
$\xi_\mathrm{hm}(r, M, z)$,
it is useful to express the average mass density profile in terms of $\xi_\mathrm{hm}(r, M, z)$:
\beqa\label{eq:rho_hm}
\rho_\mathrm{hm}(r, M, z) = \bar{\rho}_\mathrm{m0}\left[1+\xi_\mathrm{hm}(r, M, z)\right].
\eeqa
The cross-correlation is related to the cross-power spectrum $P_\mathrm{hm}(k, M, z)$ 
via the Fourier transform as
\beqa\label{eq:xi_hm}
\xi_\mathrm{hm}(r, M, z) = \bar{\rho}_\mathrm{m0}\int_{0}^{\infty}\frac{k\mathrm{d}k}{2\pi}P_\mathrm{hm}(k, M, z)j_0(kr),
\eeqa
where $j_0(x)$ is the zeroth-order spherical Bessel function.
For a given $P_\mathrm{hm}(k)$, 
the excess surface mass density profile for a single halo is computed as
\beqa\label{eq:DeltaSigma_hm}
\Delta \Sigma_h(R, M, z) = \bar{\rho}_\mathrm{m0}\int_{0}^{\infty}\frac{k\mathrm{d}k}{2\pi}P_\mathrm{hm}(k, M, z)J_2(kR), \label{eq:dSigma_single}
\eeqa
where $J_2(x)$ is the second-order Bessel function.

Taking into account the distribution of halo masses and redshifts for the clusters as well as the cluster completeness, 
we can compute a model prediction for the stacked lensing profile as
\beqa\label{eq:DeltaSigma_pred}
&&
\Delta \Sigma(R; z_\mathrm{min}, z_\mathrm{max}) = \frac{4\pi f_\mathrm{sky}}{N_\mathrm{cl}(z_\mathrm{min}, z_\mathrm{max})}\int_{z_\mathrm{min}}^{z_\mathrm{max}} \mathrm{d}z\, \frac{c\, D^2_A(z)}{H(z)} \nonumber \\
&&
\times \int \mathrm{d}\log M_\mathrm{UPP} \int \mathrm{d}M\, S(\log M_\mathrm{UPP}, z)\, {\cal P}(\log M_\mathrm{UPP} | M, z) \nonumber \\
&&
\qquad \qquad \qquad \qquad \qquad \qquad
\times
n_h(M, z)\, \Delta \Sigma_h(R, M, z). \label{eq:Dsigma_model}
\eeqa
In practice, our choice of a proxy for the cluster center might not match the true
center.
To account for this off-centering effect, we follow the model proposed in Ref.~\cite{2011PhRvD..83b3008O}
by modifying the halo-matter cross-power spectrum as
\beqa
P_\mathrm{hm}(k, M, z) &\rightarrow& {\cal F}_\mathrm{off}(k, M, z)\, P_\mathrm{hm}(k, M, z), \\
{\cal F}_\mathrm{off}(k, M, z) &=& f_\mathrm{cen}(M) + [1-f_\mathrm{cen}(M)] \nonumber \\
&&
\qquad \qquad \qquad
\times \tilde{p}_\mathrm{off}(k;R_\mathrm{off}(z)),
\eeqa
when computing Eq.~(\ref{eq:Dsigma_model}) in the redshift range of interest. 
Here $f_\mathrm{cen}$ is the fraction of clusters with no off-centering effects, while $1-f_\mathrm{cen}$ is the fraction of off-centered clusters.
Motivated by a possible mass dependence in $f_\mathrm{cen}$ (e.g. see Ref.~\cite{Murata:2017zdo} for the richness dependence), we adopt a functional form of $f_\mathrm{cen}(M) = f_0 + f_1 \ln(M/M_\mathrm{pivot})$ where $f_0$ and $f_1$ are free parameters to be varied and 
$M_\mathrm{pivot}$ is set to $3\times10^{14}\, h^{-1}M_\odot$. 
The function $p_\mathrm{off}(r;R_\mathrm{off})$ is 
the normalized radial profile of off-centered ``central" galaxies with respect to the true
center for which we assume a Gaussian distribution given by
$p_\mathrm{off}(r; R_\mathrm{off}) = \exp(-r^2/2/R^2_\mathrm{off})/[(2\pi)^{3/2}R^3_\mathrm{off}]$, 
where $R_\mathrm{off}$ is a parameter to model the typical off-centering radius. 
The Fourier transform of $p_\mathrm{off}(r;R_\mathrm{off})$ is then given by 
$\tilde{p}_\mathrm{off}(k;R_\mathrm{off}) = \exp(-k^2R^2_\mathrm{off}/2)$.
We parametrize the off-centering radius as
$R_\mathrm{off}(z) = \alpha_\mathrm{off} D_A(z) \sigma_\mathrm{off}$,
where $\alpha_\mathrm{off}$ is a free parameter and 
$\sigma_\mathrm{off}$ is a reference scale for recovered position offsets obtained from source insertion simulations as done in Ref.~\cite{ACT:2020lcv}. 
Based on the best estimate inferred from the simulations in Ref.~\cite{ACT:2020lcv}, we set $\sigma_\mathrm{off} = 0.28\, \mathrm{arcmin}$ throughout this paper.
In addition to the parameters in the mass scaling relation, 
we thus have three free parameters, $f_0$, $f_1$ and $\alpha_\mathrm{off}$, to be varied in our analysis.

\subsection{Baryonic effects}\label{subsec:baryonic_effects}

Baryonic feedback can affect the mass density distribution in and around dark matter halos. In cluster-sized halos, feedback processes from active galactic nuclei (AGN) would expel intracluster material from the cluster center, modifying cluster mass density profiles even at Mpc scales (see Ref.~\cite{Pratt:2019cnf} for a recent review).
To account for possible baryonic effects on the cluster mass profiles and to marginalize these effects in our parameter inference, we adopt a semi-analytic model developed in Refs.~\cite{Schneider:2015wta, Schneider:2018pfw, Giri:2021qin}.
The model is widely referred to as the {\it baryonification} method in the literature; it is designed to include baryonic effects in gravity-only N-body simulations by shifting positions of N-body particles with respect to centers of various dark matter halos. 
We apply the baryonification method to the outputs of the Dark Emulator, 
increasing flexibility in our model of stacked lensing signals.

The baryonification method aims at the transformation of dark-matter-only (dmo) profiles into modified, dark-matter-baryon (dmb) profiles that include the effects of dark matter, gas, and stars;
\beqa\label{eq:rho_tot}
&&
\rho_\mathrm{hm+t}(r, M) \rightarrow \rho_\mathrm{hm, dmb}(r, M) \nonumber \\
&&
\qquad \quad
= \rho_\mathrm{DM}(r, M) + \rho_\mathrm{gas}(r, M) + \rho_\mathrm{star}(r, M), 
\eeqa
where $\rho_\mathrm{hm+t}(r, M)$ represents the average mass density profile at halo mass $M$ predicted by the Dark Emulator, but applying a truncation at outer regions, 
$\rho_\mathrm{hm, dmb}(r, M)$ consists of dark matter ($\rho_\mathrm{DM}$), 
gas ($\rho_\mathrm{gas}$), and stellar density profile ($\rho_\mathrm{star}$). 
Some truncation needs to be applied to the original output of the Dark Emulator $\rho_\mathrm{hm}$, because the baryonification assumes that the density profiles of any component in a halo must be constrained to have a converged enclosed mass in the limit of an infinite boundary radius \cite{Schneider:2015wta}.
Note that the Dark Emulator provides the prediction without the common decomposition into one- and two-halo terms, making the truncation somewhat arbitrary.
In this paper, we perform a profile fitting to the Dark-Emulator output as in Ref.~\cite{Diemer:2014xya} and then find an outer radius $r_\mathrm{out}$ for which the logarithmic slope is steepest.
Using the estimate of the boundary radius $r_\mathrm{out}$, we then correct the Dark-Emulator prediction with a truncation below \cite{Baltz:2007vq, Oguri:2011vj};
\beqa
\rho_\mathrm{hm+t}(r) = \left\{
\begin{array}{ll}
\rho_\mathrm{hm}(r) & (r\le r_\mathrm{out}) \\
\rho_\mathrm{hm}(r_\mathrm{out})\left(\frac{r}{r_\mathrm{out}}\right)^{\gamma} \left(\frac{r^2_\mathrm{out}+\tau^2}{r^2+\tau^2}\right)^2 & (r > r_\mathrm{out})
\end{array}
\right.
, \label{eq:Emulator_truncation}
\eeqa
where $\gamma$ is the density slope of $\rho_\mathrm{hm}$ at $r=r_\mathrm{out}$
and we set $\tau = 3 r_\mathrm{out}$.
More details about the truncation and the explicit functional forms of $\rho_\mathrm{DM}$,
$\rho_\mathrm{gas}$ and $\rho_\mathrm{star}$ are found in Appendix~\ref{apdx:baryonification}.

Given the  corrected density profiles, we can compute the enclosed mass profile as $M_\chi(r) \equiv \int_0^{r}4\pi s^2 \mathrm{d}s\, \rho_\chi(s)$ ($\chi = \mathrm{hm+t}, \mathrm{DM}, \mathrm{gas}, \mathrm{star}$). The baryonification method then defines a Lagrangian mapping from 
the initial (dmo) to the final (dmb) density profile by shifting the location of a spherical mass shell as
\beqa
\frac{r_f}{r_i} -1 &=& {\cal A}\left[\left(\frac{M_i}{M_f}\right)^{n}-1\right], \label{eq:mass_shell_shift}\\
M_i &=& M_\mathrm{hm}(r_i) \\
M_f &=& M_\mathrm{DM}(r_i) + M_\mathrm{gas}(r_f) + M_\mathrm{star}(r_f),
\eeqa
where $r_i$ and $r_f$ represent the initial and final locations of the mass shell from the halo center, respectively.
In Eq.~(\ref{eq:mass_shell_shift}), 
we have two parameters, ${\cal A}$ and $n$, governing the adiabatic contraction and expansion of the mass shells.
Throughout this paper, we set ${\cal A}=0.3$ and $n=2$ as in Ref.~\cite{Schneider:2018pfw} (Note that these values can provide a fit to the simulation results in Ref.~\cite{Abadi:2009ve}). 
One can solve Eq.~(\ref{eq:mass_shell_shift}) for $\zeta=r_f / r_i$ iteratively,
allowing to have the displacement length (i.e. $d[r_i] = r_f - r_i$) at a given $r_i$.
In the end, the halo-matter cross-correlation with baryonic effects can be expressed as
\beqa
\xi_\mathrm{hm+b}(r, M, z, \mathbf{p}) &=& \xi_\mathrm{1h}(r, M, z, \mathbf{p}) + \xi_\mathrm{2h}(r, M, z), \label{eq:xi_hm_with_baryon}\\
\xi_\mathrm{1h}(r, M, z, \mathrm{p}) &=& \frac{\rho_\mathrm{hm,dmb}(r, M, z, \mathbf{p})}{\bar{\rho}_\mathrm{m0}}, \\
\xi_\mathrm{2h}(r, M, z) &=& \frac{\rho_\mathrm{hm}(r,M,z)-\rho_\mathrm{hm+t}(r,M,z)}{\bar{\rho}_\mathrm{m0}}\nonumber \\
&&
\qquad \qquad \qquad \qquad \qquad
-1
\eeqa
where $\mathbf{p}$ represents a set of parameters in the baryonification method. 
We here assume that an effective two-halo term of $\rho_\mathrm{hm}-\rho_\mathrm{hm+t}$ is not affected by baryonic effects. When including the baryonic effects in the model of $\Delta \Sigma$, we first perform the Fourier transform of Eq.~(\ref{eq:xi_hm_with_baryon}) to have the corresponding halo-matter power spectrum $P_\mathrm{hm+b}$ and then plug in $P_\mathrm{hm+b}$ for $P_\mathrm{hm}$ in Eq.~(\ref{eq:dSigma_single}).
It is worth noting that the baryonification method does not introduce any changes in the halo mass function $n_h(M)$ as long as we distinguish each halo by a single halo mass $M$.
After the baryonification is applied, the halo mass $M$ does not mean the spherical halo mass of $M_\mathrm{200b}$ as designed in the Dark Emulator, but should be regarded as an auxiliary variable to identify individual halos.

The baryonification method introduces various parameters.
Because our primary interest is to measure the masses of cluster-sized halos, 
it is reasonable to vary only a small set of the model parameters which are relevant to baryonic effects on mass density fields at Mpc scales.
As examined in Refs.~\cite{vanDaalen:2019pst, Giri:2021qin}, the baryon mass fraction inside halos is most relevant to baryonic effects on Mpc-scale matter clustering (also see Refs.~\cite{Debackere:2021ado, Cromer:2021iyp} for baryonic effects on cluster mass estimates). 
Motivated by these previous findings, we allow three parameters of gas physics to vary in our analysis. These include 
(i) the mass dependence $\mu$ in the inner density slope of $\rho_\mathrm{gas}$,
(ii) the pivot mass scale $M_c$ at which the inner density slope becomes 3/2,
(iii) and the dimensionless outer scale radius of $\rho_\mathrm{gas}$ expressed as $\theta_\mathrm{ej}$.
See Appendix~\ref{apdx:baryonification} for details.

\begin{figure*}[!t]
\includegraphics[clip, width=2.1\columnwidth]{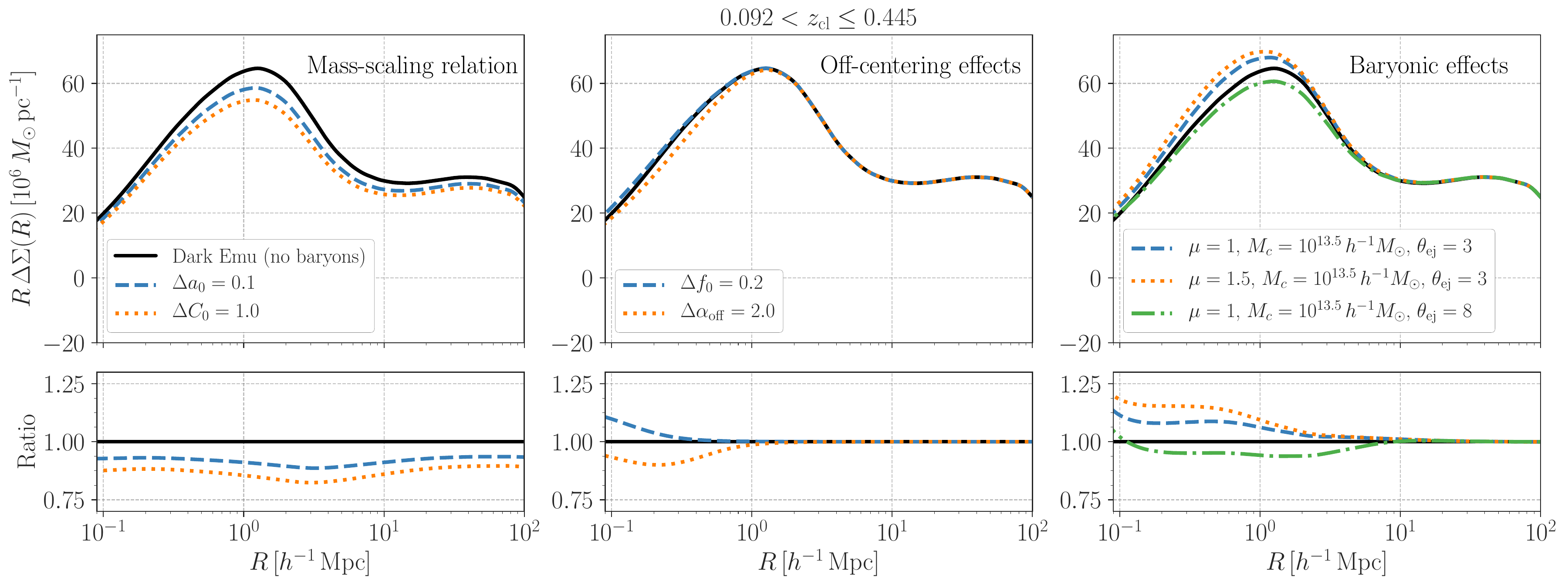}
\caption{\label{fig:Dsigma_example} 
Parameter dependence of stacked lensing signals $\Delta \Sigma$ for the ACT clusters at redshifts of $0.092<z\le 0.445$. 
The left, middle, and right panels show $\Delta \Sigma$ 
when we vary the parameters related to the mass-scaling relation, off-centering effects, 
and baryonic effects, respectively.
The top left panel shows the change in $\Delta \Sigma$ caused by varying the parameter in the mean scaling relation $a_0$ and the scatter $C_0$, while the top middle panel summarizes the change by the off-centering parameters $f_0$ and $\alpha_\mathrm{off}$ (see Section~\ref{subsec:halo_model} for the definition of those parameters.).
The top right panel represents $\Delta \Sigma$ for several baryon models based on 
the {\it baryonification} method as described in Section~\ref{subsec:baryonic_effects}.
Note that we plot $R\, \Delta \Sigma$ in the top panels for visualization purposes.
The baseline model is shown as the black solid line in each panel: it does not include any baryonic effects.
See Section~\ref{subsec:param_dependence} for the choice of model parameters in this figure.
\rev{The model parameters are 
comparable to the adopted prior ranges
in our likelihood analysis as described in Section~\ref{subsec:likelihood}}.
The bottom panels show the ratio of $\Delta \Sigma$  to our baseline model.
}
\end{figure*}

\subsection{Parameter dependence of $\Delta \Sigma$}\label{subsec:param_dependence}

To provide intuition, the dependence of stacked lensing signals on the model parameters are briefly summarized here.
Fig.~\ref{fig:Dsigma_example} shows our model of stacked lensing signal $\Delta \Sigma$ for the ACT clusters in the redshift interval of $0.092<z\le 0.445$ as we vary some model parameters.
In this figure, we consider eight cases as follows:
\begin{itemize}
\item[(i)] The fiducial model, assuming a mass-scaling relation with $a_0 = 0.5$ and $a_1 = 0$; scatter in the scaling relation set at $C_0 = 1$; and the off-centering effects with $f_0 = 0.75$, $f_1 = 0.25$, and $\alpha_\mathrm{off}=1$. This model does not include any baryonic effects;
\item[(sl-i)] the model assuming a mass-scaling relation set to a higher $a_0$ by 0.1;
\item[(sl-ii)] the model with a higher scatter in the mass scaling relation, changing $C_0$ by 1.0;
\item[(oc-i)] the model with fewer off-centered clusters, with a higher $f_0$ by 0.2
\item[(oc-ii)] the model with a larger off-centering length with a higher $\alpha_\mathrm{off}$ by 2.0
\item[(b-i)] The fiducial model of the mass-scaling relation and off-centering effects, but including baryonic effects with $\mu=1.0$, $M_c = 10^{13.5}\, [h^{-1}M_\odot]$, and $\theta_\mathrm{ej}=3$;
\item[(b-ii)] The fiducial model of the mass-scaling relation and off-centering effects, and
the baryonic effects with  $\mu$ increased by 0.5;
\item[(b-iii)] The fiducial model of the mass-scaling relation and off-centering effects, and
baryonic effects with  $\theta_\mathrm{ej}$ increased by 5;
\end{itemize}

The left panel in Fig.~\ref{fig:Dsigma_example} displays the models labeled above as (sl-i) and (sl-ii); 
the middle panel shows models (oc-i) and (oc-ii);
and models (b-i), (b-ii), and (b-iii) are summarized in the right panel.
Note that the fiducial model of (i) is shown in every panel 
by the black solid line for ease of comparison.

\subsubsection*{Mass-scaling relation}

The left panel in Fig.~\ref{fig:Dsigma_example} highlights 
the dependence of $\Delta \Sigma$ on the mass-scaling relation. 
In our model, the scaling relation between SZ and total masses 
is set by three parameters: $a_0$, $a_1$, and $C_0$.
The stacked lensing signal is most sensitive to two parameters, $a_0$ and $C_0$. 
Because the completeness of our ACT clusters is fixed as shown in Fig.~\ref{fig:ACT_selecfunc}, the 
average SZ mass in the cluster sample in a given redshift range is not affected by the choice of our model parameters. Hence, increasing $a_0$ indicates that smaller halo masses become 
more relevant to the average SZ mass, making the amplitude of $\Delta \Sigma$ smaller.
Also, increasing the scatter in the mass-scaling relation (i.e. a higher $C_0$ in our model) 
can lead to a smaller $\Delta \Sigma$, because the larger scatter can allow less massive clusters to contribute to the stacked lensing signal. 
As shown in the left panel, there exists a strong degeneracy between the two parameters 
$a_0$ and $C_0$ in our model, 
which has been mentioned in the literature \cite{Murata:2017zdo, Wu:2020xxy}.
It is worth noting that both $a_0$ and $C_0$ can affect the signal not only at $R\simlt 10\, h^{-1}\mathrm{Mpc}$ but also at a large separation of $R\simgt 10\, h^{-1}\mathrm{Mpc}$. This effect can be explained by the change in the linear halo bias as one varies the model parameter \cite{Wu:2020xxy}.

\subsubsection*{Off-centering effects}

The middle panel in Fig.~\ref{fig:Dsigma_example} summarizes 
off-centering effects in our model of stacked lensing measurements.
Our model characterizes the off-centering effects by three parameters $f_0$, $f_1$ and $\alpha_\mathrm{off}$; of these,
$f_0$ and $\alpha_\mathrm{off}$ are most relevant in our case.
The parameter $f_0$ is the fraction of clusters with no off-centering at halo mass of $3\times10^{14}\, h^{-1}M_{\odot}$.
Hence, increasing $f_0$ can make the prediction of $\Delta \Sigma$ at $R\simlt1\, h^{-1}\mathrm{Mpc}$ higher.
At a fixed $f_0$, we can see how a typical off-centering length affects the stacked lensing signal by varying the parameter $\alpha_\mathrm{off}$.
Increasing $\alpha_\mathrm{off}$ means the stacked lensing signal can be affected by off-centering effects to larger radii.

\subsubsection*{Baryonic effects}

The right panel in Fig.~\ref{fig:Dsigma_example} represents baryonic effects in the stacked lensing signal $\Delta \Sigma$ in our model.
By comparing the solid line and the blue dashed line in the figure, we can see the impact of baryonic effects; including baryons can lead to a more concentrated halo mass profile
(see also Ref.~\cite{Shirasaki:2017jlt}), but it does not affect the signal 
at large radii which are governed by gravity alone.
We have three parameters in the model of baryonic effects, 
enabling the prediction to be highly flexible.
For instance, increasing the parameter $\mu$ at a given $M_c$ steepens the inner slope of the gas density profile, leading to an increase in the mass density  concentration.
Also, increasing the parameter $\theta_\mathrm{ej}$ indicates that 
the gas in a cluster can be distributed more in outer regions beyond the cluster's virial radius, 
allowing the mass density profile to experience some adiabatic expansion.
As shown in the figure, baryonic effects may affect the mass density profile at Mpc scales, which can not be ignored when infering the average halo mass from 
the stacked lensing measurements.
Nevertheless, the information on $\Delta \Sigma$ at large radii can be helpful to constrain the halo mass even if our model is subject to uncertainty in baryonic physics.

\begin{figure*}[!t]
\includegraphics[clip, width=2.1\columnwidth]{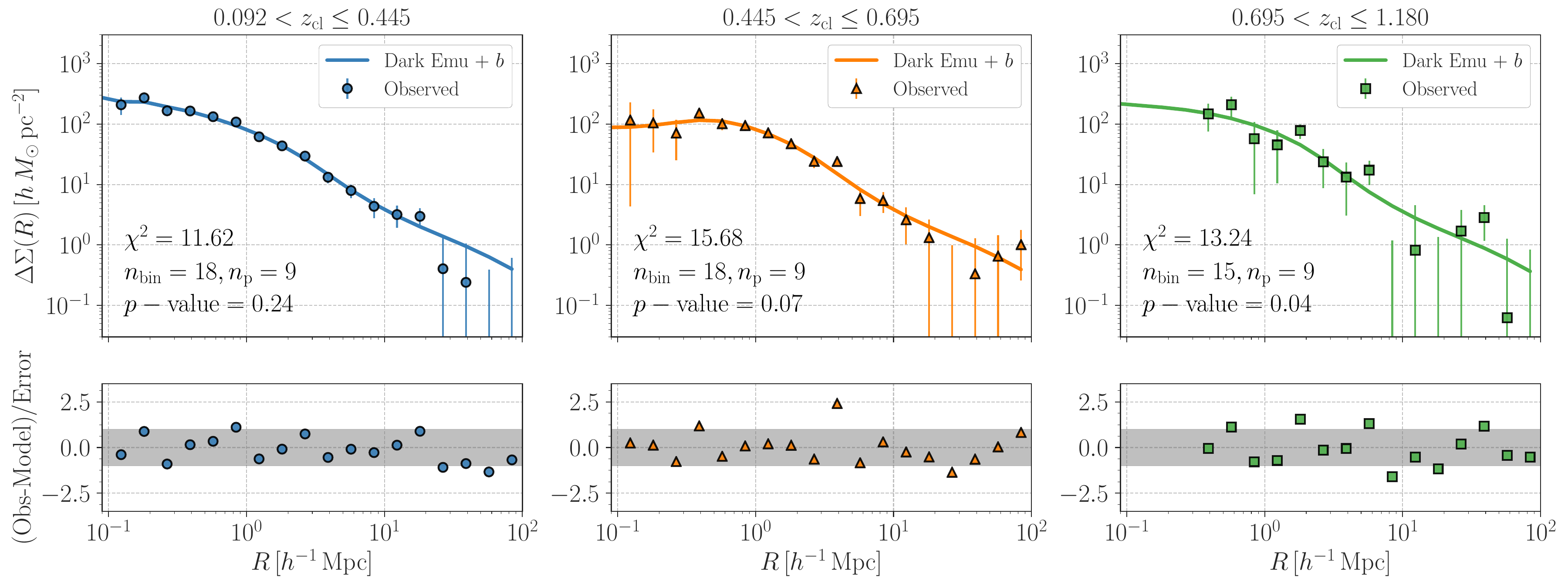}
\caption{\label{fig:Dsigma_bestfit} 
The observed stacked lensing signals around ACT SZ-selected clusters and the best-fit model derived from  the MCMC analysis. We work with three different cluster redshift bins (increasing from left to right). In the upper panels, the colored points represent the observed signal of $\Delta \Sigma$, while the line corresponds to the best-fit model.
The error bars in each upper panel are estimated with a combination of two mock lensing analyses (see Section~\ref{subsec:covariance} for details). In the lower panels, we plot the difference between the observed $\Delta \Sigma$ signal and the model counterpart. There, we normalize the difference by the diagonal component of the covariance matrix.
The gray region in the lower panels highlights a $\pm1\sigma$-level difference.
For reference, we note the chi-square value $\chi^2$ as defined in Eq.~(\ref{eq:chisq_bestfit}), the number of the data elements $n_\mathrm{bin}$,
the number of free parameters in the model $n_\mathrm{p}$,
and the $p$ value of the $\chi^2$ value with respect to the number of degrees of freedom (i.e.~$n_\mathrm{bin}-n_\mathrm{p}$) in each upper panel.
Our model adopts the Dark Emulator \cite{Nishimichi:2018etk} for a baseline prediction of statistics of cluster-sized halos, and it also takes into account possible baryonic effects with the prescription of {\it baryonification} \cite{Schneider:2015wta, Schneider:2018pfw, Giri:2021qin}. See Section~\ref{subsec:halo_model} and \ref{subsec:baryonic_effects} for further details on our halo-based model. 
}
\end{figure*}

\subsection{Likelihood}\label{subsec:likelihood}

To constrain the model parameters with the stacked lensing measurements, we define a log-likelihood function as
\beqa
- 2 \ln {\cal L}(\bd{\theta}) &=& \sum_{i,j}  d_i \mathrm{Cov}^{-1}_{ij} d_j
+
\left(\frac{N_\mathrm{cl, obs}-N_\mathrm{cl}(\bd{\theta})}{2N^{1/2}_\mathrm{cl, obs}}\right)^2, \label{eq:logL} \\
d_i &\equiv& \Delta \Sigma_\mathrm{obs}(R_i) - \Delta \Sigma(R_i; \bd{\theta})
\eeqa
where $\Delta \Sigma_\mathrm{obs}(R_i)$ represents the stacked lensing signal at the i-th radial bin from the cluster center, 
$N_\mathrm{cl, obs}=32$ is the observed cluster number count,
$\Delta \Sigma (R_i)$ and $N_\mathrm{cl}$ are the cluster number counts predicted by our halo-based model (see Eq.~[\ref{eq:Ncl}]),
and $\bd{\theta}$ describes a set of our model parameters.
In Eq.~(\ref{eq:logL}), we assume a double Poisson error for the cluster number count to account for possible uncertainties in our model (e.g. our model ignores a variation of the survey depth in the ACT observation \cite{ACT:2020lcv}).
Note that the statistical errors on the cluster abundance estimated by our mock catalogs are consistent with the Poisson expectation with a level of 10\%, while the cross covariance between $\Delta \Sigma_\mathrm{obs}$ and $N_\mathrm{cl}$ in the mock catalogs is found to be less than 0.1 after we normalize the cross covariance with the variances of $\Delta \Sigma_\mathrm{obs}$ and $N_\mathrm{cl}$.
Hence, our likelihood function primarily uses the information of lensing signals, but it has a useful contribution from the cluster number counts to break possible degeneracy among our model parameters.

In the log-likelihood function, we have nine free parameters apart from cosmological parameters. Six of these parameters set the mass-scaling relation ($a_0$, $a_1$, and $C_0$)
and possible off-centering effects in the ACT cluster sample ($f_0$, $f_1$ and $\alpha_\mathrm{off}$). The other three parameters control baryonic effects on the stacked lensing signal ($\mu$, $M_c$, and $\theta_\mathrm{ej}$).
We set flat priors in the ranges of 
$0.01 \le a_0 \le 0.75$, $-0.2 \le a_1 \le 0.2$, $0.2 \le C_0 \le 5.0$, 
$0 \le f_0 \le 1$, $0 \le f_1 \le 1$, $0 \le \alpha_\mathrm{off} \le 3$,
$0 \le \mu \le 2$, $12.0 \le \log(M_c \,[h^{-1}M_\odot]) \le 15.0$
and
$1 \le \theta_\mathrm{ej} \le 10$
in the likelihood analysis. To compute the log-likelihood function in our nine-parameter space, we use the Markov Chain Monte Carlo (MCMC) sampler ${\tt emcee}$ \cite{Foreman-Mackey:2012any}. 
We set 45 walkers, ran 32000 steps of burn-in, and
then 32000 steps to sample the likelihood function. 
We confirmed that sampling within and among chains has converged to a level of $1.5\%$.
In this paper, we perform an MCMC analysis for the cluster subsample at each of the three redshift bins separately.

\section{Results}\label{sec:results}

\subsection{Parameter inference}

\begin{figure*}[!t]
\includegraphics[clip, width=2.1\columnwidth]{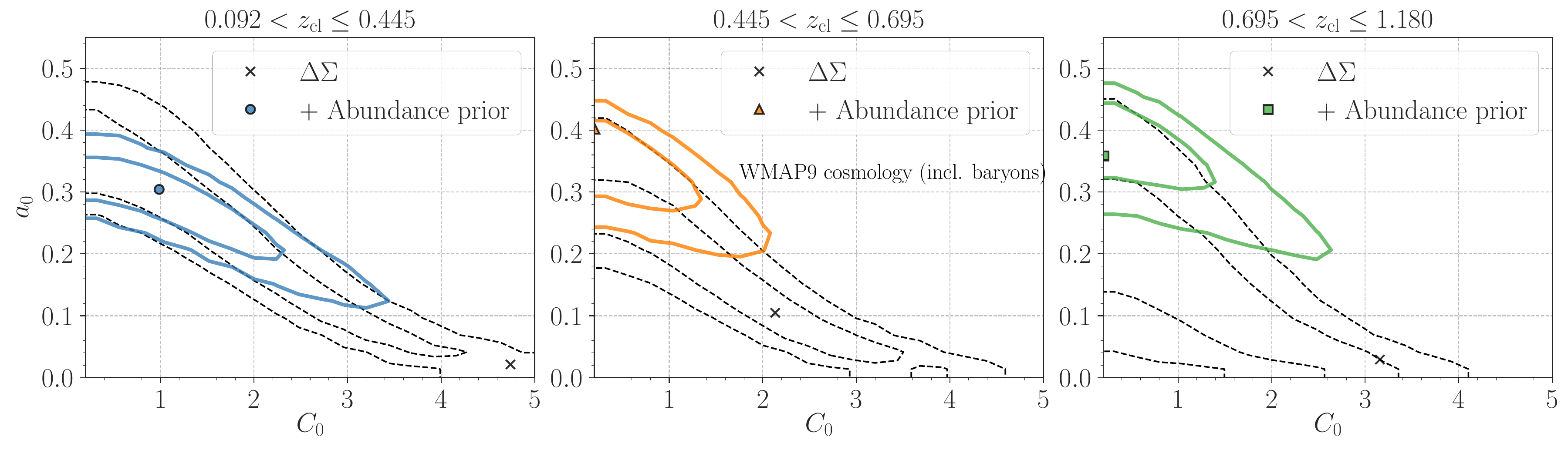}
\caption{\label{fig:impact_abundance_prior} 
Impact of the cluster abundance on parameter inference. In each panel, we show the 68\% and 95\% credible intervals in the two-dimensional parameter space ($a_0$ and $C_0$) for three different cluster redshift bins. The solid colored lines show our fiducial MCMC results including cluster abundance information, while the dashed ones are the counterparts without information on the cluster number count.
The colored symbol in each panel highlights the best-fit parameters in our fiducial analysis,
and the cross symbol corresponds to the best-fit values when we do not use the abundance information in the parameter inference.
The strong degeneracy between $a_0$ and $C_0$ can make the best-fit values off-center in the contours after marginalization in our high-dimensional parameter space when we work on the MCMC analysis without the abundance prior.
}
\end{figure*}

We here summarize our parameter inference on the mass density profile around the ACT clusters. 
Fig.~\ref{fig:Dsigma_bestfit} shows the comparison of stacked lensing signals 
$\Delta \Sigma$ with our best-fit models. 
To assess the goodness of fit, we define a chi-square quantity as
\beqa
\chi^2 &=& \sum_{i,j} d_{i, \mathrm{bf}} \mathrm{Cov}_{ij}^{-1} d_{j, \mathrm{bf}}, \label{eq:chisq_bestfit}\\
d_{i,\mathrm{bf}} &=& \Delta \Sigma_\mathrm{obs}(R_i) - \Delta \Sigma(R_i;\bd{\theta}_\mathrm{bf}),
\eeqa
where $\bd{\theta}_\mathrm{bf}$ represents the best-fit parameters obtained 
by our MCMC analysis.
We find the chi-square values of $\chi^2 = 11.62, 15.68$ and $13.24$ at the cluster redshifts of $0.092 < z \le 0.445$, $0.445 < z \le 0.695$, and $0.685 < z \le 1.180$, respectively.
Our model consists of nine free parameters and the number of data elements is
18 for the first two redshift bins and 15 for the highest redshift bin. 
Hence, our best-fit model provides a reasonable chi-square value with 
respect to the number of degrees of freedom.
We expect that an effective number of degrees of freedom would be larger than the naive expectation of $n_\mathrm{bin} - n_\mathrm{p}$ where $n_\mathrm{bin}$ and $n_\mathrm{p}$ are the number of data elements and free parameters, because 
our measurements are not informative enough to constrain all the free parameters. 
For instance, the parameters of the baryonic effects ($\mu$, $\log M_c$, and $\theta_\mathrm{ej}$) are not well constrained with our analysis alone.
We provide credible intervals for the nine free parameters in Appendix~\ref{apdx:details_mcmc}, highlighting that our analysis can put meaningful constraints on only two parameters, $a_0$ and $C_0$. 
Other parameters can be regarded as basically nuisance parameters in our analysis, accounting for systematic uncertainties in our modeling of the stacked lensing measurements. 
As a cross check, we compute a probability-to-exceed (PTE) for our minimum $\chi^2$ by using $10^{6}$ random data vectors with mean zero and the data covariance. We find that the PTE values are 0.87, 0.62, and 0.58 in order of increasing redshifts, showing our best-fit model provides a good fit to the data.

\begin{figure*}[!t]
\includegraphics[clip, width=2.1\columnwidth]{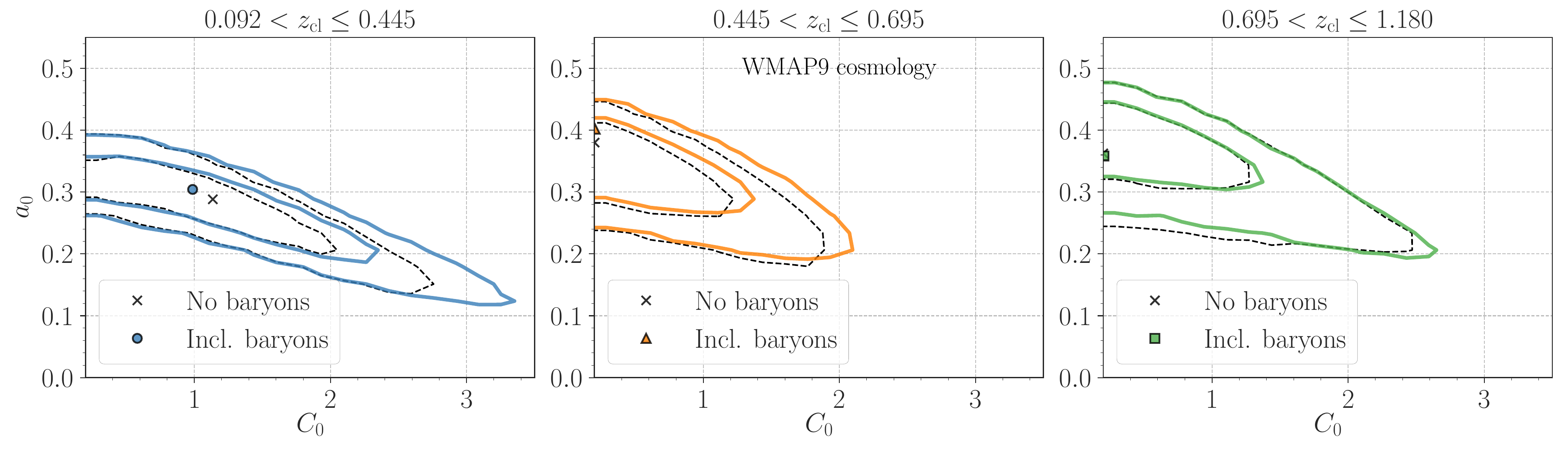}
\caption{\label{fig:impact_baryons} 
Similar to Fig.~\ref{fig:impact_abundance_prior}.  We compare credible intervals when accounting for the baryonic effects in the model or not. The dashed lines in each panel represent the 68\% and 95\% credible intervals when our model does not include baryonic effects, and the cross symbol shows the best-fit parameters in this case.
}
\end{figure*}

For reference purposes, we compute the cumulative signal-to-noise ratio of our stacked lensing measurements as
\beqa
\left(\frac{\mathrm{S}}{\mathrm{N}}\right)^2 = \sum_{i,j} \Delta \Sigma_\mathrm{obs}(R_i)\mathrm{Cov}_{ij}^{-1} \Delta \Sigma_\mathrm{obs}(R_j),
\eeqa
where the indices $i,j$ run over the range of $0.1<R\, [h^{-1}\mathrm{Mpc}]< 100$.
We find $\mathrm{S/N} = 14.6, 12.0$ and $6.6$ at the cluster redshifts of $0.092 < z \le 0.445$, $0.445 < z \le 0.695$, and $0.685 < z \le 1.180$, respectively.
The cumulative signal-to-noise ratios indicate that our measurements can constrain the amplitude of $\Delta \Sigma$ with a level of $6.8\%$, $8.3\%$, and $15\%$ 
from lower to higher redshift bins.
Our MCMC analysis provides a 68\% credible interval of the primary parameters ($a_0$ and $C_0$) as
\beqa
a_0 = 
\left\{
\begin{array}{ll}
0.27^{+0.05}_{-0.07} & (0.092 < z \le 0.445) \\
0.32\pm{0.05} & (0.445 < z \le 0.695) \\
0.35^{+0.05}_{-0.06} & (0.695 < z \le 1.180)
\end{array}
\right.,
\eeqa
and 
\beqa
C_0 = 
\left\{
\begin{array}{ll}
1.3^{+0.9}_{-0.7} & (0.092 < z \le 0.445) \\
0.7^{+0.6}_{-0.4} & (0.445 < z \le 0.695) \\
0.7^{+0.7}_{-0.4} & (0.695 < z \le 1.180)
\end{array}
\right.,
\eeqa
where we have a $15\%$-level constraints on $a_0$ with our lensing analysis in each redshift bin. Note the peak in the posterior distribution of $C_0$ is found to be close to the lowest level in the prior ($C_0=0.2$) except for the lowest redshift bin.
Hence, our estimate of the 68\% credible interval for $C_0$ is meaningful in the lowest redshift bin alone.
We also examined a wide prior of $0.01\le C_0 \le 5.0$ in our likelihood analysis, confirming that our constraints of lensing masses are not affected by the lower bound of the $C_0$ prior.

\begin{figure*}[!t]
\includegraphics[clip, width=2.1\columnwidth]{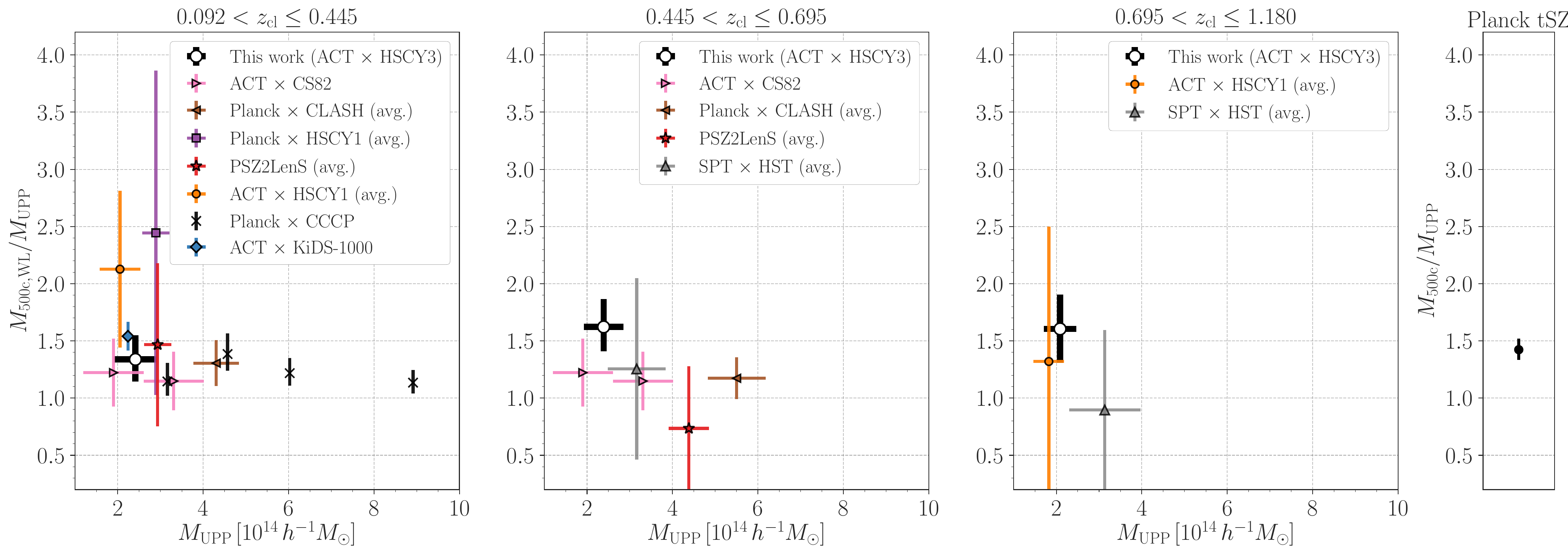}
\caption{\label{fig:mean_M500c_Msz} 
Constraints of the mass bias $B = M_\mathrm{500c}/M_\mathrm{UPP}$ from our stacked lensing measurements. In three left panels, the open circle with error bars represents the 68\% credible interval of $B$ in the different cluster redshift bins (lower to higher redshift from left to right).
Note that we define the mass bias in logarithmic space as $\log B \equiv \langle \log M_\mathrm{500c} | \log M_\mathrm{UPP} \rangle - \log M_\mathrm{UPP}$ where $\langle \log M_\mathrm{500c} | \log M_\mathrm{UPP} \rangle$ is the average of $\log M_\mathrm{500c}$ at the average  $M_\mathrm{UPP}$ (see Section~\ref{subsec:mass_bias} for details).
For comparison, we also show the constraints on the mass bias by previous lensing analyses in the literature; 
ACT $\times$ CS82 \cite{Battaglia:2015zfa}, 
Planck $\times$ CLASH \cite{Penna-Lima:2016tvo},
Planck $\times$ HSCY1 \cite{2018PASJ...70S..28M},
PSZ2LenS \cite{Sereno:2017zcn},
ACT $\times$ HSCY1 \cite{Miyatake:2018lpb},
Planck $\times$ CCCP \cite{Herbonnet:2019byy}, 
SPT $\times$ HST \cite{Schrabback:2020hxx},
and ACT $\times$ KiDS-1000 \cite{Robertson:2023gxe}.
Caution that the limits of Planck $\times$ CLASH,
Planck $\times$ HSCY1, PSZ2LenS,
ACT $\times$ HSCY1 and SPT $\times$ HST are set by the average masses of individual lensing measurements. The limits of ACT $\times$ CS82, Planck $\times$ CCCP and ACT $\times$ KiDS-1000 should be used for a fair comparison.
For reference purposes, the right panel displays the constraint on $B$ derived from the auto power spectrum of the Planck thermal SZ map \cite{Bolliet:2017lha}.
Note that some previous studies have used the inverse convention of quoting $B^{-1} \equiv 1-b$.
}
\end{figure*}

Our baseline analysis adopts a loose prior in the cluster abundance as seen in the second term on the right-hand side of Eq.~(\ref{eq:logL}).
It is beneficial to see how important the abundance prior is in our parameter inference.
Fig.~\ref{fig:impact_abundance_prior} summarizes the 68\% and 95\% credible intervals in the two-dimensional plane of $a_0$ versus $C_0$ to highlight the impact of the abundance prior.
In the figure, the solid colored lines represent the constraints when we include the abundance prior, while the dashed lines stand for the corresponding constraints without the abundance information.
For the MCMC analysis made without the abundance prior, we double the number of 
MCMC steps to have a stable parameter inference.
We find that the abundance prior is efficient in rejecting the parameter space at lower $a_0$ values in the MCMC analysis. Note that a lower $a_0$ value in our model can predict a higher mean halo mass at a given SZ mass, providing smaller cluster number counts in each redshift bin.
Hence, information on the cluster abundance can be effective to break the inherent parameter degeneracy between $a_0$ and $C_0$ in the $\Delta \Sigma$ model.
We also observe that some parameter regions with large $a_0$ and small $C_0$ values are incompatible with the cluster abundance in the lowest cluster redshift bin of $0.092 < z \le 0.445$, because those parameters provide an overestimate of the cluster abundance. 

We also study the importance of including baryonic effects in the $\Delta \Sigma$ model in our analysis. Fig.~\ref{fig:impact_baryons} shows the 68\% and 95\% credible intervals in the $a_0-C_0$ plane when we include or exclude baryonic effects in our modeling. 
We find that our measurements are not very sensitive to details of the baryonic effects at present. Nevertheless, we observe a trend: including baryonic effects increase the credible intervals in the two lowest cluster redshift bins, implying that it will be more important to account for baryons in parameter inference as the cumulative signal-to-noise ratio increases.

\subsection{Total mass at a given SZ mass}\label{subsec:mass_bias}

The SZ mass $M_\mathrm{UPP}$ is not necessarily equal to the underlying cluster mass $M_\mathrm{500c}$, because the scaling relation between the electron pressure and the SZ mass is derived under the assumption of hydrostatic equilibrium.
The UPP scaling relation was also derived from X-ray observations of $z = 0.1$ clusters \cite{Arnaud:2009tt}.
Also, any unknown systematic errors in X-ray analyses induce deviations of $M_\mathrm{UPP}$ from the true mass of $M_\mathrm{500c}$ even if hydrostatic equilibrium is a good approximation on cluster scales. 
We here study the mass bias of $B\equiv M_\mathrm{500c}/M_\mathrm{UPP}$ at a fixed SZ mass
with our MCMC parameter inference based on the stacked lensing measurements. 

In our halo-based forward model, we can estimate the probability distribution of $M_\mathrm{500c}$ at a given $\log M_\mathrm{UPP}$ and redshift $z$ as
\beqa
&&\mathrm{Prob}(M_\mathrm{500c}=m | \log M_\mathrm{UPP}, z) 
= {\cal N}^{-1}\int \mathrm{d}M \, n_h(M,z) \nonumber \\
&&
\qquad \qquad
\times {\cal P}(\log M_\mathrm{UPP}|M) \, \delta_\mathrm{D}\left(f_\mathrm{500c}(M,z)-m\right),
\label{eq:prob_m500c_mUPP} \\
&&{\cal N} = \int \mathrm{d}M \, n_h(M,z) \, {\cal P}(\log M_\mathrm{UPP}|M),
\eeqa
where $\delta_\mathrm{D}(x)$ is the one-dimensional Dirac function and 
we find the function of $f_\mathrm{500c}(M,z)$ as a  
solution of Eqs.~(\ref{eq:def_M500c_base}) and (\ref{eq:def_M500c}).
Note that the probability density of Eq.~(\ref{eq:prob_m500c_mUPP}) can be derived for a given set of our model parameters. 
The parameters $a_0$, $a_1$ and $C_0$ set the function ${\cal P}(\log M_\mathrm{UPP}|M)$ and 
the parameters involved in baryonic effects ($\mu$, $\log M_c$ and $\theta_\mathrm{ej}$) 
affect the function $f_\mathrm{500c}$ by changing the relation between the spherical overdensity mass $M_\mathrm{500c}$ and the baseline halo mass $M$ as given by the Dark Emulator. 
Using Eq.~(\ref{eq:prob_m500c_mUPP}), we then derive the average logarithmic mass at a given SZ mass as
\beqa
&& \langle \log M_\mathrm{500c} | \log M_\mathrm{UPP} \rangle 
\equiv \int \mathrm{d}M_\mathrm{500c}\, \log M_\mathrm{500c} \nonumber \\
&& 
\qquad \qquad \qquad \qquad 
\times \mathrm{Prob}(M_\mathrm{500c}|\log M_\mathrm{UPP}, z), \label{eq:avg_logm500c}
\eeqa
where we use the average redshift over the 32 clusters in each subsample to compute Eq.~(\ref{eq:avg_logm500c}). To be specific, we set $z=0.324, 0.567$ and $0.878$ 
for the cluster sample in the redshift ranges $0.092<z\le0.445$, $0.445<z \le 0.695$ 
and $0.695<z\le1.180$, respectively.

Fig.~\ref{fig:mean_M500c_Msz} summarizes the 68\% credible interval of the mass bias\footnote{We compute the ratio of $M_\mathrm{500c}/M_\mathrm{UPP}$ to be $10^{\langle \log M_\mathrm{500c} | \log M_\mathrm{UPP}\rangle - \log M_\mathrm{UPP}}$ for our constraints.}
inferred from the posterior distribution 
of our model parameters. 
Our stacked lensing measurements are able to constrain the mass bias at 
the average SZ mass as
\beqa \label{eq:B}
B = \left\{
\begin{array}{ll}
1.3 \pm 0.2 & (0.092 < z \le 0.445) \\
1.6 \pm 0.2 & (0.445 < z \le 0.695) \\
1.6 \pm 0.3 & (0.695 < z \le 1.180)
\end{array}
\right.,
\eeqa
so we find marginal evidence of the mass bias ($B \neq 1$) at every redshift bin in our analysis. 

It is worth noting that our limits of Eq.~(\ref{eq:B}) depend on the choice of prior in our model parameters. As shown in Appendix~\ref{apdx:details_mcmc}, we observe that the mass bias parameter can correlate with not only $a_0$ (the baseline mean scaling relation) but also $C_0$ (the scatter in the scaling relation) in our posterior distribution. 
This indicates that any additional information of the scatter 
can improve the limits of $B$.
Apart from $C_0$, $a_1$ (describing the mass dependence in the scaling relation) has a weak correlation with the derived parameter $B$ in our inference. Future precise 
stacked-lensing measurements would require to break the degeneracy between $B$ and $a_1$ by further dividing cluster samples with $M_\mathrm{UPP}$ or another mass proxy.


For comparison, we overlay cluster mass estimates of $M_\mathrm{500c}$ from other weak lensing analyses in Fig.~\ref{fig:mean_M500c_Msz}. 
Those include 
the stacked lensing analysis of 9 ACT-selected clusters \cite{Battaglia:2015zfa},
the lensing analysis from the Cluster Lensing And Supernova survey with Hubble (CLASH) for 21 galaxy clusters selected by their Planck thermal SZ signal  \cite{Penna-Lima:2016tvo},
the lensing mass measurements of 5 Planck-selected clusters using HSC year-1 shear data \cite{2018PASJ...70S..28M},
the lensing analysis of 35 Planck-selected clusters employing galaxy shapes from CFHTLenS and RCSLenS \cite{Sereno:2017zcn}, 
the lensing mass measurements of 8 ACT-selected clusters employing HSC year-1 shear data \cite{Miyatake:2018lpb},
the scaling relation between the lensing and the Planck SZ masses derived by the Canadian Cluster Comparison Project with the help of the Multi Epoch Nearby Cluster Survey \cite{Herbonnet:2019byy},
the lensing mass measurements of 30 galaxy clusters from the South Pole Telescope Sunyaev-Zel'dovich (SPT-SZ) Survey with galaxy shape measurements by Hubble Space Telescope \cite{Schrabback:2020hxx},
and
the stacked lensing analysis of 157 ACT-selected clusters using KiDS-1000 observations \cite{Robertson:2023gxe}.
In the figure, we compute the average masses from individual lensing mass measurements in Refs.~\cite{Penna-Lima:2016tvo, 2018PASJ...70S..28M, Sereno:2017zcn, Miyatake:2018lpb, Schrabback:2020hxx} for visualization purposes.
We find that our inference of the mass bias is consistent with the previous constraints
at a $\sim 1\sigma$ level.

Another important constraint on the mass bias has been derived from the power spectrum analysis of the Planck thermal SZ map \cite{Bolliet:2017lha}.
The right panel in Fig.~\ref{fig:mean_M500c_Msz} highlights the constraint on the mass bias from this analysis,
assuming our fiducial $\Lambda$CDM cosmology.
We caution that Ref.~\cite{Bolliet:2017lha} assumes a deterministic scaling relation between the SZ mass $M_\mathrm{UPP}$ and the spherical overdensity mass $M_\mathrm{500c}$,
suggesting that a direct comparison with our results may be inappropriate.

\subsection{Different choices of cluster forward modeling}\label{subsec:model_choice}

We evaluate the impact of our modeling choices by implementing a simpler forward model with several different assumptions. Here we assume the density profile to be an NFW profile \citep{Navarro:1996gj},\begin{equation}
    \rho_\mathrm{1h}(r) = \frac{\delta_c\rho_c}{(r/r_\mathrm{s})\left(1+r/r_\mathrm{s}\right)^2}
\end{equation}
instead of Eq.\ (\ref{eq:rho_tot}), using the expression for the NFW excess surface density given by \cite{Wright2000}. Here, $\rho_c$ is the normalization and $r_\mathrm{s}$ is the scale radius, and
\begin{equation}
    \delta_c = \frac{500}{3}\frac{c_{500}^3}{\left[\ln(1+c_{500})-c_{500}/(1+c_{500})\right]}.
\end{equation}
with $c_{500}\equiv r_{500}/r_s$. We assume the mass-concentration relation of Ref.~\cite{Ishiyama:2020vao} as implemented in \texttt{colossus} \citep{Diemer2018}. We rewrite Eq.\ (\ref{eq:mean}) as
\begin{equation}\label{eq:mor_alt}
    \log M'_\mathrm{UPP} = a_0' + a_1'\log \left(\frac{M_\mathrm{500c}}{5\times10^{14}\,M_\odot}\right),
\end{equation}
where for simplicity $a_1'=1$ while $a_0'$ is fit to each of the redshift bins with a uniform prior, $a_0'\sim\mathcal{U}(13, 16)$.
We assume a log-normal scatter (Eq.~\ref{eq:scatter}) independent of mass and redshift, $\sigma_{\log M_\mathrm{UPP}}\sim\mathcal{U}(0.01, 1)$, in each bin.

The 2-halo term is calculated explicitly as in Ref.~\cite{Robertson:2023gxe}, i.e., assuming the cluster power spectrum is biased with respect to the dark matter power spectrum,
\begin{equation}
    P_\mathrm{2h}(k|M,z) = \mathcal{A}_\mathrm{2h}b_\mathrm{h}(M)P_\mathrm{mm}(k|M,z)
\end{equation}
with $b_\mathrm{h}(M)$ the halo bias prescription of Ref.~\cite{Tinker2010}, and using Eqs.~(\ref{eq:xi_hm})--(\ref{eq:DeltaSigma_hm}) to calculate the lensing 2-halo term from the power spectrum. The predicted signal for each redshift bin is calculated using Eq.~(\ref{eq:DeltaSigma_pred}), adopting the halo mass function of Ref.~\cite{Tinker:2008ff}. In this implementation we calculate $b_\mathrm{h}$, $P_\mathrm{mm}$, and $n_h$ using the Core Cosmology Library \cite{Chisari2019}. We adopt a flat prior 
$0.5 \le \mathcal{A}_\mathrm{2h} \le 2.0$
to give the model some freedom over these choices.

Finally, we assume that a fraction 
$f_\mathrm{off}$ of clusters are off-centered following a Rayleigh distribution \citep{Johnston2007} with a characteristic width $\sigma_\mathrm{mis}$, independent of mass and redshift. Consistent with the fiducial model, we set wide priors for these parameters, making them effective parameters capturing any small-scale deviations from our model: $f_\mathrm{off}\sim\mathcal{U}(0, 1)$ and $\sigma_\mathrm{off}\sim\mathcal{U}(0, 1)\,h^{-1}$Mpc.

As in the fiducial models, we fit the model above independently to each of the redshift bins. 

\begin{table}
\centering
\begin{tabular}{c|ccc}
\hline
Redshift bin &  $a_0'$ & $\sigma_{\log M_\mathrm{UPP}}$$^\dag$ & $\mathcal{A}_\mathrm{2h}$ \\
\hline
 $0.092 < z \le 0.445$ & $13.82_{-0.57}^{+0.56}$ & $\leq0.48$ & $1.0_{-0.3}^{+0.4}$ \\
 $0.445 < z \le 0.695$ & $13.86_{-0.59}^{+0.53}$ & $\leq0.44$ & $1.1_{-0.4}^{+0.5}$ \\
 $0.695 < z \le 1.180$ & $13.86_{-0.60}^{+0.44}$ & $\leq0.40$ & $1.0_{-0.3}^{+0.5}$\\
 \hline
\end{tabular}
\caption{Marginalized posterior constraints on the alternative model of Sec.\ \ref{subsec:model_choice}.
$^\dag$95\% upper limits.}
\label{t:alt}
\end{table}


We list the best-fit parameters in Table \ref{t:alt}. Miscentering parameters are unconstrained in all cases.
The resulting mean mass bias at the average $M_\mathrm{UPP}$ at each redshift bin are 
\begin{equation}\label{eq:Balt}
B^\mathrm{alt} = 
\begin{cases}
1.3 \pm 0.2 & (0.092 < z \le 0.445) \\
1.3 \pm 0.2 & (0.445 < z \le 0.695) \\
1.6_{-0.4}^{+0.5} & (0.695 < z \le 1.180),
\end{cases}
\end{equation}
where the minimum $\chi^2$ of 
each redshift bin are 14.2, 20.0, 12.9, resulting in PTEs of 0.72, 0.33, 0.61.
Eq.~\ref{eq:Balt} should be directly compared with our fiducial limits of Eq.~\ref{eq:B}. 
Hence, we find that the mass-bias limits by our stacked lensing measurements can be affected by choices of cluster modeling with a $1\sigma$ level at largest.
A similar systematic effect has been reported by our previous analysis \cite{Miyatake:2018lpb}.


\section{Limitation}\label{sec:limit}

Before concluding, we here summarize the major limitations of our analysis.
The results in Section~\ref{sec:results} depend strongly on our forward model of galaxy clusters selected by the thermal SZ effect in CMB observations.
We note that our model relies on three assumptions; 
(i) that the completeness of the SZ-selected clusters is determined by SZ masses and redshifts alone, (ii) that the stacked lensing signals around galaxy clusters beyond their virial radii are determined by total cluster masses, redshifts, and cosmology,
and (iii) that the lensing effect induced by a single cluster 
is independent of the SZ effect of that cluster.
The validity of these assumptions will be addressed in future studies.
In the following, we briefly discuss what can break the assumptions and 
how to improve our halo-based forward modeling of the SZ-selected clusters.

\subsection{Completeness of SZ-selected clusters}

The completeness of the list of galaxy clusters in a given SZ observation plays 
a critical role in developing a precise prediction of stacked lensing 
measurements around the selected clusters. 
In this paper, we adopt the method developed in Ref.~\cite{ACT:2020lcv} 
to estimate the completeness. 
This is a simulation-based method assuming the thermal SZ scaling relation in Ref.~\cite{Arnaud:2009tt} with a log-normal scatter and random Gaussian noise in accord with the ACT observation.
Note that the electron-pressure model in Ref.~\cite{Arnaud:2009tt} has been constructed with a cluster sample at $z<0.3$; recent X-ray and thermal SZ observations, however, have confirmed that it can provide a reasonable fit to the average pressure profiles in cluster outskirts even for higher redshifts up to $z\sim1$ \cite{Sayers:2012vs, SPT:2014sco, Sayers:2016sro, Romero:2016kac, Bourdin:2017ufp, Ghirardini:2017ugk, SPT:2017ees, Ghirardini:2018byi, Ghirardini:2020vyv, Pointecouteau:2021huw, Sayers:2022hue}.

In addition to its average, the scatter in the thermal SZ effect of individual clusters
can induce possible systematic effects in estimating the completeness.
For instance, Ref.~\cite{Green:2020vak} developed a semi-analytic model of spherical-symmetric pressure profiles of intracluster media in the presence of mass accretion, finding that the scatter in the thermal SZ-mass scaling relation on cluster scales can be closely related to the amount of assembly-driven, non-thermal pressure.
Because the non-thermal pressure in cluster outskirts becomes more significant as the clusters experience mass assembly at a faster accretion rate \cite{Nelson:2014vda},
the thermal SZ observation can preferentially select galaxy clusters with {\it slower} 
accretion rates (i.e. less non-thermal pressure support) at a given halo mass.
In the end, stacked lensing signals around the SZ-selected clusters can thus depend on the mass accretion history of the clusters \cite{Diemer:2014xya, 2019ApJ...883...36S}, which our halo-based model does not take into account.
Also, compact radio sources can cause addtional scatter in the cluster scaling relations.
Deep radio-wavelength observations would help to quantify contaminations from the point sources \cite{Coble:2006ff, Dicker:2021ijw}. 

To examine the completeness of SZ-selected clusters in detail, it would be useful to have samples of galaxy clusters selected by weak-lensing observations 
of large-scale structures \cite{Hamana:2003ts, Hennawi:2004ai, Maturi:2004rn}.
Lensing-selected clusters are predominantly sensitive to the underlying mass distributions around the clusters \cite{2012MNRAS.425.2287H, 2020ApJ...891..139C}, providing comprehensive information about the selection function of galaxy clusters.
As a matter of fact, the HSC survey is aimed at constructing a large sample of 
lensing-selected clusters \cite{Miyazaki:2018dpu, Oguri:2021vro}. 
Measurements of thermodynamic properties of lensing-selected clusters can 
give insight into cluster astrophysics \cite{Giles:2014bea}.

\subsection{Secondary bias effects on galaxy-cluster scales}

Our inference of a representative halo mass of the SZ-selected clusters 
by the stacked lensing measurements partly relies on a theoretical model of 
the mass distribution around the clusters beyond their virial radii. 
In a halo-model language (see Ref.~\cite{Cooray:2002dia} for a review of the halo model),
the stacked lensing signal at large radial distances can be explained by the two-halo term,
describing clustering effects of mass density between neighboring clusters \cite{Hayashi:2007uk}.
At the lowest order, the two-halo term is proportional to the linear biasing factor and the correlation function of mass density fields, which are merely a function of the cluster masses, redshifts, and cosmological models.

However, numerical simulations have shown that the linear biases of dark matter halos can depend not only on their masses but also on other properties defined in their inner density profiles at a given redshift \cite{Gao:2006qz, Faltenbacher:2009fj}.
These are commonly referred to as secondary bias effects, and can potentially lead to a biased estimate of the cluster mass from stacked lensing analyses.
Given the correlation between the thermal SZ effect and the mass accretion history,  
secondary bias on cluster scales can be important for future lensing analyses 
with high precision. 
Note that a detailed characterization of the secondary bias of cluster-sized halos requires a careful assessment of the mass assembly history \cite{Mao:2017aym, Chue:2018hxk, 2019ApJ...883...36S}.
A large-volume numerical simulation with a fine-time sampling
is urgently needed to improve our understanding of the secondary bias on cluster scales.

\subsection{Correlated scatter between stacked lensing signals and cluster observables}

Our halo-based model of the stacked lensing signals implicitly assumes 
that the SZ mass of a single cluster does not correlate with 
the inner mass density profile of the cluster.
This is just an assumption and further investigations into possible correlations between the thermal SZ and lensing signals may be needed.
In fact, numerical simulations have shown that the volume-integrated Compton $Y$ parameter of a galaxy cluster can correlate with their cluster lensing mass when one infers those from projected pressure and mass density fields \cite{Shirasaki:2016buk}.
The correlation can arise from coherent structures of gas pressure and mass density along a line of sight. The importance of such possible correlations in stacked lensing analyses around SZ-selected clusters is still uncertain, whereas 
recent studies of optically-selected galaxy clusters have pointed out that the correlated scatter between the lensing signal and the cluster richness can provide a biased estimate of the linear halo bias \cite{Sunayama:2020wvq} and the halo mass \cite{DES:2022qkn}.
To examine the correlated scatter for SZ-selected clusters, we require a careful analysis based on realistic mock observations of CMB anisotropies as well as weak lensing shear.

\section{Conclusion}\label{sec:conclusion}

In this paper, we performed a stacked lensing analysis of galaxy clusters selected 
through thermal Sunyaev-Zel'dovich (SZ) effects 
in observations of the cosmic microwave background by the Atacama Cosmology Telescope.
For the analysis, we used the galaxy shape catalog constructed from 
the Subaru Hyper-Suprime Cam (HSC) Year 3 data.
The large number density in the HSC galaxy catalog allows us to study 
the stacked lensing signal of galaxy clusters as a function of cluster redshift.
We improved upon our previous analysis \cite{Miyatake:2018lpb} by 
increasing the number of galaxy clusters, 
using a precise estimation of statistical errors in our measurements 
based on mock galaxy and cluster catalogs,
adopting a simulation-based prior on the scatter in an observable-mass scaling relation, and accounting for possible baryonic effects on the stacked lensing signals with a semi-analytic model as motivated by Refs.~\cite{Schneider:2015wta, Schneider:2018pfw, Giri:2021qin}.
We worked with a halo-based forward model 
to infer statistical relationships between cluster masses estimated 
by the thermal SZ analysis (i.e. the SZ masses) and weak-lensing counterparts.
Our model relies on an accurate emulation tool, the Dark Emulator \cite{Nishimichi:2018etk}, for a baseline prediction of statistics of cluster-sized dark matter halos, while modifying the emulator's output so that the model includes possible changes in the mass density distribution due to the existence of stellar and gas components in the clusters.
Our findings in this study are summarized below.

\begin{enumerate}

\item[(i)] Imposing a selection of foreground clusters 
with $>5\sigma$ significance in the thermal SZ analysis and background source galaxies with secure redshift cuts, we found that the cumulative signal-to-noise ratio of our stacked lensing measurements in the range of $0.1<R\, [h^{-1}\mathrm{Mpc}]<100$ reaches 14.6, 12.0 and 6.6 at the cluster redshifts of $0.092 < z \le 0.445$, $0.445 < z \le 0.695$, and $0.695 < z \le 1.180$, respectively. 
Our lensing analysis passed several null tests about B-mode signals and boost factors.
We also confirmed that the measurements are only minimally affected by the choice of cluster centers and details in photometric redshift estimations. 

\item[(ii)] We inferred the statistical relationship between the SZ and lensing masses with the MCMC analysis including a loose prior about the cluster abundance. Our best-fit model provided a reasonable chi-squared value with a p-value larger than 0.01. We found that the abundance prior helps to break the inherent parameter degeneracy between the mean and the scatter in the mass scaling relation. Our stacked lensing measurements are able to constrain a parameter describing the mean scaling relation at a level of 15\% at different cluster redshifts, while the scatter is not well constrained and is mostly determined by the prior. We also observed that baryonic effects are not important in our parameter inference at this stage.

\item[(iii)] Our analysis put a constraint of the mass bias parameter $B = M_\mathrm{500c} / M_\mathrm{UPP}$, where $M_\mathrm{500c}$ is the spherical overdensity mass with respect to 500 times critical density and $M_\mathrm{UPP}$ is the counterpart estimated by the thermal SZ analysis.
For the average $M_\mathrm{UPP}$ of $2.1-2.4\times10^{14}\, h^{-1}M_\odot$ in our cluster selection, we found that 
$B = 1.3 \pm 0.2,
1.6 \pm 0.2$
and $1.6 \pm 0.3$
at the redshifts of $0.092 < z \le 0.445$, $0.445 < z \le 0.695$, and $0.695 < z \le 1.180$, respectively (the errors are set at 68\% credible intervals).
The constraints were found to be consistent with ones by previous lensing analyses at a $1\sigma$ level, whereas the choice of the cluster forward modeling can shift the mass bias by $\sim1\sigma$ at largest.

\end{enumerate}

Despite this careful analysis, we realize that the results in this paper depend on 
details of our forward modeling approach.
We discussed some limitations of our model in Section~\ref{sec:limit}, prompting further investigation of possible biases in the thermal SZ-based selection of galaxy clusters and of secondary bias effects of cluster-sized dark matter halos, as well as detailed characterization of stacked lensing signals around the thermal SZ-selected clusters with realistic mock catalogs.
The HSC survey will provide their final release of weak lensing shear data, increasing the survey area by a factor of $\sim3$ in a timely manner.
A detailed mass calibration of the thermal SZ-selected clusters by a stacked lensing analysis with the HSC final data can deepen our understanding of the deficit in cluster masses inferred from visible and dark elements, unlocking clues about structure formation of the most massive objects in the universe.

\begin{acknowledgments}

Support for ACT was through the U.S.~National Science Foundation through awards AST-0408698, AST-0965625, and AST-1440226 for the ACT project, as well as awards PHY-0355328, PHY-0855887 and PHY-1214379. Funding was also provided by Princeton University, the University of Pennsylvania, and a Canada Foundation for Innovation (CFI) award to UBC. ACT operated in the Parque Astron\'omico Atacama in northern Chile under the auspices of the Agencia Nacional de Investigaci\'on y Desarrollo (ANID). The development of multichroic detectors and lenses was supported by NASA grants NNX13AE56G and NNX14AB58G. Detector research at NIST was supported by the NIST Innovations in Measurement Science program.  Computing for ACT was performed using the Princeton Research Computing resources at Princeton University, the National Energy Research Scientific Computing Center (NERSC),  and the Niagara supercomputer at the SciNet HPC Consortium. SciNet is funded by the CFI under the auspices of Compute Canada, the Government of Ontario, the Ontario Research Fund-Research Excellence, and the University of Toronto. We thank the Republic of Chile for hosting ACT in the northern Atacama, and the local indigenous Licanantay communities whom we follow in observing and learning from the night sky.

The Hyper Suprime-Cam (HSC) collaboration includes
the astronomical communities of Japan and Taiwan, and
Princeton University. The HSC instrumentation and software 
were developed by the National Astronomical
Observatory of Japan (NAOJ), the Kavli Institute for
the Physics and Mathematics of the Universe (Kavli
IPMU), the University of Tokyo, the High Energy Accelerator Research Organization (KEK), 
the Academia Sinica Institute for Astronomy and Astrophysics 
in Taiwan (ASIAA), and Princeton University. Funding was
contributed by the FIRST program from Japanese Cabinet Office, 
the Ministry of Education, Culture, Sports,
Science and Technology (MEXT), the Japan Society for
the Promotion of Science (JSPS), Japan Science and
Technology Agency (JST), the Toray Science Foundation, 
NAOJ, Kavli IPMU, KEK, ASIAA, and Princeton University.

This paper makes use of software developed for the
Large Synoptic Survey Telescope. We thank the LSST
Project for making their code available as free software
at \url{http://dm.lsst.org}.

The Pan-STARRS1 Surveys (PS1) have been made
possible through contributions of the Institute for Astronomy, 
the University of Hawaii, the Pan-STARRS
Project Office, the Max-Planck Society, and its participating institutes, 
the Max Planck Institute for Astronomy, Heidelberg, and the Max Planck Institute 
for Extraterrestrial Physics, Garching, The Johns Hopkins University, 
Durham University, the University of Edinburgh,
Queen’s University Belfast, the Harvard-Smithsonian
Center for Astrophysics, the Las Cumbres Observatory
Global Telescope Network Incorporated, the National
Central University of Taiwan, the Space Telescope Science Institute, 
the National Aeronautics and Space Administration under Grant No. NNX08AR22G issued
through the Planetary Science Division of 
the NASA Science Mission Directorate, 
the National Science Foundation under Grant No. AST-1238877, 
the University of Maryland, Eotvos Lorand University (ELTE) and
the Los Alamos National Laboratory.

Based on data collected at the Subaru Telescope and
retrieved from the HSC data archive system, 
which is operated by Subaru Telescope and Astronomy Data Center,
National Astronomical Observatory of Japan.

We thank the referee and Bruce Partridge for providing useful comments on the manuscript.
This work is supported by MEXT KAKENHI Grant Numbers,
19K14767, 20H05861, 20H01932, 20H05855, 21H05456, 22K21349, 23H00108 and 23K19070.
This work is also supported by Tokai Pathways to Global Excellence (T-GEx), 
part of MEXT Strategic Professional Development Program for Young Researchers.
CS acknowledges support from the Agencia Nacional de Investigaci\'on y Desarrollo (ANID) through Basal project FB210003.
DN is supported by NSF (AST-2206055 \& 2307280) and NASA (80NSSC22K0821 \& TM3-24007X) grants.
MH acknowledges support from the National Research Foundation of South Africa (grant no. 137975).
A part of numerical computations was carried out on Cray XC50 at the Center for Computational Astrophysics in NAOJ.
\end{acknowledgments}


\bibliography{ref}

\begin{thebibliography}{137}%
\makeatletter
\providecommand \@ifxundefined [1]{%
 \@ifx{#1\undefined}
}%
\providecommand \@ifnum [1]{%
 \ifnum #1\expandafter \@firstoftwo
 \else \expandafter \@secondoftwo
 \fi
}%
\providecommand \@ifx [1]{%
 \ifx #1\expandafter \@firstoftwo
 \else \expandafter \@secondoftwo
 \fi
}%
\providecommand \natexlab [1]{#1}%
\providecommand \enquote  [1]{``#1''}%
\providecommand \bibnamefont  [1]{#1}%
\providecommand \bibfnamefont [1]{#1}%
\providecommand \citenamefont [1]{#1}%
\providecommand \href@noop [0]{\@secondoftwo}%
\providecommand \href [0]{\begingroup \@sanitize@url \@href}%
\providecommand \@href[1]{\@@startlink{#1}\@@href}%
\providecommand \@@href[1]{\endgroup#1\@@endlink}%
\providecommand \@sanitize@url [0]{\catcode `\\12\catcode `\$12\catcode
  `\&12\catcode `\#12\catcode `\^12\catcode `\_12\catcode `\%12\relax}%
\providecommand \@@startlink[1]{}%
\providecommand \@@endlink[0]{}%
\providecommand \url  [0]{\begingroup\@sanitize@url \@url }%
\providecommand \@url [1]{\endgroup\@href {#1}{\urlprefix }}%
\providecommand \urlprefix  [0]{URL }%
\providecommand \Eprint [0]{\href }%
\providecommand \doibase [0]{https://doi.org/}%
\providecommand \selectlanguage [0]{\@gobble}%
\providecommand \bibinfo  [0]{\@secondoftwo}%
\providecommand \bibfield  [0]{\@secondoftwo}%
\providecommand \translation [1]{[#1]}%
\providecommand \BibitemOpen [0]{}%
\providecommand \bibitemStop [0]{}%
\providecommand \bibitemNoStop [0]{.\EOS\space}%
\providecommand \EOS [0]{\spacefactor3000\relax}%
\providecommand \BibitemShut  [1]{\csname bibitem#1\endcsname}%
\let\auto@bib@innerbib\@empty
\bibitem [{\citenamefont {Press}\ and\ \citenamefont
  {Schechter}(1974)}]{Press:1973iz}%
  \BibitemOpen
  \bibfield  {author} {\bibinfo {author} {\bibfnamefont {W.~H.}\ \bibnamefont
  {Press}}\ and\ \bibinfo {author} {\bibfnamefont {P.}~\bibnamefont
  {Schechter}},\ }\bibfield  {title} {\bibinfo {title} {{Formation of galaxies
  and clusters of galaxies by selfsimilar gravitational condensation}},\ }\href
  {https://doi.org/10.1086/152650} {\bibfield  {journal} {\bibinfo  {journal}
  {Astrophys. J.}\ }\textbf {\bibinfo {volume} {187}},\ \bibinfo {pages} {425}
  (\bibinfo {year} {1974})}\BibitemShut {NoStop}%
\bibitem [{\citenamefont {Sheth}(1998)}]{Sheth:1998ew}%
  \BibitemOpen
  \bibfield  {author} {\bibinfo {author} {\bibfnamefont {R.~K.}\ \bibnamefont
  {Sheth}},\ }\bibfield  {title} {\bibinfo {title} {{An Excursion set model for
  the distribution of dark matter and dark matter halos}},\ }\href
  {https://doi.org/10.1046/j.1365-8711.1998.01976.x} {\bibfield  {journal}
  {\bibinfo  {journal} {Mon. Not. Roy. Astron. Soc.}\ }\textbf {\bibinfo
  {volume} {300}},\ \bibinfo {pages} {1057} (\bibinfo {year} {1998})},\ \Eprint
  {https://arxiv.org/abs/astro-ph/9805319} {arXiv:astro-ph/9805319}
  \BibitemShut {NoStop}%
\bibitem [{\citenamefont {Tinker}\ \emph {et~al.}(2008)\citenamefont {Tinker},
  \citenamefont {Kravtsov}, \citenamefont {Klypin}, \citenamefont {Abazajian},
  \citenamefont {Warren}, \citenamefont {Yepes}, \citenamefont {Gottlober},\
  and\ \citenamefont {Holz}}]{Tinker:2008ff}%
  \BibitemOpen
  \bibfield  {author} {\bibinfo {author} {\bibfnamefont {J.~L.}\ \bibnamefont
  {Tinker}}, \bibinfo {author} {\bibfnamefont {A.~V.}\ \bibnamefont
  {Kravtsov}}, \bibinfo {author} {\bibfnamefont {A.}~\bibnamefont {Klypin}},
  \bibinfo {author} {\bibfnamefont {K.}~\bibnamefont {Abazajian}}, \bibinfo
  {author} {\bibfnamefont {M.~S.}\ \bibnamefont {Warren}}, \bibinfo {author}
  {\bibfnamefont {G.}~\bibnamefont {Yepes}}, \bibinfo {author} {\bibfnamefont
  {S.}~\bibnamefont {Gottlober}},\ and\ \bibinfo {author} {\bibfnamefont
  {D.~E.}\ \bibnamefont {Holz}},\ }\bibfield  {title} {\bibinfo {title}
  {{Toward a halo mass function for precision cosmology: The Limits of
  universality}},\ }\href {https://doi.org/10.1086/591439} {\bibfield
  {journal} {\bibinfo  {journal} {Astrophys. J.}\ }\textbf {\bibinfo {volume}
  {688}},\ \bibinfo {pages} {709} (\bibinfo {year} {2008})},\ \Eprint
  {https://arxiv.org/abs/0803.2706} {arXiv:0803.2706 [astro-ph]} \BibitemShut
  {NoStop}%
\bibitem [{\citenamefont {{Bhattacharya}}\ \emph {et~al.}(2011)\citenamefont
  {{Bhattacharya}}, \citenamefont {{Heitmann}}, \citenamefont {{White}},
  \citenamefont {{Luki{\'c}}}, \citenamefont {{Wagner}},\ and\ \citenamefont
  {{Habib}}}]{2011ApJ...732..122B}%
  \BibitemOpen
  \bibfield  {author} {\bibinfo {author} {\bibfnamefont {S.}~\bibnamefont
  {{Bhattacharya}}}, \bibinfo {author} {\bibfnamefont {K.}~\bibnamefont
  {{Heitmann}}}, \bibinfo {author} {\bibfnamefont {M.}~\bibnamefont {{White}}},
  \bibinfo {author} {\bibfnamefont {Z.}~\bibnamefont {{Luki{\'c}}}}, \bibinfo
  {author} {\bibfnamefont {C.}~\bibnamefont {{Wagner}}},\ and\ \bibinfo
  {author} {\bibfnamefont {S.}~\bibnamefont {{Habib}}},\ }\bibfield  {title}
  {\bibinfo {title} {{Mass Function Predictions Beyond
  {\ensuremath{\Lambda}}CDM}},\ }\href
  {https://doi.org/10.1088/0004-637X/732/2/122} {\bibfield  {journal} {\bibinfo
   {journal} {\apj}\ }\textbf {\bibinfo {volume} {732}},\ \bibinfo {eid} {122}
  (\bibinfo {year} {2011})},\ \Eprint {https://arxiv.org/abs/1005.2239}
  {arXiv:1005.2239 [astro-ph.CO]} \BibitemShut {NoStop}%
\bibitem [{\citenamefont {Watson}\ \emph {et~al.}(2013)\citenamefont {Watson},
  \citenamefont {Iliev}, \citenamefont {D'Aloisio}, \citenamefont {Knebe},
  \citenamefont {Shapiro},\ and\ \citenamefont {Yepes}}]{Watson:2012mt}%
  \BibitemOpen
  \bibfield  {author} {\bibinfo {author} {\bibfnamefont {W.~A.}\ \bibnamefont
  {Watson}}, \bibinfo {author} {\bibfnamefont {I.~T.}\ \bibnamefont {Iliev}},
  \bibinfo {author} {\bibfnamefont {A.}~\bibnamefont {D'Aloisio}}, \bibinfo
  {author} {\bibfnamefont {A.}~\bibnamefont {Knebe}}, \bibinfo {author}
  {\bibfnamefont {P.~R.}\ \bibnamefont {Shapiro}},\ and\ \bibinfo {author}
  {\bibfnamefont {G.}~\bibnamefont {Yepes}},\ }\bibfield  {title} {\bibinfo
  {title} {{The halo mass function through the cosmic ages}},\ }\href
  {https://doi.org/10.1093/mnras/stt791} {\bibfield  {journal} {\bibinfo
  {journal} {Mon. Not. Roy. Astron. Soc.}\ }\textbf {\bibinfo {volume} {433}},\
  \bibinfo {pages} {1230} (\bibinfo {year} {2013})},\ \Eprint
  {https://arxiv.org/abs/1212.0095} {arXiv:1212.0095 [astro-ph.CO]}
  \BibitemShut {NoStop}%
\bibitem [{\citenamefont {Despali}\ \emph {et~al.}(2016)\citenamefont
  {Despali}, \citenamefont {Giocoli}, \citenamefont {Angulo}, \citenamefont
  {Tormen}, \citenamefont {Sheth}, \citenamefont {Baso},\ and\ \citenamefont
  {Moscardini}}]{Despali:2015yla}%
  \BibitemOpen
  \bibfield  {author} {\bibinfo {author} {\bibfnamefont {G.}~\bibnamefont
  {Despali}}, \bibinfo {author} {\bibfnamefont {C.}~\bibnamefont {Giocoli}},
  \bibinfo {author} {\bibfnamefont {R.~E.}\ \bibnamefont {Angulo}}, \bibinfo
  {author} {\bibfnamefont {G.}~\bibnamefont {Tormen}}, \bibinfo {author}
  {\bibfnamefont {R.~K.}\ \bibnamefont {Sheth}}, \bibinfo {author}
  {\bibfnamefont {G.}~\bibnamefont {Baso}},\ and\ \bibinfo {author}
  {\bibfnamefont {L.}~\bibnamefont {Moscardini}},\ }\bibfield  {title}
  {\bibinfo {title} {{The universality of the virial halo mass function and
  models for non-universality of other halo definitions}},\ }\href
  {https://doi.org/10.1093/mnras/stv2842} {\bibfield  {journal} {\bibinfo
  {journal} {Mon. Not. Roy. Astron. Soc.}\ }\textbf {\bibinfo {volume} {456}},\
  \bibinfo {pages} {2486} (\bibinfo {year} {2016})},\ \Eprint
  {https://arxiv.org/abs/1507.05627} {arXiv:1507.05627 [astro-ph.CO]}
  \BibitemShut {NoStop}%
\bibitem [{\citenamefont {Shirasaki}\ \emph {et~al.}(2021)\citenamefont
  {Shirasaki}, \citenamefont {Ishiyama},\ and\ \citenamefont
  {Ando}}]{Shirasaki:2021orc}%
  \BibitemOpen
  \bibfield  {author} {\bibinfo {author} {\bibfnamefont {M.}~\bibnamefont
  {Shirasaki}}, \bibinfo {author} {\bibfnamefont {T.}~\bibnamefont
  {Ishiyama}},\ and\ \bibinfo {author} {\bibfnamefont {S.}~\bibnamefont
  {Ando}},\ }\bibfield  {title} {\bibinfo {title} {{Virial Halo Mass Function
  in the Planck Cosmology}},\ }\href {https://doi.org/10.3847/1538-4357/ac214b}
  {\bibfield  {journal} {\bibinfo  {journal} {Astrophys. J.}\ }\textbf
  {\bibinfo {volume} {922}},\ \bibinfo {pages} {89} (\bibinfo {year} {2021})},\
  \Eprint {https://arxiv.org/abs/2108.11038} {arXiv:2108.11038 [astro-ph.CO]}
  \BibitemShut {NoStop}%
\bibitem [{\citenamefont {Allen}\ \emph {et~al.}(2011)\citenamefont {Allen},
  \citenamefont {Evrard},\ and\ \citenamefont {Mantz}}]{Allen:2011zs}%
  \BibitemOpen
  \bibfield  {author} {\bibinfo {author} {\bibfnamefont {S.~W.}\ \bibnamefont
  {Allen}}, \bibinfo {author} {\bibfnamefont {A.~E.}\ \bibnamefont {Evrard}},\
  and\ \bibinfo {author} {\bibfnamefont {A.~B.}\ \bibnamefont {Mantz}},\
  }\bibfield  {title} {\bibinfo {title} {{Cosmological Parameters from
  Observations of Galaxy Clusters}},\ }\href
  {https://doi.org/10.1146/annurev-astro-081710-102514} {\bibfield  {journal}
  {\bibinfo  {journal} {Ann. Rev. Astron. Astrophys.}\ }\textbf {\bibinfo
  {volume} {49}},\ \bibinfo {pages} {409} (\bibinfo {year} {2011})},\ \Eprint
  {https://arxiv.org/abs/1103.4829} {arXiv:1103.4829 [astro-ph.CO]}
  \BibitemShut {NoStop}%
\bibitem [{\citenamefont {Sunyaev}\ and\ \citenamefont
  {Zeldovich}(1972)}]{Sunyaev:1972eq}%
  \BibitemOpen
  \bibfield  {author} {\bibinfo {author} {\bibfnamefont {R.~A.}\ \bibnamefont
  {Sunyaev}}\ and\ \bibinfo {author} {\bibfnamefont {Y.~B.}\ \bibnamefont
  {Zeldovich}},\ }\bibfield  {title} {\bibinfo {title} {{The Observations of
  relic radiation as a test of the nature of X-Ray radiation from the clusters
  of galaxies}},\ }\href@noop {} {\bibfield  {journal} {\bibinfo  {journal}
  {Comments Astrophys. Space Phys.}\ }\textbf {\bibinfo {volume} {4}},\
  \bibinfo {pages} {173} (\bibinfo {year} {1972})}\BibitemShut {NoStop}%
\bibitem [{\citenamefont {Ade}\ \emph {et~al.}(2016{\natexlab{a}})\citenamefont
  {Ade} \emph {et~al.}}]{Planck:2015koh}%
  \BibitemOpen
  \bibfield  {author} {\bibinfo {author} {\bibfnamefont {P.~A.~R.}\
  \bibnamefont {Ade}} \emph {et~al.} (\bibinfo {collaboration} {Planck}),\
  }\bibfield  {title} {\bibinfo {title} {{Planck 2015 results. XXVII. The
  Second Planck Catalogue of Sunyaev-Zeldovich Sources}},\ }\href
  {https://doi.org/10.1051/0004-6361/201525823} {\bibfield  {journal} {\bibinfo
   {journal} {Astron. Astrophys.}\ }\textbf {\bibinfo {volume} {594}},\
  \bibinfo {pages} {A27} (\bibinfo {year} {2016}{\natexlab{a}})},\ \Eprint
  {https://arxiv.org/abs/1502.01598} {arXiv:1502.01598 [astro-ph.CO]}
  \BibitemShut {NoStop}%
\bibitem [{\citenamefont {Bocquet}\ \emph {et~al.}(2019)\citenamefont {Bocquet}
  \emph {et~al.}}]{SPT:2018njh}%
  \BibitemOpen
  \bibfield  {author} {\bibinfo {author} {\bibfnamefont {S.}~\bibnamefont
  {Bocquet}} \emph {et~al.} (\bibinfo {collaboration} {SPT}),\ }\bibfield
  {title} {\bibinfo {title} {{Cluster Cosmology Constraints from the 2500
  deg$^2$ SPT-SZ Survey: Inclusion of Weak Gravitational Lensing Data from
  Magellan and the Hubble Space Telescope}},\ }\href
  {https://doi.org/10.3847/1538-4357/ab1f10} {\bibfield  {journal} {\bibinfo
  {journal} {Astrophys. J.}\ }\textbf {\bibinfo {volume} {878}},\ \bibinfo
  {pages} {55} (\bibinfo {year} {2019})},\ \Eprint
  {https://arxiv.org/abs/1812.01679} {arXiv:1812.01679 [astro-ph.CO]}
  \BibitemShut {NoStop}%
\bibitem [{\citenamefont {Hilton}\ \emph {et~al.}(2021)\citenamefont {Hilton}
  \emph {et~al.}}]{ACT:2020lcv}%
  \BibitemOpen
  \bibfield  {author} {\bibinfo {author} {\bibfnamefont {M.}~\bibnamefont
  {Hilton}} \emph {et~al.} (\bibinfo {collaboration} {ACT, DES}),\ }\bibfield
  {title} {\bibinfo {title} {{The Atacama Cosmology Telescope: A Catalog of
  \ensuremath{>}4000 Sunyaev\textendash{}Zel\textquoteright{}dovich Galaxy
  Clusters}},\ }\href {https://doi.org/10.3847/1538-4365/abd023} {\bibfield
  {journal} {\bibinfo  {journal} {Astrophys. J. Suppl.}\ }\textbf {\bibinfo
  {volume} {253}},\ \bibinfo {pages} {3} (\bibinfo {year} {2021})},\ \Eprint
  {https://arxiv.org/abs/2009.11043} {arXiv:2009.11043 [astro-ph.CO]}
  \BibitemShut {NoStop}%
\bibitem [{\citenamefont {Madhavacheril}\ \emph {et~al.}(2017)\citenamefont
  {Madhavacheril}, \citenamefont {Battaglia},\ and\ \citenamefont
  {Miyatake}}]{Madhavacheril:2017onh}%
  \BibitemOpen
  \bibfield  {author} {\bibinfo {author} {\bibfnamefont {M.~S.}\ \bibnamefont
  {Madhavacheril}}, \bibinfo {author} {\bibfnamefont {N.}~\bibnamefont
  {Battaglia}},\ and\ \bibinfo {author} {\bibfnamefont {H.}~\bibnamefont
  {Miyatake}},\ }\bibfield  {title} {\bibinfo {title} {{Fundamental physics
  from future weak-lensing calibrated Sunyaev-Zel\textquoteright{}dovich galaxy
  cluster counts}},\ }\href {https://doi.org/10.1103/PhysRevD.96.103525}
  {\bibfield  {journal} {\bibinfo  {journal} {Phys. Rev. D}\ }\textbf {\bibinfo
  {volume} {96}},\ \bibinfo {pages} {103525} (\bibinfo {year} {2017})},\
  \Eprint {https://arxiv.org/abs/1708.07502} {arXiv:1708.07502 [astro-ph.CO]}
  \BibitemShut {NoStop}%
\bibitem [{\citenamefont {Hasselfield}\ \emph {et~al.}(2013)\citenamefont
  {Hasselfield} \emph {et~al.}}]{Hasselfield:2013wf}%
  \BibitemOpen
  \bibfield  {author} {\bibinfo {author} {\bibfnamefont {M.}~\bibnamefont
  {Hasselfield}} \emph {et~al.},\ }\bibfield  {title} {\bibinfo {title} {{The
  Atacama Cosmology Telescope: Sunyaev-Zel'dovich selected galaxyclusters at
  148 GHz from three seasons of data}},\ }\href
  {https://doi.org/10.1088/1475-7516/2013/07/008} {\bibfield  {journal}
  {\bibinfo  {journal} {JCAP}\ }\textbf {\bibinfo {volume} {07}},\ \bibinfo
  {pages} {008}},\ \Eprint {https://arxiv.org/abs/1301.0816} {arXiv:1301.0816
  [astro-ph.CO]} \BibitemShut {NoStop}%
\bibitem [{\citenamefont {Ade}\ \emph {et~al.}(2016{\natexlab{b}})\citenamefont
  {Ade} \emph {et~al.}}]{Planck:2015lwi}%
  \BibitemOpen
  \bibfield  {author} {\bibinfo {author} {\bibfnamefont {P.~A.~R.}\
  \bibnamefont {Ade}} \emph {et~al.} (\bibinfo {collaboration} {Planck}),\
  }\bibfield  {title} {\bibinfo {title} {{Planck 2015 results. XXIV. Cosmology
  from Sunyaev-Zeldovich cluster counts}},\ }\href
  {https://doi.org/10.1051/0004-6361/201525833} {\bibfield  {journal} {\bibinfo
   {journal} {Astron. Astrophys.}\ }\textbf {\bibinfo {volume} {594}},\
  \bibinfo {pages} {A24} (\bibinfo {year} {2016}{\natexlab{b}})},\ \Eprint
  {https://arxiv.org/abs/1502.01597} {arXiv:1502.01597 [astro-ph.CO]}
  \BibitemShut {NoStop}%
\bibitem [{\citenamefont {Salvati}\ \emph {et~al.}(2022)\citenamefont {Salvati}
  \emph {et~al.}}]{Salvati:2021gkt}%
  \BibitemOpen
  \bibfield  {author} {\bibinfo {author} {\bibfnamefont {L.}~\bibnamefont
  {Salvati}} \emph {et~al.},\ }\bibfield  {title} {\bibinfo {title} {{Combining
  Planck and SPT Cluster Catalogs: Cosmological Analysis and Impact on the
  Planck Scaling Relation Calibration}},\ }\href
  {https://doi.org/10.3847/1538-4357/ac7ab4} {\bibfield  {journal} {\bibinfo
  {journal} {Astrophys. J.}\ }\textbf {\bibinfo {volume} {934}},\ \bibinfo
  {pages} {129} (\bibinfo {year} {2022})},\ \Eprint
  {https://arxiv.org/abs/2112.03606} {arXiv:2112.03606 [astro-ph.CO]}
  \BibitemShut {NoStop}%
\bibitem [{\citenamefont {Rosati}\ \emph {et~al.}(2002)\citenamefont {Rosati},
  \citenamefont {Borgani},\ and\ \citenamefont {Norman}}]{Rosati:2002ts}%
  \BibitemOpen
  \bibfield  {author} {\bibinfo {author} {\bibfnamefont {P.}~\bibnamefont
  {Rosati}}, \bibinfo {author} {\bibfnamefont {S.}~\bibnamefont {Borgani}},\
  and\ \bibinfo {author} {\bibfnamefont {C.}~\bibnamefont {Norman}},\
  }\bibfield  {title} {\bibinfo {title} {{The evolution of x-ray clusters of
  galaxies}},\ }\href {https://doi.org/10.1146/annurev.astro.40.120401.150547}
  {\bibfield  {journal} {\bibinfo  {journal} {Ann. Rev. Astron. Astrophys.}\
  }\textbf {\bibinfo {volume} {40}},\ \bibinfo {pages} {539} (\bibinfo {year}
  {2002})},\ \Eprint {https://arxiv.org/abs/astro-ph/0209035}
  {arXiv:astro-ph/0209035} \BibitemShut {NoStop}%
\bibitem [{\citenamefont {Pratt}\ \emph {et~al.}(2019)\citenamefont {Pratt},
  \citenamefont {Arnaud}, \citenamefont {Biviano}, \citenamefont {Eckert},
  \citenamefont {Ettori}, \citenamefont {Nagai}, \citenamefont {Okabe},\ and\
  \citenamefont {Reiprich}}]{Pratt:2019cnf}%
  \BibitemOpen
  \bibfield  {author} {\bibinfo {author} {\bibfnamefont {G.~W.}\ \bibnamefont
  {Pratt}}, \bibinfo {author} {\bibfnamefont {M.}~\bibnamefont {Arnaud}},
  \bibinfo {author} {\bibfnamefont {A.}~\bibnamefont {Biviano}}, \bibinfo
  {author} {\bibfnamefont {D.}~\bibnamefont {Eckert}}, \bibinfo {author}
  {\bibfnamefont {S.}~\bibnamefont {Ettori}}, \bibinfo {author} {\bibfnamefont
  {D.}~\bibnamefont {Nagai}}, \bibinfo {author} {\bibfnamefont
  {N.}~\bibnamefont {Okabe}},\ and\ \bibinfo {author} {\bibfnamefont {T.~H.}\
  \bibnamefont {Reiprich}},\ }\bibfield  {title} {\bibinfo {title} {{The galaxy
  cluster mass scale and its impact on cosmological constraints from the
  cluster population}},\ }\href {https://doi.org/10.1007/s11214-019-0591-0}
  {\bibfield  {journal} {\bibinfo  {journal} {Space Sci. Rev.}\ }\textbf
  {\bibinfo {volume} {215}},\ \bibinfo {pages} {25} (\bibinfo {year} {2019})},\
  \Eprint {https://arxiv.org/abs/1902.10837} {arXiv:1902.10837 [astro-ph.CO]}
  \BibitemShut {NoStop}%
\bibitem [{\citenamefont {Umetsu}(2020)}]{Umetsu:2020wlf}%
  \BibitemOpen
  \bibfield  {author} {\bibinfo {author} {\bibfnamefont {K.}~\bibnamefont
  {Umetsu}},\ }\bibfield  {title} {\bibinfo {title}
  {{Cluster\textendash{}galaxy weak lensing}},\ }\href
  {https://doi.org/10.1007/s00159-020-00129-w} {\bibfield  {journal} {\bibinfo
  {journal} {Astron. Astrophys. Rev.}\ }\textbf {\bibinfo {volume} {28}},\
  \bibinfo {pages} {7} (\bibinfo {year} {2020})},\ \Eprint
  {https://arxiv.org/abs/2007.00506} {arXiv:2007.00506 [astro-ph.CO]}
  \BibitemShut {NoStop}%
\bibitem [{\citenamefont {Okabe}\ \emph {et~al.}(2010)\citenamefont {Okabe},
  \citenamefont {Takada}, \citenamefont {Umetsu}, \citenamefont {Futamase},\
  and\ \citenamefont {Smith}}]{Okabe:2009pf}%
  \BibitemOpen
  \bibfield  {author} {\bibinfo {author} {\bibfnamefont {N.}~\bibnamefont
  {Okabe}}, \bibinfo {author} {\bibfnamefont {M.}~\bibnamefont {Takada}},
  \bibinfo {author} {\bibfnamefont {K.}~\bibnamefont {Umetsu}}, \bibinfo
  {author} {\bibfnamefont {T.}~\bibnamefont {Futamase}},\ and\ \bibinfo
  {author} {\bibfnamefont {G.~P.}\ \bibnamefont {Smith}},\ }\bibfield  {title}
  {\bibinfo {title} {{LoCuSS: Subaru Weak Lensing Study of 30 Galaxy
  Clusters}},\ }\href {https://doi.org/10.1093/pasj/62.3.811} {\bibfield
  {journal} {\bibinfo  {journal} {Publ. Astron. Soc. Jap.}\ }\textbf {\bibinfo
  {volume} {62}},\ \bibinfo {pages} {811} (\bibinfo {year} {2010})},\ \Eprint
  {https://arxiv.org/abs/0903.1103} {arXiv:0903.1103 [astro-ph.CO]}
  \BibitemShut {NoStop}%
\bibitem [{\citenamefont {Li}\ \emph {et~al.}(2022)\citenamefont {Li} \emph
  {et~al.}}]{Li:2021mvq}%
  \BibitemOpen
  \bibfield  {author} {\bibinfo {author} {\bibfnamefont {X.}~\bibnamefont {Li}}
  \emph {et~al.},\ }\bibfield  {title} {\bibinfo {title} {{The three-year shear
  catalog of the Subaru Hyper Suprime-Cam SSP Survey}},\ }\href
  {https://doi.org/10.1093/pasj/psac006} {\bibfield  {journal} {\bibinfo
  {journal} {Publ. Astron. Soc. Jap.}\ }\textbf {\bibinfo {volume} {74}},\
  \bibinfo {pages} {421} (\bibinfo {year} {2022})},\ \Eprint
  {https://arxiv.org/abs/2107.00136} {arXiv:2107.00136 [astro-ph.CO]}
  \BibitemShut {NoStop}%
\bibitem [{\citenamefont {Miyatake}\ \emph {et~al.}(2019)\citenamefont
  {Miyatake} \emph {et~al.}}]{Miyatake:2018lpb}%
  \BibitemOpen
  \bibfield  {author} {\bibinfo {author} {\bibfnamefont {H.}~\bibnamefont
  {Miyatake}} \emph {et~al.},\ }\bibfield  {title} {\bibinfo {title}
  {{Weak-lensing Mass Calibration of ACTPol
  Sunyaev\textendash{}Zel\textquoteright{}dovich Clusters with the Hyper
  Suprime-Cam Survey}},\ }\href {https://doi.org/10.3847/1538-4357/ab0af0}
  {\bibfield  {journal} {\bibinfo  {journal} {Astrophys. J.}\ }\textbf
  {\bibinfo {volume} {875}},\ \bibinfo {pages} {63} (\bibinfo {year} {2019})},\
  \Eprint {https://arxiv.org/abs/1804.05873} {arXiv:1804.05873 [astro-ph.CO]}
  \BibitemShut {NoStop}%
\bibitem [{\citenamefont {Takahashi}\ \emph {et~al.}(2017)\citenamefont
  {Takahashi}, \citenamefont {Hamana}, \citenamefont {Shirasaki}, \citenamefont
  {Namikawa}, \citenamefont {Nishimichi}, \citenamefont {Osato},\ and\
  \citenamefont {Shiroyama}}]{Takahashi:2017hjr}%
  \BibitemOpen
  \bibfield  {author} {\bibinfo {author} {\bibfnamefont {R.}~\bibnamefont
  {Takahashi}}, \bibinfo {author} {\bibfnamefont {T.}~\bibnamefont {Hamana}},
  \bibinfo {author} {\bibfnamefont {M.}~\bibnamefont {Shirasaki}}, \bibinfo
  {author} {\bibfnamefont {T.}~\bibnamefont {Namikawa}}, \bibinfo {author}
  {\bibfnamefont {T.}~\bibnamefont {Nishimichi}}, \bibinfo {author}
  {\bibfnamefont {K.}~\bibnamefont {Osato}},\ and\ \bibinfo {author}
  {\bibfnamefont {K.}~\bibnamefont {Shiroyama}},\ }\bibfield  {title} {\bibinfo
  {title} {{Full-sky Gravitational Lensing Simulation for Large-area Galaxy
  Surveys and Cosmic Microwave Background Experiments}},\ }\href
  {https://doi.org/10.3847/1538-4357/aa943d} {\bibfield  {journal} {\bibinfo
  {journal} {Astrophys. J.}\ }\textbf {\bibinfo {volume} {850}},\ \bibinfo
  {pages} {24} (\bibinfo {year} {2017})},\ \Eprint
  {https://arxiv.org/abs/1706.01472} {arXiv:1706.01472 [astro-ph.CO]}
  \BibitemShut {NoStop}%
\bibitem [{\citenamefont {Springel}\ \emph {et~al.}(2018)\citenamefont
  {Springel} \emph {et~al.}}]{Springel:2017tpz}%
  \BibitemOpen
  \bibfield  {author} {\bibinfo {author} {\bibfnamefont {V.}~\bibnamefont
  {Springel}} \emph {et~al.},\ }\bibfield  {title} {\bibinfo {title} {{First
  results from the IllustrisTNG simulations: matter and galaxy clustering}},\
  }\href {https://doi.org/10.1093/mnras/stx3304} {\bibfield  {journal}
  {\bibinfo  {journal} {Mon. Not. Roy. Astron. Soc.}\ }\textbf {\bibinfo
  {volume} {475}},\ \bibinfo {pages} {676} (\bibinfo {year} {2018})},\ \Eprint
  {https://arxiv.org/abs/1707.03397} {arXiv:1707.03397 [astro-ph.GA]}
  \BibitemShut {NoStop}%
\bibitem [{\citenamefont {Nishimichi}\ \emph {et~al.}(2019)\citenamefont
  {Nishimichi} \emph {et~al.}}]{Nishimichi:2018etk}%
  \BibitemOpen
  \bibfield  {author} {\bibinfo {author} {\bibfnamefont {T.}~\bibnamefont
  {Nishimichi}} \emph {et~al.},\ }\bibfield  {title} {\bibinfo {title} {{Dark
  Quest. I. Fast and Accurate Emulation of Halo Clustering Statistics and Its
  Application to Galaxy Clustering}},\ }\href
  {https://doi.org/10.3847/1538-4357/ab3719} {\bibfield  {journal} {\bibinfo
  {journal} {Astrophys. J.}\ }\textbf {\bibinfo {volume} {884}},\ \bibinfo
  {pages} {29} (\bibinfo {year} {2019})},\ \Eprint
  {https://arxiv.org/abs/1811.09504} {arXiv:1811.09504 [astro-ph.CO]}
  \BibitemShut {NoStop}%
\bibitem [{\citenamefont {Robertson}\ \emph {et~al.}(2024)\citenamefont
  {Robertson} \emph {et~al.}}]{Robertson:2023gxe}%
  \BibitemOpen
  \bibfield  {author} {\bibinfo {author} {\bibfnamefont {N.~C.}\ \bibnamefont
  {Robertson}} \emph {et~al.},\ }\bibfield  {title} {\bibinfo {title} {{ACT-DR5
  Sunyaev-Zel\textquoteright{}dovich clusters: Weak lensing mass calibration
  with KiDS}},\ }\href {https://doi.org/10.1051/0004-6361/202346712} {\bibfield
   {journal} {\bibinfo  {journal} {Astron. Astrophys.}\ }\textbf {\bibinfo
  {volume} {681}},\ \bibinfo {pages} {A87} (\bibinfo {year} {2024})},\ \Eprint
  {https://arxiv.org/abs/2304.10219} {arXiv:2304.10219 [astro-ph.CO]}
  \BibitemShut {NoStop}%
\bibitem [{\citenamefont {Navarro}\ \emph {et~al.}(1997)\citenamefont
  {Navarro}, \citenamefont {Frenk},\ and\ \citenamefont
  {White}}]{Navarro:1996gj}%
  \BibitemOpen
  \bibfield  {author} {\bibinfo {author} {\bibfnamefont {J.~F.}\ \bibnamefont
  {Navarro}}, \bibinfo {author} {\bibfnamefont {C.~S.}\ \bibnamefont {Frenk}},\
  and\ \bibinfo {author} {\bibfnamefont {S.~D.~M.}\ \bibnamefont {White}},\
  }\bibfield  {title} {\bibinfo {title} {{A Universal density profile from
  hierarchical clustering}},\ }\href {https://doi.org/10.1086/304888}
  {\bibfield  {journal} {\bibinfo  {journal} {Astrophys. J.}\ }\textbf
  {\bibinfo {volume} {490}},\ \bibinfo {pages} {493} (\bibinfo {year}
  {1997})},\ \Eprint {https://arxiv.org/abs/astro-ph/9611107}
  {arXiv:astro-ph/9611107} \BibitemShut {NoStop}%
\bibitem [{\citenamefont {Hinshaw}\ \emph {et~al.}(2013)\citenamefont {Hinshaw}
  \emph {et~al.}}]{WMAP:2012nax}%
  \BibitemOpen
  \bibfield  {author} {\bibinfo {author} {\bibfnamefont {G.}~\bibnamefont
  {Hinshaw}} \emph {et~al.} (\bibinfo {collaboration} {WMAP}),\ }\bibfield
  {title} {\bibinfo {title} {{Nine-Year Wilkinson Microwave Anisotropy Probe
  (WMAP) Observations: Cosmological Parameter Results}},\ }\href
  {https://doi.org/10.1088/0067-0049/208/2/19} {\bibfield  {journal} {\bibinfo
  {journal} {Astrophys. J. Suppl.}\ }\textbf {\bibinfo {volume} {208}},\
  \bibinfo {pages} {19} (\bibinfo {year} {2013})},\ \Eprint
  {https://arxiv.org/abs/1212.5226} {arXiv:1212.5226 [astro-ph.CO]}
  \BibitemShut {NoStop}%
\bibitem [{\citenamefont {Aghanim}\ \emph {et~al.}(2020)\citenamefont {Aghanim}
  \emph {et~al.}}]{Planck:2018vyg}%
  \BibitemOpen
  \bibfield  {author} {\bibinfo {author} {\bibfnamefont {N.}~\bibnamefont
  {Aghanim}} \emph {et~al.} (\bibinfo {collaboration} {Planck}),\ }\bibfield
  {title} {\bibinfo {title} {{Planck 2018 results. VI. Cosmological
  parameters}},\ }\href {https://doi.org/10.1051/0004-6361/201833910}
  {\bibfield  {journal} {\bibinfo  {journal} {Astron. Astrophys.}\ }\textbf
  {\bibinfo {volume} {641}},\ \bibinfo {pages} {A6} (\bibinfo {year} {2020})},\
  \bibinfo {note} {[Erratum: Astron.Astrophys. 652, C4 (2021)]},\ \Eprint
  {https://arxiv.org/abs/1807.06209} {arXiv:1807.06209 [astro-ph.CO]}
  \BibitemShut {NoStop}%
\bibitem [{\citenamefont {Naess}\ \emph {et~al.}(2020)\citenamefont {Naess}
  \emph {et~al.}}]{Naess:2020wgi}%
  \BibitemOpen
  \bibfield  {author} {\bibinfo {author} {\bibfnamefont {S.}~\bibnamefont
  {Naess}} \emph {et~al.},\ }\bibfield  {title} {\bibinfo {title} {{The Atacama
  Cosmology Telescope: arcminute-resolution maps of 18 000 square degrees of
  the microwave sky from ACT 2008\textendash{}2018 data combined with
  Planck}},\ }\href {https://doi.org/10.1088/1475-7516/2020/12/046} {\bibfield
  {journal} {\bibinfo  {journal} {JCAP}\ }\textbf {\bibinfo {volume} {12}},\
  \bibinfo {pages} {046}},\ \Eprint {https://arxiv.org/abs/2007.07290}
  {arXiv:2007.07290 [astro-ph.IM]} \BibitemShut {NoStop}%
\bibitem [{\citenamefont {Melin}\ \emph {et~al.}(2006)\citenamefont {Melin},
  \citenamefont {Bartlett},\ and\ \citenamefont {Delabrouille}}]{Melin:2006qq}%
  \BibitemOpen
  \bibfield  {author} {\bibinfo {author} {\bibfnamefont {J.-B.}\ \bibnamefont
  {Melin}}, \bibinfo {author} {\bibfnamefont {J.~G.}\ \bibnamefont
  {Bartlett}},\ and\ \bibinfo {author} {\bibfnamefont {J.}~\bibnamefont
  {Delabrouille}},\ }\bibfield  {title} {\bibinfo {title} {{Catalog extraction
  in sz cluster surveys: a matched filter approach}},\ }\href
  {https://doi.org/10.1051/0004-6361:20065034} {\bibfield  {journal} {\bibinfo
  {journal} {Astron. Astrophys.}\ }\textbf {\bibinfo {volume} {459}},\ \bibinfo
  {pages} {341} (\bibinfo {year} {2006})},\ \Eprint
  {https://arxiv.org/abs/astro-ph/0602424} {arXiv:astro-ph/0602424}
  \BibitemShut {NoStop}%
\bibitem [{\citenamefont {Williamson}\ \emph {et~al.}(2011)\citenamefont
  {Williamson} \emph {et~al.}}]{Williamson:2011jz}%
  \BibitemOpen
  \bibfield  {author} {\bibinfo {author} {\bibfnamefont {R.}~\bibnamefont
  {Williamson}} \emph {et~al.},\ }\bibfield  {title} {\bibinfo {title} {{An
  SZ-selected sample of the most massive galaxy clusters in the
  2500-square-degree South Pole Telescope survey}},\ }\href
  {https://doi.org/10.1088/0004-637X/738/2/139} {\bibfield  {journal} {\bibinfo
   {journal} {Astrophys. J.}\ }\textbf {\bibinfo {volume} {738}},\ \bibinfo
  {pages} {139} (\bibinfo {year} {2011})},\ \Eprint
  {https://arxiv.org/abs/1101.1290} {arXiv:1101.1290 [astro-ph.CO]}
  \BibitemShut {NoStop}%
\bibitem [{\citenamefont {Arnaud}\ \emph {et~al.}(2010)\citenamefont {Arnaud},
  \citenamefont {Pratt}, \citenamefont {Piffaretti}, \citenamefont
  {Boehringer}, \citenamefont {Croston},\ and\ \citenamefont
  {Pointecouteau}}]{Arnaud:2009tt}%
  \BibitemOpen
  \bibfield  {author} {\bibinfo {author} {\bibfnamefont {M.}~\bibnamefont
  {Arnaud}}, \bibinfo {author} {\bibfnamefont {G.~W.}\ \bibnamefont {Pratt}},
  \bibinfo {author} {\bibfnamefont {R.}~\bibnamefont {Piffaretti}}, \bibinfo
  {author} {\bibfnamefont {H.}~\bibnamefont {Boehringer}}, \bibinfo {author}
  {\bibfnamefont {J.~H.}\ \bibnamefont {Croston}},\ and\ \bibinfo {author}
  {\bibfnamefont {E.}~\bibnamefont {Pointecouteau}},\ }\bibfield  {title}
  {\bibinfo {title} {{The universal galaxy cluster pressure profile from a
  representative sample of nearby systems (REXCESS) and the Y\_SZ-M\_500
  relation}},\ }\href {https://doi.org/10.1051/0004-6361/200913416} {\bibfield
  {journal} {\bibinfo  {journal} {Astron. Astrophys.}\ }\textbf {\bibinfo
  {volume} {517}},\ \bibinfo {pages} {A92} (\bibinfo {year} {2010})},\ \Eprint
  {https://arxiv.org/abs/0910.1234} {arXiv:0910.1234 [astro-ph.CO]}
  \BibitemShut {NoStop}%
\bibitem [{\citenamefont {Blake}\ \emph {et~al.}(2016)\citenamefont {Blake}
  \emph {et~al.}}]{Blake:2016fcm}%
  \BibitemOpen
  \bibfield  {author} {\bibinfo {author} {\bibfnamefont {C.}~\bibnamefont
  {Blake}} \emph {et~al.},\ }\bibfield  {title} {\bibinfo {title} {{The
  2-degree Field Lensing Survey: design and clustering measurements}},\ }\href
  {https://doi.org/10.1093/mnras/stw1990} {\bibfield  {journal} {\bibinfo
  {journal} {Mon. Not. Roy. Astron. Soc.}\ }\textbf {\bibinfo {volume} {462}},\
  \bibinfo {pages} {4240} (\bibinfo {year} {2016})},\ \Eprint
  {https://arxiv.org/abs/1608.02668} {arXiv:1608.02668 [astro-ph.CO]}
  \BibitemShut {NoStop}%
\bibitem [{\citenamefont {Childress}\ \emph {et~al.}(2017)\citenamefont
  {Childress} \emph {et~al.}}]{DES:2017ndx}%
  \BibitemOpen
  \bibfield  {author} {\bibinfo {author} {\bibfnamefont {M.~J.}\ \bibnamefont
  {Childress}} \emph {et~al.} (\bibinfo {collaboration} {DES}),\ }\bibfield
  {title} {\bibinfo {title} {{OzDES multifibre spectroscopy for the Dark Energy
  Survey: 3-yr results and first data release}},\ }\href
  {https://doi.org/10.1093/mnras/stx1872} {\bibfield  {journal} {\bibinfo
  {journal} {Mon. Not. Roy. Astron. Soc.}\ }\textbf {\bibinfo {volume} {472}},\
  \bibinfo {pages} {273} (\bibinfo {year} {2017})},\ \Eprint
  {https://arxiv.org/abs/1708.04526} {arXiv:1708.04526 [astro-ph.CO]}
  \BibitemShut {NoStop}%
\bibitem [{\citenamefont {Ahumada}\ \emph {et~al.}(2020)\citenamefont {Ahumada}
  \emph {et~al.}}]{SDSS-IV:2019txh}%
  \BibitemOpen
  \bibfield  {author} {\bibinfo {author} {\bibfnamefont {R.}~\bibnamefont
  {Ahumada}} \emph {et~al.} (\bibinfo {collaboration} {SDSS-IV}),\ }\bibfield
  {title} {\bibinfo {title} {{The 16th Data Release of the Sloan Digital Sky
  Surveys: First Release from the APOGEE-2 Southern Survey and Full Release of
  eBOSS Spectra}},\ }\href {https://doi.org/10.3847/1538-4365/ab929e}
  {\bibfield  {journal} {\bibinfo  {journal} {Astrophys. J. Suppl.}\ }\textbf
  {\bibinfo {volume} {249}},\ \bibinfo {pages} {3} (\bibinfo {year} {2020})},\
  \Eprint {https://arxiv.org/abs/1912.02905} {arXiv:1912.02905 [astro-ph.GA]}
  \BibitemShut {NoStop}%
\bibitem [{\citenamefont {Scodeggio}\ \emph {et~al.}(2018)\citenamefont
  {Scodeggio} \emph {et~al.}}]{Scodeggio:2016jaf}%
  \BibitemOpen
  \bibfield  {author} {\bibinfo {author} {\bibfnamefont {M.}~\bibnamefont
  {Scodeggio}} \emph {et~al.},\ }\bibfield  {title} {\bibinfo {title} {{The
  VIMOS Public Extragalactic Redshift Survey (VIPERS). Full spectroscopic data
  and auxiliary information release (PDR-2)}},\ }\href
  {https://doi.org/10.1051/0004-6361/201630114} {\bibfield  {journal} {\bibinfo
   {journal} {Astron. Astrophys.}\ }\textbf {\bibinfo {volume} {609}},\
  \bibinfo {pages} {A84} (\bibinfo {year} {2018})},\ \Eprint
  {https://arxiv.org/abs/1611.07048} {arXiv:1611.07048 [astro-ph.GA]}
  \BibitemShut {NoStop}%
\bibitem [{\citenamefont {Rykoff}\ \emph {et~al.}(2014)\citenamefont {Rykoff}
  \emph {et~al.}}]{SDSS:2013jmz}%
  \BibitemOpen
  \bibfield  {author} {\bibinfo {author} {\bibfnamefont {E.~S.}\ \bibnamefont
  {Rykoff}} \emph {et~al.} (\bibinfo {collaboration} {SDSS}),\ }\bibfield
  {title} {\bibinfo {title} {{redMaPPer I: Algorithm and SDSS DR8 Catalog}},\
  }\href {https://doi.org/10.1088/0004-637X/785/2/104} {\bibfield  {journal}
  {\bibinfo  {journal} {Astrophys. J.}\ }\textbf {\bibinfo {volume} {785}},\
  \bibinfo {pages} {104} (\bibinfo {year} {2014})},\ \Eprint
  {https://arxiv.org/abs/1303.3562} {arXiv:1303.3562 [astro-ph.CO]}
  \BibitemShut {NoStop}%
\bibitem [{\citenamefont {Rykoff}\ \emph {et~al.}(2016)\citenamefont {Rykoff}
  \emph {et~al.}}]{DES:2016dkd}%
  \BibitemOpen
  \bibfield  {author} {\bibinfo {author} {\bibfnamefont {E.~S.}\ \bibnamefont
  {Rykoff}} \emph {et~al.} (\bibinfo {collaboration} {DES}),\ }\bibfield
  {title} {\bibinfo {title} {{The redMaPPer Galaxy Cluster Catalog From DES
  Science Verification Data}},\ }\href
  {https://doi.org/10.3847/0067-0049/224/1/1} {\bibfield  {journal} {\bibinfo
  {journal} {Astrophys. J. Suppl.}\ }\textbf {\bibinfo {volume} {224}},\
  \bibinfo {pages} {1} (\bibinfo {year} {2016})},\ \Eprint
  {https://arxiv.org/abs/1601.00621} {arXiv:1601.00621 [astro-ph.CO]}
  \BibitemShut {NoStop}%
\bibitem [{\citenamefont {Oguri}\ \emph {et~al.}(2018)\citenamefont {Oguri}
  \emph {et~al.}}]{Oguri:2017khw}%
  \BibitemOpen
  \bibfield  {author} {\bibinfo {author} {\bibfnamefont {M.}~\bibnamefont
  {Oguri}} \emph {et~al.},\ }\bibfield  {title} {\bibinfo {title} {{An
  optically-selected cluster catalog at redshift 0.1\ensuremath{<}
  z\ensuremath{<} 1.1 from the Hyper Suprime-Cam Subaru Strategic Program S16A
  data}},\ }\href {https://doi.org/10.1093/pasj/psx042} {\bibfield  {journal}
  {\bibinfo  {journal} {Publ. Astron. Soc. Jap.}\ }\textbf {\bibinfo {volume}
  {70}},\ \bibinfo {pages} {S20} (\bibinfo {year} {2018})},\ \Eprint
  {https://arxiv.org/abs/1701.00818} {arXiv:1701.00818 [astro-ph.CO]}
  \BibitemShut {NoStop}%
\bibitem [{\citenamefont {Hilton}\ \emph {et~al.}(2018)\citenamefont {Hilton}
  \emph {et~al.}}]{ACT:2017dgj}%
  \BibitemOpen
  \bibfield  {author} {\bibinfo {author} {\bibfnamefont {M.}~\bibnamefont
  {Hilton}} \emph {et~al.} (\bibinfo {collaboration} {ACT}),\ }\bibfield
  {title} {\bibinfo {title} {{The Atacama Cosmology Telescope: The Two-Season
  ACTPol Sunyaev-Zel'dovich Effect Selected Cluster Catalog}},\ }\href
  {https://doi.org/10.3847/1538-4365/aaa6cb} {\bibfield  {journal} {\bibinfo
  {journal} {Astrophys. J. Suppl.}\ }\textbf {\bibinfo {volume} {235}},\
  \bibinfo {pages} {20} (\bibinfo {year} {2018})},\ \Eprint
  {https://arxiv.org/abs/1709.05600} {arXiv:1709.05600 [astro-ph.CO]}
  \BibitemShut {NoStop}%
\bibitem [{\citenamefont {{Furusawa}}\ \emph {et~al.}(2018)\citenamefont
  {{Furusawa}}, \citenamefont {{Koike}}, \citenamefont {{Takata}},
  \citenamefont {{Okura}}, \citenamefont {{Miyatake}}, \citenamefont
  {{Lupton}}, \citenamefont {{Bickerton}}, \citenamefont {{Price}},
  \citenamefont {{Bosch}}, \citenamefont {{Yasuda}}, \citenamefont {{Mineo}},
  \citenamefont {{Yamada}}, \citenamefont {{Miyazaki}}, \citenamefont
  {{Nakata}}, \citenamefont {{Koshida}}, \citenamefont {{Komiyama}},
  \citenamefont {{Utsumi}}, \citenamefont {{Kawanomoto}}, \citenamefont
  {{Jeschke}}, \citenamefont {{Noumaru}}, \citenamefont {{Schubert}},
  \citenamefont {{Iwata}}, \citenamefont {{Finet}}, \citenamefont
  {{Fujiyoshi}}, \citenamefont {{Tajitsu}}, \citenamefont {{Terai}},\ and\
  \citenamefont {{Lee}}}]{2018PASJ...70S...3F}%
  \BibitemOpen
  \bibfield  {author} {\bibinfo {author} {\bibfnamefont {H.}~\bibnamefont
  {{Furusawa}}}, \bibinfo {author} {\bibfnamefont {M.}~\bibnamefont {{Koike}}},
  \bibinfo {author} {\bibfnamefont {T.}~\bibnamefont {{Takata}}}, \bibinfo
  {author} {\bibfnamefont {Y.}~\bibnamefont {{Okura}}}, \bibinfo {author}
  {\bibfnamefont {H.}~\bibnamefont {{Miyatake}}}, \bibinfo {author}
  {\bibfnamefont {R.~H.}\ \bibnamefont {{Lupton}}}, \bibinfo {author}
  {\bibfnamefont {S.}~\bibnamefont {{Bickerton}}}, \bibinfo {author}
  {\bibfnamefont {P.~A.}\ \bibnamefont {{Price}}}, \bibinfo {author}
  {\bibfnamefont {J.}~\bibnamefont {{Bosch}}}, \bibinfo {author} {\bibfnamefont
  {N.}~\bibnamefont {{Yasuda}}}, \bibinfo {author} {\bibfnamefont
  {S.}~\bibnamefont {{Mineo}}}, \bibinfo {author} {\bibfnamefont
  {Y.}~\bibnamefont {{Yamada}}}, \bibinfo {author} {\bibfnamefont
  {S.}~\bibnamefont {{Miyazaki}}}, \bibinfo {author} {\bibfnamefont
  {F.}~\bibnamefont {{Nakata}}}, \bibinfo {author} {\bibfnamefont
  {S.}~\bibnamefont {{Koshida}}}, \bibinfo {author} {\bibfnamefont
  {Y.}~\bibnamefont {{Komiyama}}}, \bibinfo {author} {\bibfnamefont
  {Y.}~\bibnamefont {{Utsumi}}}, \bibinfo {author} {\bibfnamefont
  {S.}~\bibnamefont {{Kawanomoto}}}, \bibinfo {author} {\bibfnamefont
  {E.}~\bibnamefont {{Jeschke}}}, \bibinfo {author} {\bibfnamefont
  {J.}~\bibnamefont {{Noumaru}}}, \bibinfo {author} {\bibfnamefont
  {K.}~\bibnamefont {{Schubert}}}, \bibinfo {author} {\bibfnamefont
  {I.}~\bibnamefont {{Iwata}}}, \bibinfo {author} {\bibfnamefont
  {F.}~\bibnamefont {{Finet}}}, \bibinfo {author} {\bibfnamefont
  {T.}~\bibnamefont {{Fujiyoshi}}}, \bibinfo {author} {\bibfnamefont
  {A.}~\bibnamefont {{Tajitsu}}}, \bibinfo {author} {\bibfnamefont
  {T.}~\bibnamefont {{Terai}}},\ and\ \bibinfo {author} {\bibfnamefont {C.-H.}\
  \bibnamefont {{Lee}}},\ }\bibfield  {title} {\bibinfo {title} {{The on-site
  quality-assurance system for Hyper Suprime-Cam: OSQAH}},\ }\href
  {https://doi.org/10.1093/pasj/psx079} {\bibfield  {journal} {\bibinfo
  {journal} {\pasj}\ }\textbf {\bibinfo {volume} {70}},\ \bibinfo {eid} {S3}
  (\bibinfo {year} {2018})}\BibitemShut {NoStop}%
\bibitem [{\citenamefont {{Miyazaki}}\ \emph {et~al.}(2018)\citenamefont
  {{Miyazaki}}, \citenamefont {{Komiyama}}, \citenamefont {{Kawanomoto}},
  \citenamefont {{Doi}}, \citenamefont {{Furusawa}}, \citenamefont {{Hamana}},
  \citenamefont {{Hayashi}}, \citenamefont {{Ikeda}}, \citenamefont {{Kamata}},
  \citenamefont {{Karoji}}, \citenamefont {{Koike}}, \citenamefont
  {{Kurakami}}, \citenamefont {{Miyama}}, \citenamefont {{Morokuma}},
  \citenamefont {{Nakata}}, \citenamefont {{Namikawa}}, \citenamefont
  {{Nakaya}}, \citenamefont {{Nariai}}, \citenamefont {{Obuchi}}, \citenamefont
  {{Oishi}}, \citenamefont {{Okada}}, \citenamefont {{Okura}}, \citenamefont
  {{Tait}}, \citenamefont {{Takata}}, \citenamefont {{Tanaka}}, \citenamefont
  {{Tanaka}}, \citenamefont {{Terai}}, \citenamefont {{Tomono}}, \citenamefont
  {{Uraguchi}}, \citenamefont {{Usuda}}, \citenamefont {{Utsumi}},
  \citenamefont {{Yamada}}, \citenamefont {{Yamanoi}}, \citenamefont
  {{Aihara}}, \citenamefont {{Fujimori}}, \citenamefont {{Mineo}},
  \citenamefont {{Miyatake}}, \citenamefont {{Oguri}}, \citenamefont
  {{Uchida}}, \citenamefont {{Tanaka}}, \citenamefont {{Yasuda}}, \citenamefont
  {{Takada}}, \citenamefont {{Murayama}}, \citenamefont {{Nishizawa}},
  \citenamefont {{Sugiyama}}, \citenamefont {{Chiba}}, \citenamefont
  {{Futamase}}, \citenamefont {{Wang}}, \citenamefont {{Chen}}, \citenamefont
  {{Ho}}, \citenamefont {{Liaw}}, \citenamefont {{Chiu}}, \citenamefont {{Ho}},
  \citenamefont {{Lai}}, \citenamefont {{Lee}}, \citenamefont {{Jeng}},
  \citenamefont {{Iwamura}}, \citenamefont {{Armstrong}}, \citenamefont
  {{Bickerton}}, \citenamefont {{Bosch}}, \citenamefont {{Gunn}}, \citenamefont
  {{Lupton}}, \citenamefont {{Loomis}}, \citenamefont {{Price}}, \citenamefont
  {{Smith}}, \citenamefont {{Strauss}}, \citenamefont {{Turner}}, \citenamefont
  {{Suzuki}}, \citenamefont {{Miyazaki}}, \citenamefont {{Muramatsu}},
  \citenamefont {{Yamamoto}}, \citenamefont {{Endo}}, \citenamefont {{Ezaki}},
  \citenamefont {{Ito}}, \citenamefont {{Kawaguchi}}, \citenamefont {{Sofuku}},
  \citenamefont {{Taniike}}, \citenamefont {{Akutsu}}, \citenamefont {{Dojo}},
  \citenamefont {{Kasumi}}, \citenamefont {{Matsuda}}, \citenamefont {{Imoto}},
  \citenamefont {{Miwa}}, \citenamefont {{Suzuki}}, \citenamefont {{Takeshi}},\
  and\ \citenamefont {{Yokota}}}]{2018PASJ...70S...1M}%
  \BibitemOpen
  \bibfield  {author} {\bibinfo {author} {\bibfnamefont {S.}~\bibnamefont
  {{Miyazaki}}}, \bibinfo {author} {\bibfnamefont {Y.}~\bibnamefont
  {{Komiyama}}}, \bibinfo {author} {\bibfnamefont {S.}~\bibnamefont
  {{Kawanomoto}}}, \bibinfo {author} {\bibfnamefont {Y.}~\bibnamefont {{Doi}}},
  \bibinfo {author} {\bibfnamefont {H.}~\bibnamefont {{Furusawa}}}, \bibinfo
  {author} {\bibfnamefont {T.}~\bibnamefont {{Hamana}}}, \bibinfo {author}
  {\bibfnamefont {Y.}~\bibnamefont {{Hayashi}}}, \bibinfo {author}
  {\bibfnamefont {H.}~\bibnamefont {{Ikeda}}}, \bibinfo {author} {\bibfnamefont
  {Y.}~\bibnamefont {{Kamata}}}, \bibinfo {author} {\bibfnamefont
  {H.}~\bibnamefont {{Karoji}}}, \bibinfo {author} {\bibfnamefont
  {M.}~\bibnamefont {{Koike}}}, \bibinfo {author} {\bibfnamefont
  {T.}~\bibnamefont {{Kurakami}}}, \bibinfo {author} {\bibfnamefont
  {S.}~\bibnamefont {{Miyama}}}, \bibinfo {author} {\bibfnamefont
  {T.}~\bibnamefont {{Morokuma}}}, \bibinfo {author} {\bibfnamefont
  {F.}~\bibnamefont {{Nakata}}}, \bibinfo {author} {\bibfnamefont
  {K.}~\bibnamefont {{Namikawa}}}, \bibinfo {author} {\bibfnamefont
  {H.}~\bibnamefont {{Nakaya}}}, \bibinfo {author} {\bibfnamefont
  {K.}~\bibnamefont {{Nariai}}}, \bibinfo {author} {\bibfnamefont
  {Y.}~\bibnamefont {{Obuchi}}}, \bibinfo {author} {\bibfnamefont
  {Y.}~\bibnamefont {{Oishi}}}, \bibinfo {author} {\bibfnamefont
  {N.}~\bibnamefont {{Okada}}}, \bibinfo {author} {\bibfnamefont
  {Y.}~\bibnamefont {{Okura}}}, \bibinfo {author} {\bibfnamefont
  {P.}~\bibnamefont {{Tait}}}, \bibinfo {author} {\bibfnamefont
  {T.}~\bibnamefont {{Takata}}}, \bibinfo {author} {\bibfnamefont
  {Y.}~\bibnamefont {{Tanaka}}}, \bibinfo {author} {\bibfnamefont
  {M.}~\bibnamefont {{Tanaka}}}, \bibinfo {author} {\bibfnamefont
  {T.}~\bibnamefont {{Terai}}}, \bibinfo {author} {\bibfnamefont
  {D.}~\bibnamefont {{Tomono}}}, \bibinfo {author} {\bibfnamefont
  {F.}~\bibnamefont {{Uraguchi}}}, \bibinfo {author} {\bibfnamefont
  {T.}~\bibnamefont {{Usuda}}}, \bibinfo {author} {\bibfnamefont
  {Y.}~\bibnamefont {{Utsumi}}}, \bibinfo {author} {\bibfnamefont
  {Y.}~\bibnamefont {{Yamada}}}, \bibinfo {author} {\bibfnamefont
  {H.}~\bibnamefont {{Yamanoi}}}, \bibinfo {author} {\bibfnamefont
  {H.}~\bibnamefont {{Aihara}}}, \bibinfo {author} {\bibfnamefont
  {H.}~\bibnamefont {{Fujimori}}}, \bibinfo {author} {\bibfnamefont
  {S.}~\bibnamefont {{Mineo}}}, \bibinfo {author} {\bibfnamefont
  {H.}~\bibnamefont {{Miyatake}}}, \bibinfo {author} {\bibfnamefont
  {M.}~\bibnamefont {{Oguri}}}, \bibinfo {author} {\bibfnamefont
  {T.}~\bibnamefont {{Uchida}}}, \bibinfo {author} {\bibfnamefont {M.~M.}\
  \bibnamefont {{Tanaka}}}, \bibinfo {author} {\bibfnamefont {N.}~\bibnamefont
  {{Yasuda}}}, \bibinfo {author} {\bibfnamefont {M.}~\bibnamefont {{Takada}}},
  \bibinfo {author} {\bibfnamefont {H.}~\bibnamefont {{Murayama}}}, \bibinfo
  {author} {\bibfnamefont {A.~J.}\ \bibnamefont {{Nishizawa}}}, \bibinfo
  {author} {\bibfnamefont {N.}~\bibnamefont {{Sugiyama}}}, \bibinfo {author}
  {\bibfnamefont {M.}~\bibnamefont {{Chiba}}}, \bibinfo {author} {\bibfnamefont
  {T.}~\bibnamefont {{Futamase}}}, \bibinfo {author} {\bibfnamefont {S.-Y.}\
  \bibnamefont {{Wang}}}, \bibinfo {author} {\bibfnamefont {H.-Y.}\
  \bibnamefont {{Chen}}}, \bibinfo {author} {\bibfnamefont {P.~T.~P.}\
  \bibnamefont {{Ho}}}, \bibinfo {author} {\bibfnamefont {E.~J.~Y.}\
  \bibnamefont {{Liaw}}}, \bibinfo {author} {\bibfnamefont {C.-F.}\
  \bibnamefont {{Chiu}}}, \bibinfo {author} {\bibfnamefont {C.-L.}\
  \bibnamefont {{Ho}}}, \bibinfo {author} {\bibfnamefont {T.-C.}\ \bibnamefont
  {{Lai}}}, \bibinfo {author} {\bibfnamefont {Y.-C.}\ \bibnamefont {{Lee}}},
  \bibinfo {author} {\bibfnamefont {D.-Z.}\ \bibnamefont {{Jeng}}}, \bibinfo
  {author} {\bibfnamefont {S.}~\bibnamefont {{Iwamura}}}, \bibinfo {author}
  {\bibfnamefont {R.}~\bibnamefont {{Armstrong}}}, \bibinfo {author}
  {\bibfnamefont {S.}~\bibnamefont {{Bickerton}}}, \bibinfo {author}
  {\bibfnamefont {J.}~\bibnamefont {{Bosch}}}, \bibinfo {author} {\bibfnamefont
  {J.~E.}\ \bibnamefont {{Gunn}}}, \bibinfo {author} {\bibfnamefont {R.~H.}\
  \bibnamefont {{Lupton}}}, \bibinfo {author} {\bibfnamefont {C.}~\bibnamefont
  {{Loomis}}}, \bibinfo {author} {\bibfnamefont {P.}~\bibnamefont {{Price}}},
  \bibinfo {author} {\bibfnamefont {S.}~\bibnamefont {{Smith}}}, \bibinfo
  {author} {\bibfnamefont {M.~A.}\ \bibnamefont {{Strauss}}}, \bibinfo {author}
  {\bibfnamefont {E.~L.}\ \bibnamefont {{Turner}}}, \bibinfo {author}
  {\bibfnamefont {H.}~\bibnamefont {{Suzuki}}}, \bibinfo {author}
  {\bibfnamefont {Y.}~\bibnamefont {{Miyazaki}}}, \bibinfo {author}
  {\bibfnamefont {M.}~\bibnamefont {{Muramatsu}}}, \bibinfo {author}
  {\bibfnamefont {K.}~\bibnamefont {{Yamamoto}}}, \bibinfo {author}
  {\bibfnamefont {M.}~\bibnamefont {{Endo}}}, \bibinfo {author} {\bibfnamefont
  {Y.}~\bibnamefont {{Ezaki}}}, \bibinfo {author} {\bibfnamefont
  {N.}~\bibnamefont {{Ito}}}, \bibinfo {author} {\bibfnamefont
  {N.}~\bibnamefont {{Kawaguchi}}}, \bibinfo {author} {\bibfnamefont
  {S.}~\bibnamefont {{Sofuku}}}, \bibinfo {author} {\bibfnamefont
  {T.}~\bibnamefont {{Taniike}}}, \bibinfo {author} {\bibfnamefont
  {K.}~\bibnamefont {{Akutsu}}}, \bibinfo {author} {\bibfnamefont
  {N.}~\bibnamefont {{Dojo}}}, \bibinfo {author} {\bibfnamefont
  {K.}~\bibnamefont {{Kasumi}}}, \bibinfo {author} {\bibfnamefont
  {T.}~\bibnamefont {{Matsuda}}}, \bibinfo {author} {\bibfnamefont
  {K.}~\bibnamefont {{Imoto}}}, \bibinfo {author} {\bibfnamefont
  {Y.}~\bibnamefont {{Miwa}}}, \bibinfo {author} {\bibfnamefont
  {M.}~\bibnamefont {{Suzuki}}}, \bibinfo {author} {\bibfnamefont
  {K.}~\bibnamefont {{Takeshi}}},\ and\ \bibinfo {author} {\bibfnamefont
  {H.}~\bibnamefont {{Yokota}}},\ }\bibfield  {title} {\bibinfo {title} {{Hyper
  Suprime-Cam: System design and verification of image quality}},\ }\href
  {https://doi.org/10.1093/pasj/psx063} {\bibfield  {journal} {\bibinfo
  {journal} {\pasj}\ }\textbf {\bibinfo {volume} {70}},\ \bibinfo {eid} {S1}
  (\bibinfo {year} {2018})}\BibitemShut {NoStop}%
\bibitem [{\citenamefont {Aihara}\ \emph {et~al.}(2018)\citenamefont {Aihara}
  \emph {et~al.}}]{Aihara:2017paw}%
  \BibitemOpen
  \bibfield  {author} {\bibinfo {author} {\bibfnamefont {H.}~\bibnamefont
  {Aihara}} \emph {et~al.},\ }\bibfield  {title} {\bibinfo {title} {{The Hyper
  Suprime-Cam SSP Survey: Overview and Survey Design}},\ }\href
  {https://doi.org/10.1093/pasj/psx066} {\bibfield  {journal} {\bibinfo
  {journal} {Publ. Astron. Soc. Jap.}\ }\textbf {\bibinfo {volume} {70}},\
  \bibinfo {pages} {S4} (\bibinfo {year} {2018})},\ \Eprint
  {https://arxiv.org/abs/1704.05858} {arXiv:1704.05858 [astro-ph.IM]}
  \BibitemShut {NoStop}%
\bibitem [{\citenamefont {Hirata}\ and\ \citenamefont
  {Seljak}(2003)}]{Hirata:2003cv}%
  \BibitemOpen
  \bibfield  {author} {\bibinfo {author} {\bibfnamefont {C.~M.}\ \bibnamefont
  {Hirata}}\ and\ \bibinfo {author} {\bibfnamefont {U.}~\bibnamefont
  {Seljak}},\ }\bibfield  {title} {\bibinfo {title} {{Shear calibration biases
  in weak lensing surveys}},\ }\href
  {https://doi.org/10.1046/j.1365-8711.2003.06683.x} {\bibfield  {journal}
  {\bibinfo  {journal} {Mon. Not. Roy. Astron. Soc.}\ }\textbf {\bibinfo
  {volume} {343}},\ \bibinfo {pages} {459} (\bibinfo {year} {2003})},\ \Eprint
  {https://arxiv.org/abs/astro-ph/0301054} {arXiv:astro-ph/0301054}
  \BibitemShut {NoStop}%
\bibitem [{\citenamefont {Mandelbaum}\ \emph {et~al.}(2005)\citenamefont
  {Mandelbaum}, \citenamefont {Hirata}, \citenamefont {Seljak}, \citenamefont
  {Guzik}, \citenamefont {Padmanabhan}, \citenamefont {Blake}, \citenamefont
  {Blanton}, \citenamefont {Lupton},\ and\ \citenamefont
  {Brinkmann}}]{Mandelbaum:2005wv}%
  \BibitemOpen
  \bibfield  {author} {\bibinfo {author} {\bibfnamefont {R.}~\bibnamefont
  {Mandelbaum}}, \bibinfo {author} {\bibfnamefont {C.~M.}\ \bibnamefont
  {Hirata}}, \bibinfo {author} {\bibfnamefont {U.}~\bibnamefont {Seljak}},
  \bibinfo {author} {\bibfnamefont {J.}~\bibnamefont {Guzik}}, \bibinfo
  {author} {\bibfnamefont {N.}~\bibnamefont {Padmanabhan}}, \bibinfo {author}
  {\bibfnamefont {C.}~\bibnamefont {Blake}}, \bibinfo {author} {\bibfnamefont
  {M.~R.}\ \bibnamefont {Blanton}}, \bibinfo {author} {\bibfnamefont
  {R.}~\bibnamefont {Lupton}},\ and\ \bibinfo {author} {\bibfnamefont
  {J.}~\bibnamefont {Brinkmann}},\ }\bibfield  {title} {\bibinfo {title}
  {{Systematic errors in weak lensing: Application to SDSS galaxy-galaxy weak
  lensing}},\ }\href {https://doi.org/10.1111/j.1365-2966.2005.09282.x}
  {\bibfield  {journal} {\bibinfo  {journal} {Mon. Not. Roy. Astron. Soc.}\
  }\textbf {\bibinfo {volume} {361}},\ \bibinfo {pages} {1287} (\bibinfo {year}
  {2005})},\ \Eprint {https://arxiv.org/abs/astro-ph/0501201}
  {arXiv:astro-ph/0501201} \BibitemShut {NoStop}%
\bibitem [{\citenamefont {Mandelbaum}\ \emph {et~al.}(2013)\citenamefont
  {Mandelbaum}, \citenamefont {Slosar}, \citenamefont {Baldauf}, \citenamefont
  {Seljak}, \citenamefont {Hirata}, \citenamefont {Nakajima}, \citenamefont
  {Reyes},\ and\ \citenamefont {Smith}}]{Mandelbaum:2012ay}%
  \BibitemOpen
  \bibfield  {author} {\bibinfo {author} {\bibfnamefont {R.}~\bibnamefont
  {Mandelbaum}}, \bibinfo {author} {\bibfnamefont {A.}~\bibnamefont {Slosar}},
  \bibinfo {author} {\bibfnamefont {T.}~\bibnamefont {Baldauf}}, \bibinfo
  {author} {\bibfnamefont {U.}~\bibnamefont {Seljak}}, \bibinfo {author}
  {\bibfnamefont {C.~M.}\ \bibnamefont {Hirata}}, \bibinfo {author}
  {\bibfnamefont {R.}~\bibnamefont {Nakajima}}, \bibinfo {author}
  {\bibfnamefont {R.}~\bibnamefont {Reyes}},\ and\ \bibinfo {author}
  {\bibfnamefont {R.~E.}\ \bibnamefont {Smith}},\ }\bibfield  {title} {\bibinfo
  {title} {{Cosmological parameter constraints from galaxy-galaxy lensing and
  galaxy clustering with the SDSS DR7}},\ }\href
  {https://doi.org/10.1093/mnras/stt572} {\bibfield  {journal} {\bibinfo
  {journal} {Mon. Not. Roy. Astron. Soc.}\ }\textbf {\bibinfo {volume} {432}},\
  \bibinfo {pages} {1544} (\bibinfo {year} {2013})},\ \Eprint
  {https://arxiv.org/abs/1207.1120} {arXiv:1207.1120 [astro-ph.CO]}
  \BibitemShut {NoStop}%
\bibitem [{\citenamefont {{Miralda-Escude}}(1991)}]{1991ApJ...380....1M}%
  \BibitemOpen
  \bibfield  {author} {\bibinfo {author} {\bibfnamefont {J.}~\bibnamefont
  {{Miralda-Escude}}},\ }\bibfield  {title} {\bibinfo {title} {{The Correlation
  Function of Galaxy Ellipticities Produced by Gravitational Lensing}},\ }\href
  {https://doi.org/10.1086/170555} {\bibfield  {journal} {\bibinfo  {journal}
  {\apj}\ }\textbf {\bibinfo {volume} {380}},\ \bibinfo {pages} {1} (\bibinfo
  {year} {1991})}\BibitemShut {NoStop}%
\bibitem [{\citenamefont {Bernstein}\ and\ \citenamefont
  {Jarvis}(2002)}]{Bernstein:2001nz}%
  \BibitemOpen
  \bibfield  {author} {\bibinfo {author} {\bibfnamefont {G.~M.}\ \bibnamefont
  {Bernstein}}\ and\ \bibinfo {author} {\bibfnamefont {M.}~\bibnamefont
  {Jarvis}},\ }\bibfield  {title} {\bibinfo {title} {{Shapes and shears, stars
  and smears: optimal measurements for weak lensing}},\ }\href
  {https://doi.org/10.1086/338085} {\bibfield  {journal} {\bibinfo  {journal}
  {Astron. J.}\ }\textbf {\bibinfo {volume} {123}},\ \bibinfo {pages} {583}
  (\bibinfo {year} {2002})},\ \Eprint {https://arxiv.org/abs/astro-ph/0107431}
  {arXiv:astro-ph/0107431} \BibitemShut {NoStop}%
\bibitem [{\citenamefont {{Rowe}}\ \emph {et~al.}(2015)\citenamefont {{Rowe}},
  \citenamefont {{Jarvis}}, \citenamefont {{Mandelbaum}}, \citenamefont
  {{Bernstein}}, \citenamefont {{Bosch}}, \citenamefont {{Simet}},
  \citenamefont {{Meyers}}, \citenamefont {{Kacprzak}}, \citenamefont
  {{Nakajima}}, \citenamefont {{Zuntz}}, \citenamefont {{Miyatake}},
  \citenamefont {{Dietrich}}, \citenamefont {{Armstrong}}, \citenamefont
  {{Melchior}},\ and\ \citenamefont {{Gill}}}]{2015A&C....10..121R}%
  \BibitemOpen
  \bibfield  {author} {\bibinfo {author} {\bibfnamefont {B.~T.~P.}\
  \bibnamefont {{Rowe}}}, \bibinfo {author} {\bibfnamefont {M.}~\bibnamefont
  {{Jarvis}}}, \bibinfo {author} {\bibfnamefont {R.}~\bibnamefont
  {{Mandelbaum}}}, \bibinfo {author} {\bibfnamefont {G.~M.}\ \bibnamefont
  {{Bernstein}}}, \bibinfo {author} {\bibfnamefont {J.}~\bibnamefont
  {{Bosch}}}, \bibinfo {author} {\bibfnamefont {M.}~\bibnamefont {{Simet}}},
  \bibinfo {author} {\bibfnamefont {J.~E.}\ \bibnamefont {{Meyers}}}, \bibinfo
  {author} {\bibfnamefont {T.}~\bibnamefont {{Kacprzak}}}, \bibinfo {author}
  {\bibfnamefont {R.}~\bibnamefont {{Nakajima}}}, \bibinfo {author}
  {\bibfnamefont {J.}~\bibnamefont {{Zuntz}}}, \bibinfo {author} {\bibfnamefont
  {H.}~\bibnamefont {{Miyatake}}}, \bibinfo {author} {\bibfnamefont {J.~P.}\
  \bibnamefont {{Dietrich}}}, \bibinfo {author} {\bibfnamefont
  {R.}~\bibnamefont {{Armstrong}}}, \bibinfo {author} {\bibfnamefont
  {P.}~\bibnamefont {{Melchior}}},\ and\ \bibinfo {author} {\bibfnamefont
  {M.~S.~S.}\ \bibnamefont {{Gill}}},\ }\bibfield  {title} {\bibinfo {title}
  {{GALSIM: The modular galaxy image simulation toolkit}},\ }\href
  {https://doi.org/10.1016/j.ascom.2015.02.002} {\bibfield  {journal} {\bibinfo
   {journal} {Astronomy and Computing}\ }\textbf {\bibinfo {volume} {10}},\
  \bibinfo {pages} {121} (\bibinfo {year} {2015})},\ \Eprint
  {https://arxiv.org/abs/1407.7676} {arXiv:1407.7676 [astro-ph.IM]}
  \BibitemShut {NoStop}%
\bibitem [{\citenamefont {Mandelbaum}\ \emph {et~al.}(2018)\citenamefont
  {Mandelbaum} \emph {et~al.}}]{Mandelbaum:2017ctf}%
  \BibitemOpen
  \bibfield  {author} {\bibinfo {author} {\bibfnamefont {R.}~\bibnamefont
  {Mandelbaum}} \emph {et~al.},\ }\bibfield  {title} {\bibinfo {title} {{Weak
  lensing shear calibration with simulations of the HSC survey}},\ }\href
  {https://doi.org/10.1093/mnras/sty2420} {\bibfield  {journal} {\bibinfo
  {journal} {Mon. Not. Roy. Astron. Soc.}\ }\textbf {\bibinfo {volume} {481}},\
  \bibinfo {pages} {3170} (\bibinfo {year} {2018})},\ \Eprint
  {https://arxiv.org/abs/1710.00885} {arXiv:1710.00885 [astro-ph.CO]}
  \BibitemShut {NoStop}%
\bibitem [{\citenamefont {Tanaka}\ \emph {et~al.}(2018)\citenamefont {Tanaka},
  \citenamefont {Coupon}, \citenamefont {Hsieh}, \citenamefont {Mineo},
  \citenamefont {Nishizawa}, \citenamefont {Speagle}, \citenamefont {Furusawa},
  \citenamefont {Miyazaki},\ and\ \citenamefont {Murayama}}]{Tanaka:2017lit}%
  \BibitemOpen
  \bibfield  {author} {\bibinfo {author} {\bibfnamefont {M.}~\bibnamefont
  {Tanaka}}, \bibinfo {author} {\bibfnamefont {J.}~\bibnamefont {Coupon}},
  \bibinfo {author} {\bibfnamefont {B.-C.}\ \bibnamefont {Hsieh}}, \bibinfo
  {author} {\bibfnamefont {S.}~\bibnamefont {Mineo}}, \bibinfo {author}
  {\bibfnamefont {A.~J.}\ \bibnamefont {Nishizawa}}, \bibinfo {author}
  {\bibfnamefont {J.}~\bibnamefont {Speagle}}, \bibinfo {author} {\bibfnamefont
  {H.}~\bibnamefont {Furusawa}}, \bibinfo {author} {\bibfnamefont
  {S.}~\bibnamefont {Miyazaki}},\ and\ \bibinfo {author} {\bibfnamefont
  {H.}~\bibnamefont {Murayama}},\ }\bibfield  {title} {\bibinfo {title}
  {{Photometric Redshifts for Hyper Suprime-Cam Subaru Strategic Program Data
  Release 1}},\ }\href {https://doi.org/10.1093/pasj/psx077} {\bibfield
  {journal} {\bibinfo  {journal} {Publ. Astron. Soc. Jap.}\ }\textbf {\bibinfo
  {volume} {70}},\ \bibinfo {pages} {S9} (\bibinfo {year} {2018})},\ \Eprint
  {https://arxiv.org/abs/1704.05988} {arXiv:1704.05988 [astro-ph.GA]}
  \BibitemShut {NoStop}%
\bibitem [{\citenamefont {Oguri}(2014)}]{Oguri:2014eba}%
  \BibitemOpen
  \bibfield  {author} {\bibinfo {author} {\bibfnamefont {M.}~\bibnamefont
  {Oguri}},\ }\bibfield  {title} {\bibinfo {title} {{A cluster finding
  algorithm based on the multiband identification of red sequence galaxies}},\
  }\href {https://doi.org/10.1093/mnras/stu1446} {\bibfield  {journal}
  {\bibinfo  {journal} {Mon. Not. Roy. Astron. Soc.}\ }\textbf {\bibinfo
  {volume} {444}},\ \bibinfo {pages} {147} (\bibinfo {year} {2014})},\ \Eprint
  {https://arxiv.org/abs/1407.4693} {arXiv:1407.4693 [astro-ph.CO]}
  \BibitemShut {NoStop}%
\bibitem [{\citenamefont {Mandelbaum}\ \emph {et~al.}(2006)\citenamefont
  {Mandelbaum}, \citenamefont {Seljak}, \citenamefont {Cool}, \citenamefont
  {Blanton}, \citenamefont {Hirata},\ and\ \citenamefont
  {Brinkmann}}]{Mandelbaum:2006pw}%
  \BibitemOpen
  \bibfield  {author} {\bibinfo {author} {\bibfnamefont {R.}~\bibnamefont
  {Mandelbaum}}, \bibinfo {author} {\bibfnamefont {U.}~\bibnamefont {Seljak}},
  \bibinfo {author} {\bibfnamefont {R.~J.}\ \bibnamefont {Cool}}, \bibinfo
  {author} {\bibfnamefont {M.}~\bibnamefont {Blanton}}, \bibinfo {author}
  {\bibfnamefont {C.~M.}\ \bibnamefont {Hirata}},\ and\ \bibinfo {author}
  {\bibfnamefont {J.}~\bibnamefont {Brinkmann}},\ }\bibfield  {title} {\bibinfo
  {title} {{Density profiles of galaxy groups and clusters from SDSS
  galaxy-galaxy weak lensing}},\ }\href
  {https://doi.org/10.1111/j.1365-2966.2006.10906.x} {\bibfield  {journal}
  {\bibinfo  {journal} {Mon. Not. Roy. Astron. Soc.}\ }\textbf {\bibinfo
  {volume} {372}},\ \bibinfo {pages} {758} (\bibinfo {year} {2006})},\ \Eprint
  {https://arxiv.org/abs/astro-ph/0605476} {arXiv:astro-ph/0605476}
  \BibitemShut {NoStop}%
\bibitem [{\citenamefont {Miyatake}\ \emph
  {et~al.}(2022{\natexlab{a}})\citenamefont {Miyatake} \emph
  {et~al.}}]{Miyatake:2021sdd}%
  \BibitemOpen
  \bibfield  {author} {\bibinfo {author} {\bibfnamefont {H.}~\bibnamefont
  {Miyatake}} \emph {et~al.},\ }\bibfield  {title} {\bibinfo {title}
  {{Cosmological inference from an emulator based halo model. II. Joint
  analysis of galaxy-galaxy weak lensing and galaxy clustering from HSC-Y1 and
  SDSS}},\ }\href {https://doi.org/10.1103/PhysRevD.106.083520} {\bibfield
  {journal} {\bibinfo  {journal} {Phys. Rev. D}\ }\textbf {\bibinfo {volume}
  {106}},\ \bibinfo {pages} {083520} (\bibinfo {year} {2022}{\natexlab{a}})},\
  \Eprint {https://arxiv.org/abs/2111.02419} {arXiv:2111.02419 [astro-ph.CO]}
  \BibitemShut {NoStop}%
\bibitem [{\citenamefont {Sheldon}\ \emph {et~al.}(2004)\citenamefont {Sheldon}
  \emph {et~al.}}]{SDSS:2003slg}%
  \BibitemOpen
  \bibfield  {author} {\bibinfo {author} {\bibfnamefont {E.~S.}\ \bibnamefont
  {Sheldon}} \emph {et~al.} (\bibinfo {collaboration} {SDSS}),\ }\bibfield
  {title} {\bibinfo {title} {{The Galaxy - mass correlation function measured
  from weak lensing in the SDSS}},\ }\href {https://doi.org/10.1086/383293}
  {\bibfield  {journal} {\bibinfo  {journal} {Astron. J.}\ }\textbf {\bibinfo
  {volume} {127}},\ \bibinfo {pages} {2544} (\bibinfo {year} {2004})},\ \Eprint
  {https://arxiv.org/abs/astro-ph/0312036} {arXiv:astro-ph/0312036}
  \BibitemShut {NoStop}%
\bibitem [{\citenamefont {Shirasaki}\ and\ \citenamefont
  {Takada}(2018)}]{Shirasaki:2018cgl}%
  \BibitemOpen
  \bibfield  {author} {\bibinfo {author} {\bibfnamefont {M.}~\bibnamefont
  {Shirasaki}}\ and\ \bibinfo {author} {\bibfnamefont {M.}~\bibnamefont
  {Takada}},\ }\bibfield  {title} {\bibinfo {title} {{Stacked lensing
  estimators and their covariance matrices: excess surface mass density versus
  lensing shear}},\ }\href {https://doi.org/10.1093/mnras/sty1327} {\bibfield
  {journal} {\bibinfo  {journal} {Mon. Not. Roy. Astron. Soc.}\ }\textbf
  {\bibinfo {volume} {478}},\ \bibinfo {pages} {4277} (\bibinfo {year}
  {2018})},\ \Eprint {https://arxiv.org/abs/1802.09696} {arXiv:1802.09696
  [astro-ph.CO]} \BibitemShut {NoStop}%
\bibitem [{\citenamefont {{Tanaka}}(2015)}]{2015ApJ...801...20T}%
  \BibitemOpen
  \bibfield  {author} {\bibinfo {author} {\bibfnamefont {M.}~\bibnamefont
  {{Tanaka}}},\ }\bibfield  {title} {\bibinfo {title} {{Photometric Redshift
  with Bayesian Priors on Physical Properties of Galaxies}},\ }\href
  {https://doi.org/10.1088/0004-637X/801/1/20} {\bibfield  {journal} {\bibinfo
  {journal} {\apj}\ }\textbf {\bibinfo {volume} {801}},\ \bibinfo {eid} {20}
  (\bibinfo {year} {2015})},\ \Eprint {https://arxiv.org/abs/1501.02047}
  {arXiv:1501.02047 [astro-ph.GA]} \BibitemShut {NoStop}%
\bibitem [{\citenamefont {Medezinski}\ \emph {et~al.}(2018)\citenamefont
  {Medezinski} \emph {et~al.}}]{Medezinski:2017eex}%
  \BibitemOpen
  \bibfield  {author} {\bibinfo {author} {\bibfnamefont {E.}~\bibnamefont
  {Medezinski}} \emph {et~al.},\ }\bibfield  {title} {\bibinfo {title} {{Source
  Selection for Cluster Weak Lensing Measurements in the Hyper Suprime-Cam
  Survey}},\ }\href {https://doi.org/10.1093/pasj/psy009} {\bibfield  {journal}
  {\bibinfo  {journal} {Publ. Astron. Soc. Jap.}\ }\textbf {\bibinfo {volume}
  {70}},\ \bibinfo {pages} {30} (\bibinfo {year} {2018})},\ \Eprint
  {https://arxiv.org/abs/1706.00427} {arXiv:1706.00427 [astro-ph.CO]}
  \BibitemShut {NoStop}%
\bibitem [{\citenamefont {Chang}\ \emph {et~al.}(2013)\citenamefont {Chang},
  \citenamefont {Jarvis}, \citenamefont {Jain}, \citenamefont {Kahn},
  \citenamefont {Kirkby}, \citenamefont {Connolly}, \citenamefont {Krughoff},
  \citenamefont {Peng},\ and\ \citenamefont {Peterson}}]{Chang:2013xja}%
  \BibitemOpen
  \bibfield  {author} {\bibinfo {author} {\bibfnamefont {C.}~\bibnamefont
  {Chang}}, \bibinfo {author} {\bibfnamefont {M.}~\bibnamefont {Jarvis}},
  \bibinfo {author} {\bibfnamefont {B.}~\bibnamefont {Jain}}, \bibinfo {author}
  {\bibfnamefont {S.~M.}\ \bibnamefont {Kahn}}, \bibinfo {author}
  {\bibfnamefont {D.}~\bibnamefont {Kirkby}}, \bibinfo {author} {\bibfnamefont
  {A.}~\bibnamefont {Connolly}}, \bibinfo {author} {\bibfnamefont
  {S.}~\bibnamefont {Krughoff}}, \bibinfo {author} {\bibfnamefont
  {E.}~\bibnamefont {Peng}},\ and\ \bibinfo {author} {\bibfnamefont {J.~R.}\
  \bibnamefont {Peterson}},\ }\bibfield  {title} {\bibinfo {title} {{The
  Effective Number Density of Galaxies for Weak Lensing Measurements in the
  LSST Project}},\ }\href {https://doi.org/10.1093/mnras/stt1156} {\bibfield
  {journal} {\bibinfo  {journal} {Mon. Not. Roy. Astron. Soc.}\ }\textbf
  {\bibinfo {volume} {434}},\ \bibinfo {pages} {2121} (\bibinfo {year}
  {2013})},\ \Eprint {https://arxiv.org/abs/1305.0793} {arXiv:1305.0793
  [astro-ph.CO]} \BibitemShut {NoStop}%
\bibitem [{\citenamefont {Murata}\ \emph {et~al.}(2018)\citenamefont {Murata},
  \citenamefont {Nishimichi}, \citenamefont {Takada}, \citenamefont {Miyatake},
  \citenamefont {Shirasaki}, \citenamefont {More}, \citenamefont {Takahashi},\
  and\ \citenamefont {Osato}}]{Murata:2017zdo}%
  \BibitemOpen
  \bibfield  {author} {\bibinfo {author} {\bibfnamefont {R.}~\bibnamefont
  {Murata}}, \bibinfo {author} {\bibfnamefont {T.}~\bibnamefont {Nishimichi}},
  \bibinfo {author} {\bibfnamefont {M.}~\bibnamefont {Takada}}, \bibinfo
  {author} {\bibfnamefont {H.}~\bibnamefont {Miyatake}}, \bibinfo {author}
  {\bibfnamefont {M.}~\bibnamefont {Shirasaki}}, \bibinfo {author}
  {\bibfnamefont {S.}~\bibnamefont {More}}, \bibinfo {author} {\bibfnamefont
  {R.}~\bibnamefont {Takahashi}},\ and\ \bibinfo {author} {\bibfnamefont
  {K.}~\bibnamefont {Osato}},\ }\bibfield  {title} {\bibinfo {title}
  {{Constraints on the mass-richness relation from the abundance and weak
  lensing of SDSS clusters}},\ }\href
  {https://doi.org/10.3847/1538-4357/aaaab8} {\bibfield  {journal} {\bibinfo
  {journal} {Astrophys. J.}\ }\textbf {\bibinfo {volume} {854}},\ \bibinfo
  {pages} {120} (\bibinfo {year} {2018})},\ \Eprint
  {https://arxiv.org/abs/1707.01907} {arXiv:1707.01907 [astro-ph.CO]}
  \BibitemShut {NoStop}%
\bibitem [{\citenamefont {Miyatake}\ \emph
  {et~al.}(2022{\natexlab{b}})\citenamefont {Miyatake}, \citenamefont
  {Kobayashi}, \citenamefont {Takada}, \citenamefont {Nishimichi},
  \citenamefont {Shirasaki}, \citenamefont {Sugiyama}, \citenamefont
  {Takahashi}, \citenamefont {Osato}, \citenamefont {More},\ and\ \citenamefont
  {Park}}]{Miyatake:2020uhg}%
  \BibitemOpen
  \bibfield  {author} {\bibinfo {author} {\bibfnamefont {H.}~\bibnamefont
  {Miyatake}}, \bibinfo {author} {\bibfnamefont {Y.}~\bibnamefont {Kobayashi}},
  \bibinfo {author} {\bibfnamefont {M.}~\bibnamefont {Takada}}, \bibinfo
  {author} {\bibfnamefont {T.}~\bibnamefont {Nishimichi}}, \bibinfo {author}
  {\bibfnamefont {M.}~\bibnamefont {Shirasaki}}, \bibinfo {author}
  {\bibfnamefont {S.}~\bibnamefont {Sugiyama}}, \bibinfo {author}
  {\bibfnamefont {R.}~\bibnamefont {Takahashi}}, \bibinfo {author}
  {\bibfnamefont {K.}~\bibnamefont {Osato}}, \bibinfo {author} {\bibfnamefont
  {S.}~\bibnamefont {More}},\ and\ \bibinfo {author} {\bibfnamefont
  {Y.}~\bibnamefont {Park}},\ }\bibfield  {title} {\bibinfo {title}
  {{Cosmological inference from an emulator based halo model. I. Validation
  tests with HSC and SDSS mock catalogs}},\ }\href
  {https://doi.org/10.1103/PhysRevD.106.083519} {\bibfield  {journal} {\bibinfo
   {journal} {Phys. Rev. D}\ }\textbf {\bibinfo {volume} {106}},\ \bibinfo
  {pages} {083519} (\bibinfo {year} {2022}{\natexlab{b}})},\ \Eprint
  {https://arxiv.org/abs/2101.00113} {arXiv:2101.00113 [astro-ph.CO]}
  \BibitemShut {NoStop}%
\bibitem [{\citenamefont {Takada}\ and\ \citenamefont
  {Hu}(2013)}]{Takada:2013wfa}%
  \BibitemOpen
  \bibfield  {author} {\bibinfo {author} {\bibfnamefont {M.}~\bibnamefont
  {Takada}}\ and\ \bibinfo {author} {\bibfnamefont {W.}~\bibnamefont {Hu}},\
  }\bibfield  {title} {\bibinfo {title} {{Power Spectrum Super-Sample
  Covariance}},\ }\href {https://doi.org/10.1103/PhysRevD.87.123504} {\bibfield
   {journal} {\bibinfo  {journal} {Phys. Rev. D}\ }\textbf {\bibinfo {volume}
  {87}},\ \bibinfo {pages} {123504} (\bibinfo {year} {2013})},\ \Eprint
  {https://arxiv.org/abs/1302.6994} {arXiv:1302.6994 [astro-ph.CO]}
  \BibitemShut {NoStop}%
\bibitem [{\citenamefont {Takahashi}\ \emph {et~al.}(2019)\citenamefont
  {Takahashi}, \citenamefont {Nishimichi}, \citenamefont {Takada},
  \citenamefont {Shirasaki},\ and\ \citenamefont
  {Shiroyama}}]{Takahashi:2018pze}%
  \BibitemOpen
  \bibfield  {author} {\bibinfo {author} {\bibfnamefont {R.}~\bibnamefont
  {Takahashi}}, \bibinfo {author} {\bibfnamefont {T.}~\bibnamefont
  {Nishimichi}}, \bibinfo {author} {\bibfnamefont {M.}~\bibnamefont {Takada}},
  \bibinfo {author} {\bibfnamefont {M.}~\bibnamefont {Shirasaki}},\ and\
  \bibinfo {author} {\bibfnamefont {K.}~\bibnamefont {Shiroyama}},\ }\bibfield
  {title} {\bibinfo {title} {{Covariances for cosmic shear and
  galaxy\textendash{}galaxy lensing in the response approach}},\ }\href
  {https://doi.org/10.1093/mnras/sty2962} {\bibfield  {journal} {\bibinfo
  {journal} {Mon. Not. Roy. Astron. Soc.}\ }\textbf {\bibinfo {volume} {482}},\
  \bibinfo {pages} {4253} (\bibinfo {year} {2019})},\ \Eprint
  {https://arxiv.org/abs/1805.11629} {arXiv:1805.11629 [astro-ph.CO]}
  \BibitemShut {NoStop}%
\bibitem [{\citenamefont {Schneider}\ and\ \citenamefont
  {Teyssier}(2015)}]{Schneider:2015wta}%
  \BibitemOpen
  \bibfield  {author} {\bibinfo {author} {\bibfnamefont {A.}~\bibnamefont
  {Schneider}}\ and\ \bibinfo {author} {\bibfnamefont {R.}~\bibnamefont
  {Teyssier}},\ }\bibfield  {title} {\bibinfo {title} {{A new method to
  quantify the effects of baryons on the matter power spectrum}},\ }\href
  {https://doi.org/10.1088/1475-7516/2015/12/049} {\bibfield  {journal}
  {\bibinfo  {journal} {JCAP}\ }\textbf {\bibinfo {volume} {12}},\ \bibinfo
  {pages} {049}},\ \Eprint {https://arxiv.org/abs/1510.06034} {arXiv:1510.06034
  [astro-ph.CO]} \BibitemShut {NoStop}%
\bibitem [{\citenamefont {Schneider}\ \emph {et~al.}(2019)\citenamefont
  {Schneider}, \citenamefont {Teyssier}, \citenamefont {Stadel}, \citenamefont
  {Chisari}, \citenamefont {Le~Brun}, \citenamefont {Amara},\ and\
  \citenamefont {Refregier}}]{Schneider:2018pfw}%
  \BibitemOpen
  \bibfield  {author} {\bibinfo {author} {\bibfnamefont {A.}~\bibnamefont
  {Schneider}}, \bibinfo {author} {\bibfnamefont {R.}~\bibnamefont {Teyssier}},
  \bibinfo {author} {\bibfnamefont {J.}~\bibnamefont {Stadel}}, \bibinfo
  {author} {\bibfnamefont {N.~E.}\ \bibnamefont {Chisari}}, \bibinfo {author}
  {\bibfnamefont {A.~M.~C.}\ \bibnamefont {Le~Brun}}, \bibinfo {author}
  {\bibfnamefont {A.}~\bibnamefont {Amara}},\ and\ \bibinfo {author}
  {\bibfnamefont {A.}~\bibnamefont {Refregier}},\ }\bibfield  {title} {\bibinfo
  {title} {{Quantifying baryon effects on the matter power spectrum and the
  weak lensing shear correlation}},\ }\href
  {https://doi.org/10.1088/1475-7516/2019/03/020} {\bibfield  {journal}
  {\bibinfo  {journal} {JCAP}\ }\textbf {\bibinfo {volume} {03}},\ \bibinfo
  {pages} {020}},\ \Eprint {https://arxiv.org/abs/1810.08629} {arXiv:1810.08629
  [astro-ph.CO]} \BibitemShut {NoStop}%
\bibitem [{\citenamefont {Giri}\ and\ \citenamefont
  {Schneider}(2021)}]{Giri:2021qin}%
  \BibitemOpen
  \bibfield  {author} {\bibinfo {author} {\bibfnamefont {S.~K.}\ \bibnamefont
  {Giri}}\ and\ \bibinfo {author} {\bibfnamefont {A.}~\bibnamefont
  {Schneider}},\ }\bibfield  {title} {\bibinfo {title} {{Emulation of baryonic
  effects on the matter power spectrum and constraints from galaxy cluster
  data}},\ }\href {https://doi.org/10.1088/1475-7516/2021/12/046} {\bibfield
  {journal} {\bibinfo  {journal} {JCAP}\ }\textbf {\bibinfo {volume}
  {12}}\bibfield  {number} {\bibinfo  {number} { (12)},\ \bibinfo {pages}
  {046}},\ }\Eprint {https://arxiv.org/abs/2108.08863} {arXiv:2108.08863
  [astro-ph.CO]} \BibitemShut {NoStop}%
\bibitem [{\citenamefont {Wu}\ \emph {et~al.}(2021)\citenamefont {Wu},
  \citenamefont {Weinberg}, \citenamefont {Salcedo},\ and\ \citenamefont
  {Wibking}}]{Wu:2020xxy}%
  \BibitemOpen
  \bibfield  {author} {\bibinfo {author} {\bibfnamefont {H.-Y.}\ \bibnamefont
  {Wu}}, \bibinfo {author} {\bibfnamefont {D.~H.}\ \bibnamefont {Weinberg}},
  \bibinfo {author} {\bibfnamefont {A.~N.}\ \bibnamefont {Salcedo}},\ and\
  \bibinfo {author} {\bibfnamefont {B.~D.}\ \bibnamefont {Wibking}},\
  }\bibfield  {title} {\bibinfo {title} {{Cosmology with galaxy cluster weak
  lensing: statistical limits and experimental design}},\ }\href
  {https://doi.org/10.3847/1538-4357/abdc23} {\bibfield  {journal} {\bibinfo
  {journal} {Astrophys. J.}\ }\textbf {\bibinfo {volume} {910}},\ \bibinfo
  {pages} {28} (\bibinfo {year} {2021})},\ \Eprint
  {https://arxiv.org/abs/2012.01956} {arXiv:2012.01956 [astro-ph.CO]}
  \BibitemShut {NoStop}%
\bibitem [{\citenamefont {{Farahi}}\ \emph {et~al.}(2018)\citenamefont
  {{Farahi}}, \citenamefont {{Evrard}}, \citenamefont {{McCarthy}},
  \citenamefont {{Barnes}},\ and\ \citenamefont {{Kay}}}]{kllr1}%
  \BibitemOpen
  \bibfield  {author} {\bibinfo {author} {\bibfnamefont {A.}~\bibnamefont
  {{Farahi}}}, \bibinfo {author} {\bibfnamefont {A.~E.}\ \bibnamefont
  {{Evrard}}}, \bibinfo {author} {\bibfnamefont {I.}~\bibnamefont
  {{McCarthy}}}, \bibinfo {author} {\bibfnamefont {D.~J.}\ \bibnamefont
  {{Barnes}}},\ and\ \bibinfo {author} {\bibfnamefont {S.~T.}\ \bibnamefont
  {{Kay}}},\ }\bibfield  {title} {\bibinfo {title} {{Localized massive halo
  properties in BAHAMAS and MACSIS simulations: scalings, lognormality, and
  covariance}},\ }\href {https://doi.org/10.1093/mnras/sty1179} {\bibfield
  {journal} {\bibinfo  {journal} {\mnras}\ }\textbf {\bibinfo {volume} {478}},\
  \bibinfo {pages} {2618} (\bibinfo {year} {2018})},\ \Eprint
  {https://arxiv.org/abs/1711.04922} {arXiv:1711.04922 [astro-ph.CO]}
  \BibitemShut {NoStop}%
\bibitem [{\citenamefont {{Anbajagane}}\ \emph {et~al.}(2020)\citenamefont
  {{Anbajagane}}, \citenamefont {{Evrard}}, \citenamefont {{Farahi}},
  \citenamefont {{Barnes}}, \citenamefont {{Dolag}}, \citenamefont
  {{McCarthy}}, \citenamefont {{Nelson}},\ and\ \citenamefont
  {{Pillepich}}}]{kllr2}%
  \BibitemOpen
  \bibfield  {author} {\bibinfo {author} {\bibfnamefont {D.}~\bibnamefont
  {{Anbajagane}}}, \bibinfo {author} {\bibfnamefont {A.~E.}\ \bibnamefont
  {{Evrard}}}, \bibinfo {author} {\bibfnamefont {A.}~\bibnamefont {{Farahi}}},
  \bibinfo {author} {\bibfnamefont {D.~J.}\ \bibnamefont {{Barnes}}}, \bibinfo
  {author} {\bibfnamefont {K.}~\bibnamefont {{Dolag}}}, \bibinfo {author}
  {\bibfnamefont {I.~G.}\ \bibnamefont {{McCarthy}}}, \bibinfo {author}
  {\bibfnamefont {D.}~\bibnamefont {{Nelson}}},\ and\ \bibinfo {author}
  {\bibfnamefont {A.}~\bibnamefont {{Pillepich}}},\ }\bibfield  {title}
  {\bibinfo {title} {{Stellar property statistics of massive haloes from
  cosmological hydrodynamics simulations: common kernel shapes}},\ }\href
  {https://doi.org/10.1093/mnras/staa1147} {\bibfield  {journal} {\bibinfo
  {journal} {\mnras}\ }\textbf {\bibinfo {volume} {495}},\ \bibinfo {pages}
  {686} (\bibinfo {year} {2020})},\ \Eprint {https://arxiv.org/abs/2001.02283}
  {arXiv:2001.02283 [astro-ph.GA]} \BibitemShut {NoStop}%
\bibitem [{\citenamefont {{Oguri}}\ and\ \citenamefont
  {{Takada}}(2011)}]{2011PhRvD..83b3008O}%
  \BibitemOpen
  \bibfield  {author} {\bibinfo {author} {\bibfnamefont {M.}~\bibnamefont
  {{Oguri}}}\ and\ \bibinfo {author} {\bibfnamefont {M.}~\bibnamefont
  {{Takada}}},\ }\bibfield  {title} {\bibinfo {title} {{Combining cluster
  observables and stacked weak lensing to probe dark energy: Self-calibration
  of systematic uncertainties}},\ }\href
  {https://doi.org/10.1103/PhysRevD.83.023008} {\bibfield  {journal} {\bibinfo
  {journal} {\prd}\ }\textbf {\bibinfo {volume} {83}},\ \bibinfo {eid} {023008}
  (\bibinfo {year} {2011})},\ \Eprint {https://arxiv.org/abs/1010.0744}
  {arXiv:1010.0744 [astro-ph.CO]} \BibitemShut {NoStop}%
\bibitem [{\citenamefont {Diemer}\ and\ \citenamefont
  {Kravtsov}(2014)}]{Diemer:2014xya}%
  \BibitemOpen
  \bibfield  {author} {\bibinfo {author} {\bibfnamefont {B.}~\bibnamefont
  {Diemer}}\ and\ \bibinfo {author} {\bibfnamefont {A.~V.}\ \bibnamefont
  {Kravtsov}},\ }\bibfield  {title} {\bibinfo {title} {{Dependence of the outer
  density profiles of halos on their mass accretion rate}},\ }\href
  {https://doi.org/10.1088/0004-637X/789/1/1} {\bibfield  {journal} {\bibinfo
  {journal} {Astrophys. J.}\ }\textbf {\bibinfo {volume} {789}},\ \bibinfo
  {pages} {1} (\bibinfo {year} {2014})},\ \Eprint
  {https://arxiv.org/abs/1401.1216} {arXiv:1401.1216 [astro-ph.CO]}
  \BibitemShut {NoStop}%
\bibitem [{\citenamefont {Baltz}\ \emph {et~al.}(2009)\citenamefont {Baltz},
  \citenamefont {Marshall},\ and\ \citenamefont {Oguri}}]{Baltz:2007vq}%
  \BibitemOpen
  \bibfield  {author} {\bibinfo {author} {\bibfnamefont {E.~A.}\ \bibnamefont
  {Baltz}}, \bibinfo {author} {\bibfnamefont {P.}~\bibnamefont {Marshall}},\
  and\ \bibinfo {author} {\bibfnamefont {M.}~\bibnamefont {Oguri}},\ }\bibfield
   {title} {\bibinfo {title} {{Analytic models of plausible gravitational lens
  potentials}},\ }\href {https://doi.org/10.1088/1475-7516/2009/01/015}
  {\bibfield  {journal} {\bibinfo  {journal} {JCAP}\ }\textbf {\bibinfo
  {volume} {01}},\ \bibinfo {pages} {015}},\ \Eprint
  {https://arxiv.org/abs/0705.0682} {arXiv:0705.0682 [astro-ph]} \BibitemShut
  {NoStop}%
\bibitem [{\citenamefont {Oguri}\ and\ \citenamefont
  {Hamana}(2011)}]{Oguri:2011vj}%
  \BibitemOpen
  \bibfield  {author} {\bibinfo {author} {\bibfnamefont {M.}~\bibnamefont
  {Oguri}}\ and\ \bibinfo {author} {\bibfnamefont {T.}~\bibnamefont {Hamana}},\
  }\bibfield  {title} {\bibinfo {title} {{Detailed cluster lensing profiles at
  large radii and the impact on cluster weak lensing studies}},\ }\href
  {https://doi.org/10.1111/j.1365-2966.2011.18481.x} {\bibfield  {journal}
  {\bibinfo  {journal} {Mon. Not. Roy. Astron. Soc.}\ }\textbf {\bibinfo
  {volume} {414}},\ \bibinfo {pages} {1851} (\bibinfo {year} {2011})},\ \Eprint
  {https://arxiv.org/abs/1101.0650} {arXiv:1101.0650 [astro-ph.CO]}
  \BibitemShut {NoStop}%
\bibitem [{\citenamefont {Abadi}\ \emph {et~al.}(2010)\citenamefont {Abadi},
  \citenamefont {Navarro}, \citenamefont {Fardal}, \citenamefont {Babul},\ and\
  \citenamefont {Steinmetz}}]{Abadi:2009ve}%
  \BibitemOpen
  \bibfield  {author} {\bibinfo {author} {\bibfnamefont {M.~G.}\ \bibnamefont
  {Abadi}}, \bibinfo {author} {\bibfnamefont {J.~F.}\ \bibnamefont {Navarro}},
  \bibinfo {author} {\bibfnamefont {M.}~\bibnamefont {Fardal}}, \bibinfo
  {author} {\bibfnamefont {A.}~\bibnamefont {Babul}},\ and\ \bibinfo {author}
  {\bibfnamefont {M.}~\bibnamefont {Steinmetz}},\ }\bibfield  {title} {\bibinfo
  {title} {{Galaxy-Induced Transformation of Dark Matter Halos}},\ }\href
  {https://doi.org/10.1111/j.1365-2966.2010.16912.x} {\bibfield  {journal}
  {\bibinfo  {journal} {Mon. Not. Roy. Astron. Soc.}\ }\textbf {\bibinfo
  {volume} {407}},\ \bibinfo {pages} {435} (\bibinfo {year} {2010})},\ \Eprint
  {https://arxiv.org/abs/0902.2477} {arXiv:0902.2477 [astro-ph.GA]}
  \BibitemShut {NoStop}%
\bibitem [{\citenamefont {van Daalen}\ \emph {et~al.}(2020)\citenamefont {van
  Daalen}, \citenamefont {McCarthy},\ and\ \citenamefont
  {Schaye}}]{vanDaalen:2019pst}%
  \BibitemOpen
  \bibfield  {author} {\bibinfo {author} {\bibfnamefont {M.~P.}\ \bibnamefont
  {van Daalen}}, \bibinfo {author} {\bibfnamefont {I.~G.}\ \bibnamefont
  {McCarthy}},\ and\ \bibinfo {author} {\bibfnamefont {J.}~\bibnamefont
  {Schaye}},\ }\bibfield  {title} {\bibinfo {title} {{Exploring the effects of
  galaxy formation on matter clustering through a library of simulation power
  spectra}},\ }\href {https://doi.org/10.1093/mnras/stz3199} {\bibfield
  {journal} {\bibinfo  {journal} {Mon. Not. Roy. Astron. Soc.}\ }\textbf
  {\bibinfo {volume} {491}},\ \bibinfo {pages} {2424} (\bibinfo {year}
  {2020})},\ \Eprint {https://arxiv.org/abs/1906.00968} {arXiv:1906.00968
  [astro-ph.CO]} \BibitemShut {NoStop}%
\bibitem [{\citenamefont {Debackere}\ \emph {et~al.}(2021)\citenamefont
  {Debackere}, \citenamefont {Schaye},\ and\ \citenamefont
  {Hoekstra}}]{Debackere:2021ado}%
  \BibitemOpen
  \bibfield  {author} {\bibinfo {author} {\bibfnamefont {S.~N.~B.}\
  \bibnamefont {Debackere}}, \bibinfo {author} {\bibfnamefont {J.}~\bibnamefont
  {Schaye}},\ and\ \bibinfo {author} {\bibfnamefont {H.}~\bibnamefont
  {Hoekstra}},\ }\bibfield  {title} {\bibinfo {title} {{How baryons can
  significantly bias cluster count cosmology}},\ }\href
  {https://doi.org/10.1093/mnras/stab1326} {\bibfield  {journal} {\bibinfo
  {journal} {Mon. Not. Roy. Astron. Soc.}\ }\textbf {\bibinfo {volume} {505}},\
  \bibinfo {pages} {593} (\bibinfo {year} {2021})},\ \Eprint
  {https://arxiv.org/abs/2101.07800} {arXiv:2101.07800 [astro-ph.CO]}
  \BibitemShut {NoStop}%
\bibitem [{\citenamefont {Cromer}\ \emph {et~al.}(2022)\citenamefont {Cromer},
  \citenamefont {Battaglia}, \citenamefont {Miyatake},\ and\ \citenamefont
  {Simet}}]{Cromer:2021iyp}%
  \BibitemOpen
  \bibfield  {author} {\bibinfo {author} {\bibfnamefont {D.}~\bibnamefont
  {Cromer}}, \bibinfo {author} {\bibfnamefont {N.}~\bibnamefont {Battaglia}},
  \bibinfo {author} {\bibfnamefont {H.}~\bibnamefont {Miyatake}},\ and\
  \bibinfo {author} {\bibfnamefont {M.}~\bibnamefont {Simet}},\ }\bibfield
  {title} {\bibinfo {title} {{Towards 1\% accurate galaxy cluster masses:
  including baryons in weak-lensing mass inference}},\ }\href
  {https://doi.org/10.1088/1475-7516/2022/10/034} {\bibfield  {journal}
  {\bibinfo  {journal} {JCAP}\ }\textbf {\bibinfo {volume} {10}},\ \bibinfo
  {pages} {034}},\ \Eprint {https://arxiv.org/abs/2104.06925} {arXiv:2104.06925
  [astro-ph.CO]} \BibitemShut {NoStop}%
\bibitem [{\citenamefont {Shirasaki}\ \emph {et~al.}(2018)\citenamefont
  {Shirasaki}, \citenamefont {Lau},\ and\ \citenamefont
  {Nagai}}]{Shirasaki:2017jlt}%
  \BibitemOpen
  \bibfield  {author} {\bibinfo {author} {\bibfnamefont {M.}~\bibnamefont
  {Shirasaki}}, \bibinfo {author} {\bibfnamefont {E.~T.}\ \bibnamefont {Lau}},\
  and\ \bibinfo {author} {\bibfnamefont {D.}~\bibnamefont {Nagai}},\ }\bibfield
   {title} {\bibinfo {title} {{Modelling Baryonic Effects on Galaxy Cluster
  Mass Profiles}},\ }\href {https://doi.org/10.1093/mnras/sty763} {\bibfield
  {journal} {\bibinfo  {journal} {Mon. Not. Roy. Astron. Soc.}\ }\textbf
  {\bibinfo {volume} {477}},\ \bibinfo {pages} {2804} (\bibinfo {year}
  {2018})},\ \Eprint {https://arxiv.org/abs/1711.06366} {arXiv:1711.06366
  [astro-ph.CO]} \BibitemShut {NoStop}%
\bibitem [{\citenamefont {Foreman-Mackey}\ \emph {et~al.}(2013)\citenamefont
  {Foreman-Mackey}, \citenamefont {Hogg}, \citenamefont {Lang},\ and\
  \citenamefont {Goodman}}]{Foreman-Mackey:2012any}%
  \BibitemOpen
  \bibfield  {author} {\bibinfo {author} {\bibfnamefont {D.}~\bibnamefont
  {Foreman-Mackey}}, \bibinfo {author} {\bibfnamefont {D.~W.}\ \bibnamefont
  {Hogg}}, \bibinfo {author} {\bibfnamefont {D.}~\bibnamefont {Lang}},\ and\
  \bibinfo {author} {\bibfnamefont {J.}~\bibnamefont {Goodman}},\ }\bibfield
  {title} {\bibinfo {title} {{emcee: The MCMC Hammer}},\ }\href
  {https://doi.org/10.1086/670067} {\bibfield  {journal} {\bibinfo  {journal}
  {Publ. Astron. Soc. Pac.}\ }\textbf {\bibinfo {volume} {125}},\ \bibinfo
  {pages} {306} (\bibinfo {year} {2013})},\ \Eprint
  {https://arxiv.org/abs/1202.3665} {arXiv:1202.3665 [astro-ph.IM]}
  \BibitemShut {NoStop}%
\bibitem [{\citenamefont {Battaglia}\ \emph {et~al.}(2016)\citenamefont
  {Battaglia} \emph {et~al.}}]{Battaglia:2015zfa}%
  \BibitemOpen
  \bibfield  {author} {\bibinfo {author} {\bibfnamefont {N.}~\bibnamefont
  {Battaglia}} \emph {et~al.},\ }\bibfield  {title} {\bibinfo {title}
  {{Weak-Lensing Mass Calibration of the Atacama Cosmology Telescope Equatorial
  Sunyaev-Zeldovich Cluster Sample with the Canada-France-Hawaii Telescope
  Stripe 82 Survey}},\ }\href {https://doi.org/10.1088/1475-7516/2016/08/013}
  {\bibfield  {journal} {\bibinfo  {journal} {JCAP}\ }\textbf {\bibinfo
  {volume} {08}},\ \bibinfo {pages} {013}},\ \Eprint
  {https://arxiv.org/abs/1509.08930} {arXiv:1509.08930 [astro-ph.CO]}
  \BibitemShut {NoStop}%
\bibitem [{\citenamefont {Penna-Lima}\ \emph {et~al.}(2017)\citenamefont
  {Penna-Lima}, \citenamefont {Bartlett}, \citenamefont {Rozo}, \citenamefont
  {Melin}, \citenamefont {Merten}, \citenamefont {Evrard}, \citenamefont
  {Postman},\ and\ \citenamefont {Rykoff}}]{Penna-Lima:2016tvo}%
  \BibitemOpen
  \bibfield  {author} {\bibinfo {author} {\bibfnamefont {M.}~\bibnamefont
  {Penna-Lima}}, \bibinfo {author} {\bibfnamefont {J.~G.}\ \bibnamefont
  {Bartlett}}, \bibinfo {author} {\bibfnamefont {E.}~\bibnamefont {Rozo}},
  \bibinfo {author} {\bibfnamefont {J.~B.}\ \bibnamefont {Melin}}, \bibinfo
  {author} {\bibfnamefont {J.}~\bibnamefont {Merten}}, \bibinfo {author}
  {\bibfnamefont {A.~E.}\ \bibnamefont {Evrard}}, \bibinfo {author}
  {\bibfnamefont {M.}~\bibnamefont {Postman}},\ and\ \bibinfo {author}
  {\bibfnamefont {E.}~\bibnamefont {Rykoff}},\ }\bibfield  {title} {\bibinfo
  {title} {{Calibrating the Planck Cluster Mass Scale with CLASH}},\ }\href
  {https://doi.org/10.1051/0004-6361/201629971} {\bibfield  {journal} {\bibinfo
   {journal} {Astron. Astrophys.}\ }\textbf {\bibinfo {volume} {604}},\
  \bibinfo {pages} {A89} (\bibinfo {year} {2017})},\ \Eprint
  {https://arxiv.org/abs/1608.05356} {arXiv:1608.05356 [astro-ph.CO]}
  \BibitemShut {NoStop}%
\bibitem [{\citenamefont {{Medezinski}}\ \emph {et~al.}(2018)\citenamefont
  {{Medezinski}}, \citenamefont {{Battaglia}}, \citenamefont {{Umetsu}},
  \citenamefont {{Oguri}}, \citenamefont {{Miyatake}}, \citenamefont
  {{Nishizawa}}, \citenamefont {{Sif{\'o}n}}, \citenamefont {{Spergel}},
  \citenamefont {{Chiu}}, \citenamefont {{Lin}}, \citenamefont {{Bahcall}},\
  and\ \citenamefont {{Komiyama}}}]{2018PASJ...70S..28M}%
  \BibitemOpen
  \bibfield  {author} {\bibinfo {author} {\bibfnamefont {E.}~\bibnamefont
  {{Medezinski}}}, \bibinfo {author} {\bibfnamefont {N.}~\bibnamefont
  {{Battaglia}}}, \bibinfo {author} {\bibfnamefont {K.}~\bibnamefont
  {{Umetsu}}}, \bibinfo {author} {\bibfnamefont {M.}~\bibnamefont {{Oguri}}},
  \bibinfo {author} {\bibfnamefont {H.}~\bibnamefont {{Miyatake}}}, \bibinfo
  {author} {\bibfnamefont {A.~J.}\ \bibnamefont {{Nishizawa}}}, \bibinfo
  {author} {\bibfnamefont {C.}~\bibnamefont {{Sif{\'o}n}}}, \bibinfo {author}
  {\bibfnamefont {D.~N.}\ \bibnamefont {{Spergel}}}, \bibinfo {author}
  {\bibfnamefont {I.~N.}\ \bibnamefont {{Chiu}}}, \bibinfo {author}
  {\bibfnamefont {Y.-T.}\ \bibnamefont {{Lin}}}, \bibinfo {author}
  {\bibfnamefont {N.}~\bibnamefont {{Bahcall}}},\ and\ \bibinfo {author}
  {\bibfnamefont {Y.}~\bibnamefont {{Komiyama}}},\ }\bibfield  {title}
  {\bibinfo {title} {{Planck Sunyaev-Zel'dovich cluster mass calibration using
  Hyper Suprime-Cam weak lensing}},\ }\href
  {https://doi.org/10.1093/pasj/psx128} {\bibfield  {journal} {\bibinfo
  {journal} {\pasj}\ }\textbf {\bibinfo {volume} {70}},\ \bibinfo {eid} {S28}
  (\bibinfo {year} {2018})},\ \Eprint {https://arxiv.org/abs/1706.00434}
  {arXiv:1706.00434 [astro-ph.CO]} \BibitemShut {NoStop}%
\bibitem [{\citenamefont {Sereno}\ \emph {et~al.}(2017)\citenamefont {Sereno},
  \citenamefont {Covone}, \citenamefont {Izzo}, \citenamefont {Ettori},
  \citenamefont {Coupon},\ and\ \citenamefont {Lieu}}]{Sereno:2017zcn}%
  \BibitemOpen
  \bibfield  {author} {\bibinfo {author} {\bibfnamefont {M.}~\bibnamefont
  {Sereno}}, \bibinfo {author} {\bibfnamefont {G.}~\bibnamefont {Covone}},
  \bibinfo {author} {\bibfnamefont {L.}~\bibnamefont {Izzo}}, \bibinfo {author}
  {\bibfnamefont {S.}~\bibnamefont {Ettori}}, \bibinfo {author} {\bibfnamefont
  {J.}~\bibnamefont {Coupon}},\ and\ \bibinfo {author} {\bibfnamefont
  {M.}~\bibnamefont {Lieu}},\ }\bibfield  {title} {\bibinfo {title} {{PSZ2LenS.
  Weak lensing analysis of the Planck clusters in the CFHTLenS and in the
  RCSLenS}},\ }\href {https://doi.org/10.1093/mnras/stx2085} {\bibfield
  {journal} {\bibinfo  {journal} {Mon. Not. Roy. Astron. Soc.}\ }\textbf
  {\bibinfo {volume} {472}},\ \bibinfo {pages} {1946} (\bibinfo {year}
  {2017})},\ \Eprint {https://arxiv.org/abs/1703.06886} {arXiv:1703.06886
  [astro-ph.CO]} \BibitemShut {NoStop}%
\bibitem [{\citenamefont {Herbonnet}\ \emph {et~al.}(2020)\citenamefont
  {Herbonnet}, \citenamefont {Sifon}, \citenamefont {Hoekstra}, \citenamefont
  {Bahe}, \citenamefont {van~der Burg}, \citenamefont {Melin}, \citenamefont
  {von~der Linden}, \citenamefont {Sand}, \citenamefont {Kay},\ and\
  \citenamefont {Barnes}}]{Herbonnet:2019byy}%
  \BibitemOpen
  \bibfield  {author} {\bibinfo {author} {\bibfnamefont {R.}~\bibnamefont
  {Herbonnet}}, \bibinfo {author} {\bibfnamefont {C.}~\bibnamefont {Sifon}},
  \bibinfo {author} {\bibfnamefont {H.}~\bibnamefont {Hoekstra}}, \bibinfo
  {author} {\bibfnamefont {Y.}~\bibnamefont {Bahe}}, \bibinfo {author}
  {\bibfnamefont {R.~F.~J.}\ \bibnamefont {van~der Burg}}, \bibinfo {author}
  {\bibfnamefont {J.-B.}\ \bibnamefont {Melin}}, \bibinfo {author}
  {\bibfnamefont {A.}~\bibnamefont {von~der Linden}}, \bibinfo {author}
  {\bibfnamefont {D.}~\bibnamefont {Sand}}, \bibinfo {author} {\bibfnamefont
  {S.}~\bibnamefont {Kay}},\ and\ \bibinfo {author} {\bibfnamefont
  {D.}~\bibnamefont {Barnes}},\ }\bibfield  {title} {\bibinfo {title} {{CCCP
  and MENeaCS: (updated) weak-lensing masses for 100 galaxy clusters}},\ }\href
  {https://doi.org/10.1093/mnras/staa2303} {\bibfield  {journal} {\bibinfo
  {journal} {Mon. Not. Roy. Astron. Soc.}\ }\textbf {\bibinfo {volume} {497}},\
  \bibinfo {pages} {4684} (\bibinfo {year} {2020})},\ \Eprint
  {https://arxiv.org/abs/1912.04414} {arXiv:1912.04414 [astro-ph.CO]}
  \BibitemShut {NoStop}%
\bibitem [{\citenamefont {Schrabback}\ \emph {et~al.}(2021)\citenamefont
  {Schrabback} \emph {et~al.}}]{Schrabback:2020hxx}%
  \BibitemOpen
  \bibfield  {author} {\bibinfo {author} {\bibfnamefont {T.}~\bibnamefont
  {Schrabback}} \emph {et~al.},\ }\bibfield  {title} {\bibinfo {title} {{Mass
  calibration of distant SPT galaxy clusters through expanded weak-lensing
  follow-up observations with HST, VLT, \& Gemini-South}},\ }\href
  {https://doi.org/10.1093/mnras/stab1386} {\bibfield  {journal} {\bibinfo
  {journal} {Mon. Not. Roy. Astron. Soc.}\ }\textbf {\bibinfo {volume} {505}},\
  \bibinfo {pages} {3923} (\bibinfo {year} {2021})},\ \Eprint
  {https://arxiv.org/abs/2009.07591} {arXiv:2009.07591 [astro-ph.CO]}
  \BibitemShut {NoStop}%
\bibitem [{\citenamefont {Bolliet}\ \emph {et~al.}(2018)\citenamefont
  {Bolliet}, \citenamefont {Comis}, \citenamefont {Komatsu},\ and\
  \citenamefont {Mac\'\i{}as-P\'erez}}]{Bolliet:2017lha}%
  \BibitemOpen
  \bibfield  {author} {\bibinfo {author} {\bibfnamefont {B.}~\bibnamefont
  {Bolliet}}, \bibinfo {author} {\bibfnamefont {B.}~\bibnamefont {Comis}},
  \bibinfo {author} {\bibfnamefont {E.}~\bibnamefont {Komatsu}},\ and\ \bibinfo
  {author} {\bibfnamefont {J.~F.}\ \bibnamefont {Mac\'\i{}as-P\'erez}},\
  }\bibfield  {title} {\bibinfo {title} {{Dark energy constraints from the
  thermal Sunyaev\textendash{}Zeldovich power spectrum}},\ }\href
  {https://doi.org/10.1093/mnras/sty823} {\bibfield  {journal} {\bibinfo
  {journal} {Mon. Not. Roy. Astron. Soc.}\ }\textbf {\bibinfo {volume} {477}},\
  \bibinfo {pages} {4957} (\bibinfo {year} {2018})},\ \Eprint
  {https://arxiv.org/abs/1712.00788} {arXiv:1712.00788 [astro-ph.CO]}
  \BibitemShut {NoStop}%
\bibitem [{\citenamefont {{Wright}}\ and\ \citenamefont
  {{Brainerd}}(2000)}]{Wright2000}%
  \BibitemOpen
  \bibfield  {author} {\bibinfo {author} {\bibfnamefont {C.~O.}\ \bibnamefont
  {{Wright}}}\ and\ \bibinfo {author} {\bibfnamefont {T.~G.}\ \bibnamefont
  {{Brainerd}}},\ }\bibfield  {title} {\bibinfo {title} {{Gravitational Lensing
  by NFW Halos}},\ }\href {https://doi.org/10.1086/308744} {\bibfield
  {journal} {\bibinfo  {journal} {\apj}\ }\textbf {\bibinfo {volume} {534}},\
  \bibinfo {pages} {34} (\bibinfo {year} {2000})}\BibitemShut {NoStop}%
\bibitem [{\citenamefont {Ishiyama}\ \emph {et~al.}(2021)\citenamefont
  {Ishiyama} \emph {et~al.}}]{Ishiyama:2020vao}%
  \BibitemOpen
  \bibfield  {author} {\bibinfo {author} {\bibfnamefont {T.}~\bibnamefont
  {Ishiyama}} \emph {et~al.},\ }\bibfield  {title} {\bibinfo {title} {{The
  Uchuu simulations: Data Release 1 and dark matter halo concentrations}},\
  }\href {https://doi.org/10.1093/mnras/stab1755} {\bibfield  {journal}
  {\bibinfo  {journal} {Mon. Not. Roy. Astron. Soc.}\ }\textbf {\bibinfo
  {volume} {506}},\ \bibinfo {pages} {4210} (\bibinfo {year} {2021})},\ \Eprint
  {https://arxiv.org/abs/2007.14720} {arXiv:2007.14720 [astro-ph.CO]}
  \BibitemShut {NoStop}%
\bibitem [{\citenamefont {{Diemer}}(2018)}]{Diemer2018}%
  \BibitemOpen
  \bibfield  {author} {\bibinfo {author} {\bibfnamefont {B.}~\bibnamefont
  {{Diemer}}},\ }\bibfield  {title} {\bibinfo {title} {{COLOSSUS: A Python
  Toolkit for Cosmology, Large-scale Structure, and Dark Matter Halos}},\
  }\href {https://doi.org/10.3847/1538-4365/aaee8c} {\bibfield  {journal}
  {\bibinfo  {journal} {\apjs}\ }\textbf {\bibinfo {volume} {239}},\ \bibinfo
  {eid} {35} (\bibinfo {year} {2018})},\ \Eprint
  {https://arxiv.org/abs/1712.04512} {arXiv:1712.04512 [astro-ph.CO]}
  \BibitemShut {NoStop}%
\bibitem [{\citenamefont {{Tinker}}\ \emph {et~al.}(2010)\citenamefont
  {{Tinker}}, \citenamefont {{Robertson}}, \citenamefont {{Kravtsov}},
  \citenamefont {{Klypin}}, \citenamefont {{Warren}}, \citenamefont {{Yepes}},\
  and\ \citenamefont {{Gottl{\"o}ber}}}]{Tinker2010}%
  \BibitemOpen
  \bibfield  {author} {\bibinfo {author} {\bibfnamefont {J.~L.}\ \bibnamefont
  {{Tinker}}}, \bibinfo {author} {\bibfnamefont {B.~E.}\ \bibnamefont
  {{Robertson}}}, \bibinfo {author} {\bibfnamefont {A.~V.}\ \bibnamefont
  {{Kravtsov}}}, \bibinfo {author} {\bibfnamefont {A.}~\bibnamefont
  {{Klypin}}}, \bibinfo {author} {\bibfnamefont {M.~S.}\ \bibnamefont
  {{Warren}}}, \bibinfo {author} {\bibfnamefont {G.}~\bibnamefont {{Yepes}}},\
  and\ \bibinfo {author} {\bibfnamefont {S.}~\bibnamefont {{Gottl{\"o}ber}}},\
  }\bibfield  {title} {\bibinfo {title} {{The Large-scale Bias of Dark Matter
  Halos: Numerical Calibration and Model Tests}},\ }\href
  {https://doi.org/10.1088/0004-637X/724/2/878} {\bibfield  {journal} {\bibinfo
   {journal} {\apj}\ }\textbf {\bibinfo {volume} {724}},\ \bibinfo {pages}
  {878} (\bibinfo {year} {2010})},\ \Eprint {https://arxiv.org/abs/1001.3162}
  {arXiv:1001.3162 [astro-ph.CO]} \BibitemShut {NoStop}%
\bibitem [{\citenamefont {{Chisari}}\ \emph {et~al.}(2019)\citenamefont
  {{Chisari}}, \citenamefont {{Alonso}}, \citenamefont {{Krause}},
  \citenamefont {{Leonard}}, \citenamefont {{Bull}}, \citenamefont {{Neveu}},
  \citenamefont {{Villarreal}}, \citenamefont {{Singh}}, \citenamefont
  {{McClintock}}, \citenamefont {{Ellison}}, \citenamefont {{Du}},
  \citenamefont {{Zuntz}}, \citenamefont {{Mead}}, \citenamefont {{Joudaki}},
  \citenamefont {{Lorenz}}, \citenamefont {{Tr{\"o}ster}}, \citenamefont
  {{Sanchez}}, \citenamefont {{Lanusse}}, \citenamefont {{Ishak}},
  \citenamefont {{Hlozek}}, \citenamefont {{Blazek}}, \citenamefont
  {{Campagne}}, \citenamefont {{Almoubayyed}}, \citenamefont {{Eifler}},
  \citenamefont {{Kirby}}, \citenamefont {{Kirkby}}, \citenamefont
  {{Plaszczynski}}, \citenamefont {{Slosar}}, \citenamefont {{Vrastil}},
  \citenamefont {{Wagoner}},\ and\ \citenamefont {{LSST Dark Energy Science
  Collaboration}}}]{Chisari2019}%
  \BibitemOpen
  \bibfield  {author} {\bibinfo {author} {\bibfnamefont {N.~E.}\ \bibnamefont
  {{Chisari}}}, \bibinfo {author} {\bibfnamefont {D.}~\bibnamefont {{Alonso}}},
  \bibinfo {author} {\bibfnamefont {E.}~\bibnamefont {{Krause}}}, \bibinfo
  {author} {\bibfnamefont {C.~D.}\ \bibnamefont {{Leonard}}}, \bibinfo {author}
  {\bibfnamefont {P.}~\bibnamefont {{Bull}}}, \bibinfo {author} {\bibfnamefont
  {J.}~\bibnamefont {{Neveu}}}, \bibinfo {author} {\bibfnamefont {A.~S.}\
  \bibnamefont {{Villarreal}}}, \bibinfo {author} {\bibfnamefont
  {S.}~\bibnamefont {{Singh}}}, \bibinfo {author} {\bibfnamefont
  {T.}~\bibnamefont {{McClintock}}}, \bibinfo {author} {\bibfnamefont
  {J.}~\bibnamefont {{Ellison}}}, \bibinfo {author} {\bibfnamefont
  {Z.}~\bibnamefont {{Du}}}, \bibinfo {author} {\bibfnamefont {J.}~\bibnamefont
  {{Zuntz}}}, \bibinfo {author} {\bibfnamefont {A.}~\bibnamefont {{Mead}}},
  \bibinfo {author} {\bibfnamefont {S.}~\bibnamefont {{Joudaki}}}, \bibinfo
  {author} {\bibfnamefont {C.~S.}\ \bibnamefont {{Lorenz}}}, \bibinfo {author}
  {\bibfnamefont {T.}~\bibnamefont {{Tr{\"o}ster}}}, \bibinfo {author}
  {\bibfnamefont {J.}~\bibnamefont {{Sanchez}}}, \bibinfo {author}
  {\bibfnamefont {F.}~\bibnamefont {{Lanusse}}}, \bibinfo {author}
  {\bibfnamefont {M.}~\bibnamefont {{Ishak}}}, \bibinfo {author} {\bibfnamefont
  {R.}~\bibnamefont {{Hlozek}}}, \bibinfo {author} {\bibfnamefont
  {J.}~\bibnamefont {{Blazek}}}, \bibinfo {author} {\bibfnamefont {J.-E.}\
  \bibnamefont {{Campagne}}}, \bibinfo {author} {\bibfnamefont
  {H.}~\bibnamefont {{Almoubayyed}}}, \bibinfo {author} {\bibfnamefont
  {T.}~\bibnamefont {{Eifler}}}, \bibinfo {author} {\bibfnamefont
  {M.}~\bibnamefont {{Kirby}}}, \bibinfo {author} {\bibfnamefont
  {D.}~\bibnamefont {{Kirkby}}}, \bibinfo {author} {\bibfnamefont
  {S.}~\bibnamefont {{Plaszczynski}}}, \bibinfo {author} {\bibfnamefont
  {A.}~\bibnamefont {{Slosar}}}, \bibinfo {author} {\bibfnamefont
  {M.}~\bibnamefont {{Vrastil}}}, \bibinfo {author} {\bibfnamefont {E.~L.}\
  \bibnamefont {{Wagoner}}},\ and\ \bibinfo {author} {\bibnamefont {{LSST Dark
  Energy Science Collaboration}}},\ }\bibfield  {title} {\bibinfo {title}
  {{Core Cosmology Library: Precision Cosmological Predictions for LSST}},\
  }\href {https://doi.org/10.3847/1538-4365/ab1658} {\bibfield  {journal}
  {\bibinfo  {journal} {\apjs}\ }\textbf {\bibinfo {volume} {242}},\ \bibinfo
  {eid} {2} (\bibinfo {year} {2019})},\ \Eprint
  {https://arxiv.org/abs/1812.05995} {arXiv:1812.05995 [astro-ph.CO]}
  \BibitemShut {NoStop}%
\bibitem [{\citenamefont {{Johnston}}\ \emph {et~al.}(2007)\citenamefont
  {{Johnston}}, \citenamefont {{Sheldon}}, \citenamefont {{Wechsler}},
  \citenamefont {{Rozo}}, \citenamefont {{Koester}}, \citenamefont {{Frieman}},
  \citenamefont {{McKay}}, \citenamefont {{Evrard}}, \citenamefont {{Becker}},\
  and\ \citenamefont {{Annis}}}]{Johnston2007}%
  \BibitemOpen
  \bibfield  {author} {\bibinfo {author} {\bibfnamefont {D.~E.}\ \bibnamefont
  {{Johnston}}}, \bibinfo {author} {\bibfnamefont {E.~S.}\ \bibnamefont
  {{Sheldon}}}, \bibinfo {author} {\bibfnamefont {R.~H.}\ \bibnamefont
  {{Wechsler}}}, \bibinfo {author} {\bibfnamefont {E.}~\bibnamefont {{Rozo}}},
  \bibinfo {author} {\bibfnamefont {B.~P.}\ \bibnamefont {{Koester}}}, \bibinfo
  {author} {\bibfnamefont {J.~A.}\ \bibnamefont {{Frieman}}}, \bibinfo {author}
  {\bibfnamefont {T.~A.}\ \bibnamefont {{McKay}}}, \bibinfo {author}
  {\bibfnamefont {A.~E.}\ \bibnamefont {{Evrard}}}, \bibinfo {author}
  {\bibfnamefont {M.~R.}\ \bibnamefont {{Becker}}},\ and\ \bibinfo {author}
  {\bibfnamefont {J.}~\bibnamefont {{Annis}}},\ }\bibfield  {title} {\bibinfo
  {title} {{Cross-correlation Weak Lensing of SDSS galaxy Clusters II: Cluster
  Density Profiles and the Mass--Richness Relation}},\ }\href
  {https://doi.org/10.48550/arXiv.0709.1159} {\bibfield  {journal} {\bibinfo
  {journal} {arXiv e-prints}\ ,\ \bibinfo {eid} {arXiv:0709.1159}} (\bibinfo
  {year} {2007})},\ \Eprint {https://arxiv.org/abs/0709.1159} {arXiv:0709.1159
  [astro-ph]} \BibitemShut {NoStop}%
\bibitem [{\citenamefont {Sayers}\ \emph {et~al.}(2013)\citenamefont {Sayers}
  \emph {et~al.}}]{Sayers:2012vs}%
  \BibitemOpen
  \bibfield  {author} {\bibinfo {author} {\bibfnamefont {J.}~\bibnamefont
  {Sayers}} \emph {et~al.},\ }\bibfield  {title} {\bibinfo {title}
  {{Sunyaev-Zel'dovich-Measured Pressure Profiles from the Bolocam X-ray/SZ
  Galaxy Cluster Sample}},\ }\href
  {https://doi.org/10.1088/0004-637X/768/2/177} {\bibfield  {journal} {\bibinfo
   {journal} {Astrophys. J.}\ }\textbf {\bibinfo {volume} {768}},\ \bibinfo
  {pages} {177} (\bibinfo {year} {2013})},\ \Eprint
  {https://arxiv.org/abs/1211.1632} {arXiv:1211.1632 [astro-ph.CO]}
  \BibitemShut {NoStop}%
\bibitem [{\citenamefont {McDonald}\ \emph {et~al.}(2014)\citenamefont
  {McDonald} \emph {et~al.}}]{SPT:2014sco}%
  \BibitemOpen
  \bibfield  {author} {\bibinfo {author} {\bibfnamefont {M.}~\bibnamefont
  {McDonald}} \emph {et~al.} (\bibinfo {collaboration} {SPT}),\ }\bibfield
  {title} {\bibinfo {title} {{The Redshift Evolution of the Mean Temperature,
  Pressure, and Entropy Profiles in 80 SPT-Selected Galaxy Clusters}},\ }\href
  {https://doi.org/10.1088/0004-637X/794/1/67} {\bibfield  {journal} {\bibinfo
  {journal} {Astrophys. J.}\ }\textbf {\bibinfo {volume} {794}},\ \bibinfo
  {pages} {67} (\bibinfo {year} {2014})},\ \Eprint
  {https://arxiv.org/abs/1404.6250} {arXiv:1404.6250 [astro-ph.HE]}
  \BibitemShut {NoStop}%
\bibitem [{\citenamefont {Sayers}\ \emph {et~al.}(2016)\citenamefont {Sayers},
  \citenamefont {Golwala}, \citenamefont {Mantz}, \citenamefont {Molnar},
  \citenamefont {Naka}, \citenamefont {Pailet}, \citenamefont {Pierpaoli},
  \citenamefont {Siegel},\ and\ \citenamefont {Wolman}}]{Sayers:2016sro}%
  \BibitemOpen
  \bibfield  {author} {\bibinfo {author} {\bibfnamefont {J.}~\bibnamefont
  {Sayers}}, \bibinfo {author} {\bibfnamefont {S.~R.}\ \bibnamefont {Golwala}},
  \bibinfo {author} {\bibfnamefont {A.~B.}\ \bibnamefont {Mantz}}, \bibinfo
  {author} {\bibfnamefont {S.~M.}\ \bibnamefont {Molnar}}, \bibinfo {author}
  {\bibfnamefont {M.}~\bibnamefont {Naka}}, \bibinfo {author} {\bibfnamefont
  {G.}~\bibnamefont {Pailet}}, \bibinfo {author} {\bibfnamefont
  {E.}~\bibnamefont {Pierpaoli}}, \bibinfo {author} {\bibfnamefont {S.~R.}\
  \bibnamefont {Siegel}},\ and\ \bibinfo {author} {\bibfnamefont
  {B.}~\bibnamefont {Wolman}},\ }\bibfield  {title} {\bibinfo {title} {{A
  Comparison and Joint Analysis of Sunyaev-Zel'dovich Effect Measurements from
  Planck and Bolocam for a set of 47 Massive Galaxy Clusters}},\ }\href
  {https://doi.org/10.3847/0004-637X/832/1/26} {\bibfield  {journal} {\bibinfo
  {journal} {Astrophys. J.}\ }\textbf {\bibinfo {volume} {832}},\ \bibinfo
  {pages} {26} (\bibinfo {year} {2016})},\ \Eprint
  {https://arxiv.org/abs/1605.03541} {arXiv:1605.03541 [astro-ph.CO]}
  \BibitemShut {NoStop}%
\bibitem [{\citenamefont {Romero}\ \emph {et~al.}(2017)\citenamefont {Romero}
  \emph {et~al.}}]{Romero:2016kac}%
  \BibitemOpen
  \bibfield  {author} {\bibinfo {author} {\bibfnamefont {C.~E.}\ \bibnamefont
  {Romero}} \emph {et~al.},\ }\bibfield  {title} {\bibinfo {title} {{Galaxy
  Cluster Pressure Profiles as Determined by Sunyaev Zel\textquoteright{}dovich
  Effect Observations with MUSTANG and Bolocam. II. Joint Analysis of 14
  Clusters}},\ }\href {https://doi.org/10.3847/1538-4357/aa643f} {\bibfield
  {journal} {\bibinfo  {journal} {Astrophys. J.}\ }\textbf {\bibinfo {volume}
  {838}},\ \bibinfo {pages} {86} (\bibinfo {year} {2017})},\ \Eprint
  {https://arxiv.org/abs/1608.03980} {arXiv:1608.03980 [astro-ph.CO]}
  \BibitemShut {NoStop}%
\bibitem [{\citenamefont {Bourdin}\ \emph {et~al.}(2017)\citenamefont
  {Bourdin}, \citenamefont {Mazzotta}, \citenamefont {Kozmanyan}, \citenamefont
  {Jones},\ and\ \citenamefont {Vikhlinin}}]{Bourdin:2017ufp}%
  \BibitemOpen
  \bibfield  {author} {\bibinfo {author} {\bibfnamefont {H.}~\bibnamefont
  {Bourdin}}, \bibinfo {author} {\bibfnamefont {P.}~\bibnamefont {Mazzotta}},
  \bibinfo {author} {\bibfnamefont {A.}~\bibnamefont {Kozmanyan}}, \bibinfo
  {author} {\bibfnamefont {C.}~\bibnamefont {Jones}},\ and\ \bibinfo {author}
  {\bibfnamefont {A.}~\bibnamefont {Vikhlinin}},\ }\bibfield  {title} {\bibinfo
  {title} {{Pressure Profiles of Distant Galaxy Clusters in the Planck
  Catalogue}},\ }\href {https://doi.org/10.3847/1538-4357/aa74d0} {\bibfield
  {journal} {\bibinfo  {journal} {Astrophys. J.}\ }\textbf {\bibinfo {volume}
  {843}},\ \bibinfo {pages} {72} (\bibinfo {year} {2017})},\ \Eprint
  {https://arxiv.org/abs/1707.02248} {arXiv:1707.02248 [astro-ph.CO]}
  \BibitemShut {NoStop}%
\bibitem [{\citenamefont {Ghirardini}\ \emph {et~al.}(2017)\citenamefont
  {Ghirardini}, \citenamefont {Ettori}, \citenamefont {Amodeo}, \citenamefont
  {Capasso},\ and\ \citenamefont {Sereno}}]{Ghirardini:2017ugk}%
  \BibitemOpen
  \bibfield  {author} {\bibinfo {author} {\bibfnamefont {V.}~\bibnamefont
  {Ghirardini}}, \bibinfo {author} {\bibfnamefont {S.}~\bibnamefont {Ettori}},
  \bibinfo {author} {\bibfnamefont {S.}~\bibnamefont {Amodeo}}, \bibinfo
  {author} {\bibfnamefont {R.}~\bibnamefont {Capasso}},\ and\ \bibinfo {author}
  {\bibfnamefont {M.}~\bibnamefont {Sereno}},\ }\bibfield  {title} {\bibinfo
  {title} {{On the evolution of the entropy and pressure profiles in X-ray
  luminous galaxy clusters at z \ensuremath{>} 0.4}},\ }\href
  {https://doi.org/10.1051/0004-6361/201630209} {\bibfield  {journal} {\bibinfo
   {journal} {Astron. Astrophys.}\ }\textbf {\bibinfo {volume} {604}},\
  \bibinfo {pages} {A100} (\bibinfo {year} {2017})},\ \Eprint
  {https://arxiv.org/abs/1704.01587} {arXiv:1704.01587 [astro-ph.CO]}
  \BibitemShut {NoStop}%
\bibitem [{\citenamefont {McDonald}\ \emph {et~al.}(2017)\citenamefont
  {McDonald} \emph {et~al.}}]{SPT:2017ees}%
  \BibitemOpen
  \bibfield  {author} {\bibinfo {author} {\bibfnamefont {M.}~\bibnamefont
  {McDonald}} \emph {et~al.} (\bibinfo {collaboration} {SPT}),\ }\bibfield
  {title} {\bibinfo {title} {{The Remarkable Similarity of Massive Galaxy
  Clusters from z \ensuremath{\sim} 0 to z \ensuremath{\sim} 1.9}},\ }\href
  {https://doi.org/10.3847/1538-4357/aa7740} {\bibfield  {journal} {\bibinfo
  {journal} {Astrophys. J.}\ }\textbf {\bibinfo {volume} {843}},\ \bibinfo
  {pages} {28} (\bibinfo {year} {2017})},\ \Eprint
  {https://arxiv.org/abs/1702.05094} {arXiv:1702.05094 [astro-ph.CO]}
  \BibitemShut {NoStop}%
\bibitem [{\citenamefont {Ghirardini}\ \emph {et~al.}(2019)\citenamefont
  {Ghirardini} \emph {et~al.}}]{Ghirardini:2018byi}%
  \BibitemOpen
  \bibfield  {author} {\bibinfo {author} {\bibfnamefont {V.}~\bibnamefont
  {Ghirardini}} \emph {et~al.},\ }\bibfield  {title} {\bibinfo {title}
  {{Universal thermodynamic properties of the intracluster medium over two
  decades in radius in the X-COP sample}},\ }\href
  {https://doi.org/10.1051/0004-6361/201833325} {\bibfield  {journal} {\bibinfo
   {journal} {Astron. Astrophys.}\ }\textbf {\bibinfo {volume} {621}},\
  \bibinfo {pages} {A41} (\bibinfo {year} {2019})},\ \Eprint
  {https://arxiv.org/abs/1805.00042} {arXiv:1805.00042 [astro-ph.CO]}
  \BibitemShut {NoStop}%
\bibitem [{\citenamefont {Ghirardini}\ \emph {et~al.}(2021)\citenamefont
  {Ghirardini} \emph {et~al.}}]{Ghirardini:2020vyv}%
  \BibitemOpen
  \bibfield  {author} {\bibinfo {author} {\bibfnamefont {V.}~\bibnamefont
  {Ghirardini}} \emph {et~al.},\ }\bibfield  {title} {\bibinfo {title}
  {{Evolution of the Thermodynamic Properties of Clusters of Galaxies out to
  Redshift of 1.8}},\ }\href {https://doi.org/10.3847/1538-4357/abc68d}
  {\bibfield  {journal} {\bibinfo  {journal} {Astrophys. J.}\ }\textbf
  {\bibinfo {volume} {910}},\ \bibinfo {pages} {14} (\bibinfo {year} {2021})},\
  \Eprint {https://arxiv.org/abs/2004.04747} {arXiv:2004.04747 [astro-ph.CO]}
  \BibitemShut {NoStop}%
\bibitem [{\citenamefont {Pointecouteau}\ \emph {et~al.}(2021)\citenamefont
  {Pointecouteau} \emph {et~al.}}]{Pointecouteau:2021huw}%
  \BibitemOpen
  \bibfield  {author} {\bibinfo {author} {\bibfnamefont {E.}~\bibnamefont
  {Pointecouteau}} \emph {et~al.},\ }\bibfield  {title} {\bibinfo {title}
  {{PACT - II. Pressure profiles of galaxy clusters using Planck and ACT}},\
  }\href {https://doi.org/10.1051/0004-6361/202040213} {\bibfield  {journal}
  {\bibinfo  {journal} {Astron. Astrophys.}\ }\textbf {\bibinfo {volume}
  {651}},\ \bibinfo {pages} {A73} (\bibinfo {year} {2021})},\ \Eprint
  {https://arxiv.org/abs/2105.05607} {arXiv:2105.05607 [astro-ph.CO]}
  \BibitemShut {NoStop}%
\bibitem [{\citenamefont {Sayers}\ \emph {et~al.}(2023)\citenamefont {Sayers},
  \citenamefont {Mantz}, \citenamefont {Rasia}, \citenamefont {Allen},
  \citenamefont {Cui}, \citenamefont {Golwala}, \citenamefont {Morris},\ and\
  \citenamefont {Wan}}]{Sayers:2022hue}%
  \BibitemOpen
  \bibfield  {author} {\bibinfo {author} {\bibfnamefont {J.}~\bibnamefont
  {Sayers}}, \bibinfo {author} {\bibfnamefont {A.~B.}\ \bibnamefont {Mantz}},
  \bibinfo {author} {\bibfnamefont {E.}~\bibnamefont {Rasia}}, \bibinfo
  {author} {\bibfnamefont {S.~W.}\ \bibnamefont {Allen}}, \bibinfo {author}
  {\bibfnamefont {W.}~\bibnamefont {Cui}}, \bibinfo {author} {\bibfnamefont
  {S.~R.}\ \bibnamefont {Golwala}}, \bibinfo {author} {\bibfnamefont {R.~G.}\
  \bibnamefont {Morris}},\ and\ \bibinfo {author} {\bibfnamefont {J.~T.}\
  \bibnamefont {Wan}},\ }\bibfield  {title} {\bibinfo {title} {{The Evolution
  and Mass Dependence of Galaxy Cluster Pressure Profiles at 0.05
  \ensuremath{\leq} z \ensuremath{\leq} 0.60 and 4 \texttimes{} 10$^{14}$ M
  $_{\odot}$ \ensuremath{\leq} M $_{500}$ \ensuremath{\leq} 30 \texttimes{}
  10$^{14}$ M $_{\odot}$}},\ }\href {https://doi.org/10.3847/1538-4357/acb33d}
  {\bibfield  {journal} {\bibinfo  {journal} {Astrophys. J.}\ }\textbf
  {\bibinfo {volume} {944}},\ \bibinfo {pages} {221} (\bibinfo {year}
  {2023})},\ \Eprint {https://arxiv.org/abs/2206.00091} {arXiv:2206.00091
  [astro-ph.CO]} \BibitemShut {NoStop}%
\bibitem [{\citenamefont {Green}\ \emph {et~al.}(2020)\citenamefont {Green},
  \citenamefont {Aung}, \citenamefont {Nagai},\ and\ \citenamefont {van~den
  Bosch}}]{Green:2020vak}%
  \BibitemOpen
  \bibfield  {author} {\bibinfo {author} {\bibfnamefont {S.~B.}\ \bibnamefont
  {Green}}, \bibinfo {author} {\bibfnamefont {H.}~\bibnamefont {Aung}},
  \bibinfo {author} {\bibfnamefont {D.}~\bibnamefont {Nagai}},\ and\ \bibinfo
  {author} {\bibfnamefont {F.~C.}\ \bibnamefont {van~den Bosch}},\ }\bibfield
  {title} {\bibinfo {title} {{Scatter in
  Sunyaev\textendash{}Zel\textquoteright{}dovich effect scaling relations
  explained by inter-cluster variance in mass accretion histories}},\ }\href
  {https://doi.org/10.1093/mnras/staa1712} {\bibfield  {journal} {\bibinfo
  {journal} {Mon. Not. Roy. Astron. Soc.}\ }\textbf {\bibinfo {volume} {496}},\
  \bibinfo {pages} {2743} (\bibinfo {year} {2020})},\ \Eprint
  {https://arxiv.org/abs/2002.01934} {arXiv:2002.01934 [astro-ph.CO]}
  \BibitemShut {NoStop}%
\bibitem [{\citenamefont {Nelson}\ \emph {et~al.}(2014)\citenamefont {Nelson},
  \citenamefont {Lau},\ and\ \citenamefont {Nagai}}]{Nelson:2014vda}%
  \BibitemOpen
  \bibfield  {author} {\bibinfo {author} {\bibfnamefont {K.}~\bibnamefont
  {Nelson}}, \bibinfo {author} {\bibfnamefont {E.~T.}\ \bibnamefont {Lau}},\
  and\ \bibinfo {author} {\bibfnamefont {D.}~\bibnamefont {Nagai}},\ }\bibfield
   {title} {\bibinfo {title} {{Hydrodynamic Simulation of Non-thermal Pressure
  Profiles of Galaxy Clusters}},\ }\href
  {https://doi.org/10.1088/0004-637X/792/1/25} {\bibfield  {journal} {\bibinfo
  {journal} {Astrophys. J.}\ }\textbf {\bibinfo {volume} {792}},\ \bibinfo
  {pages} {25} (\bibinfo {year} {2014})},\ \Eprint
  {https://arxiv.org/abs/1404.4636} {arXiv:1404.4636 [astro-ph.CO]}
  \BibitemShut {NoStop}%
\bibitem [{\citenamefont {{Shirasaki}}(2019)}]{2019ApJ...883...36S}%
  \BibitemOpen
  \bibfield  {author} {\bibinfo {author} {\bibfnamefont {M.}~\bibnamefont
  {{Shirasaki}}},\ }\bibfield  {title} {\bibinfo {title} {{The Pseudo-evolution
  of Galaxy-cluster Masses and Its Connection to Mass Density Profile}},\
  }\href {https://doi.org/10.3847/1538-4357/ab3855} {\bibfield  {journal}
  {\bibinfo  {journal} {\apj}\ }\textbf {\bibinfo {volume} {883}},\ \bibinfo
  {eid} {36} (\bibinfo {year} {2019})},\ \Eprint
  {https://arxiv.org/abs/1905.07895} {arXiv:1905.07895 [astro-ph.CO]}
  \BibitemShut {NoStop}%
\bibitem [{\citenamefont {Coble}\ \emph {et~al.}(2007)\citenamefont {Coble},
  \citenamefont {Carlstrom}, \citenamefont {Bonamente}, \citenamefont {Dawson},
  \citenamefont {Holzapfel}, \citenamefont {Joy}, \citenamefont {LaRoque},\
  and\ \citenamefont {Reese}}]{Coble:2006ff}%
  \BibitemOpen
  \bibfield  {author} {\bibinfo {author} {\bibfnamefont {K.}~\bibnamefont
  {Coble}}, \bibinfo {author} {\bibfnamefont {J.~E.}\ \bibnamefont
  {Carlstrom}}, \bibinfo {author} {\bibfnamefont {M.}~\bibnamefont
  {Bonamente}}, \bibinfo {author} {\bibfnamefont {K.}~\bibnamefont {Dawson}},
  \bibinfo {author} {\bibfnamefont {W.}~\bibnamefont {Holzapfel}}, \bibinfo
  {author} {\bibfnamefont {M.}~\bibnamefont {Joy}}, \bibinfo {author}
  {\bibfnamefont {S.}~\bibnamefont {LaRoque}},\ and\ \bibinfo {author}
  {\bibfnamefont {E.~D.}\ \bibnamefont {Reese}},\ }\bibfield  {title} {\bibinfo
  {title} {{Radio Point Sources Toward Galaxy Clusters at 30-GHz}},\ }\href
  {https://doi.org/10.1086/519973} {\bibfield  {journal} {\bibinfo  {journal}
  {Astron. J.}\ }\textbf {\bibinfo {volume} {134}},\ \bibinfo {pages} {897}
  (\bibinfo {year} {2007})},\ \Eprint {https://arxiv.org/abs/astro-ph/0608274}
  {arXiv:astro-ph/0608274} \BibitemShut {NoStop}%
\bibitem [{\citenamefont {Dicker}\ \emph {et~al.}(2021)\citenamefont {Dicker}
  \emph {et~al.}}]{Dicker:2021ijw}%
  \BibitemOpen
  \bibfield  {author} {\bibinfo {author} {\bibfnamefont {S.~R.}\ \bibnamefont
  {Dicker}} \emph {et~al.},\ }\bibfield  {title} {\bibinfo {title}
  {{Observations of compact sources in galaxy clusters using MUSTANG2}},\
  }\href {https://doi.org/10.1093/mnras/stab2679} {\bibfield  {journal}
  {\bibinfo  {journal} {Mon. Not. Roy. Astron. Soc.}\ }\textbf {\bibinfo
  {volume} {508}},\ \bibinfo {pages} {2600} (\bibinfo {year} {2021})},\ \Eprint
  {https://arxiv.org/abs/2107.06725} {arXiv:2107.06725 [astro-ph.CO]}
  \BibitemShut {NoStop}%
\bibitem [{\citenamefont {Hamana}\ \emph {et~al.}(2004)\citenamefont {Hamana},
  \citenamefont {Takada},\ and\ \citenamefont {Yoshida}}]{Hamana:2003ts}%
  \BibitemOpen
  \bibfield  {author} {\bibinfo {author} {\bibfnamefont {T.}~\bibnamefont
  {Hamana}}, \bibinfo {author} {\bibfnamefont {M.}~\bibnamefont {Takada}},\
  and\ \bibinfo {author} {\bibfnamefont {N.}~\bibnamefont {Yoshida}},\
  }\bibfield  {title} {\bibinfo {title} {{Searching for massive clusters in
  weak lensing surveys}},\ }\href
  {https://doi.org/10.1111/j.1365-2966.2004.07691.x} {\bibfield  {journal}
  {\bibinfo  {journal} {Mon. Not. Roy. Astron. Soc.}\ }\textbf {\bibinfo
  {volume} {350}},\ \bibinfo {pages} {893} (\bibinfo {year} {2004})},\ \Eprint
  {https://arxiv.org/abs/astro-ph/0310607} {arXiv:astro-ph/0310607}
  \BibitemShut {NoStop}%
\bibitem [{\citenamefont {Hennawi}\ and\ \citenamefont
  {Spergel}(2005)}]{Hennawi:2004ai}%
  \BibitemOpen
  \bibfield  {author} {\bibinfo {author} {\bibfnamefont {J.~F.}\ \bibnamefont
  {Hennawi}}\ and\ \bibinfo {author} {\bibfnamefont {D.~N.}\ \bibnamefont
  {Spergel}},\ }\bibfield  {title} {\bibinfo {title} {{Mass selected cluster
  cosmology. 1: Tomography and optimal filtering}},\ }\href
  {https://doi.org/10.1086/428749} {\bibfield  {journal} {\bibinfo  {journal}
  {Astrophys. J.}\ }\textbf {\bibinfo {volume} {624}},\ \bibinfo {pages} {59}
  (\bibinfo {year} {2005})},\ \Eprint {https://arxiv.org/abs/astro-ph/0404349}
  {arXiv:astro-ph/0404349} \BibitemShut {NoStop}%
\bibitem [{\citenamefont {Maturi}\ \emph {et~al.}(2005)\citenamefont {Maturi},
  \citenamefont {Meneghetti}, \citenamefont {Bartelmann}, \citenamefont
  {Dolag},\ and\ \citenamefont {Moscardini}}]{Maturi:2004rn}%
  \BibitemOpen
  \bibfield  {author} {\bibinfo {author} {\bibfnamefont {M.}~\bibnamefont
  {Maturi}}, \bibinfo {author} {\bibfnamefont {M.}~\bibnamefont {Meneghetti}},
  \bibinfo {author} {\bibfnamefont {M.}~\bibnamefont {Bartelmann}}, \bibinfo
  {author} {\bibfnamefont {K.}~\bibnamefont {Dolag}},\ and\ \bibinfo {author}
  {\bibfnamefont {L.}~\bibnamefont {Moscardini}},\ }\bibfield  {title}
  {\bibinfo {title} {{An Optimal filter for the detection of galaxy clusters
  through weak lensing}},\ }\href {https://doi.org/10.1051/0004-6361:20042600}
  {\bibfield  {journal} {\bibinfo  {journal} {Astron. Astrophys.}\ }\textbf
  {\bibinfo {volume} {442}},\ \bibinfo {pages} {851} (\bibinfo {year}
  {2005})},\ \Eprint {https://arxiv.org/abs/astro-ph/0412604}
  {arXiv:astro-ph/0412604} \BibitemShut {NoStop}%
\bibitem [{\citenamefont {{Hamana}}\ \emph {et~al.}(2012)\citenamefont
  {{Hamana}}, \citenamefont {{Oguri}}, \citenamefont {{Shirasaki}},\ and\
  \citenamefont {{Sato}}}]{2012MNRAS.425.2287H}%
  \BibitemOpen
  \bibfield  {author} {\bibinfo {author} {\bibfnamefont {T.}~\bibnamefont
  {{Hamana}}}, \bibinfo {author} {\bibfnamefont {M.}~\bibnamefont {{Oguri}}},
  \bibinfo {author} {\bibfnamefont {M.}~\bibnamefont {{Shirasaki}}},\ and\
  \bibinfo {author} {\bibfnamefont {M.}~\bibnamefont {{Sato}}},\ }\bibfield
  {title} {\bibinfo {title} {{Scatter and bias in weak lensing selected
  clusters}},\ }\href {https://doi.org/10.1111/j.1365-2966.2012.21582.x}
  {\bibfield  {journal} {\bibinfo  {journal} {\mnras}\ }\textbf {\bibinfo
  {volume} {425}},\ \bibinfo {pages} {2287} (\bibinfo {year} {2012})},\ \Eprint
  {https://arxiv.org/abs/1204.6117} {arXiv:1204.6117 [astro-ph.CO]}
  \BibitemShut {NoStop}%
\bibitem [{\citenamefont {{Chen}}\ \emph {et~al.}(2020)\citenamefont {{Chen}},
  \citenamefont {{Oguri}}, \citenamefont {{Lin}},\ and\ \citenamefont
  {{Miyazaki}}}]{2020ApJ...891..139C}%
  \BibitemOpen
  \bibfield  {author} {\bibinfo {author} {\bibfnamefont {K.-F.}\ \bibnamefont
  {{Chen}}}, \bibinfo {author} {\bibfnamefont {M.}~\bibnamefont {{Oguri}}},
  \bibinfo {author} {\bibfnamefont {Y.-T.}\ \bibnamefont {{Lin}}},\ and\
  \bibinfo {author} {\bibfnamefont {S.}~\bibnamefont {{Miyazaki}}},\ }\bibfield
   {title} {\bibinfo {title} {{Mass Bias of Weak-lensing Shear-selected Galaxy
  Cluster Samples}},\ }\href {https://doi.org/10.3847/1538-4357/ab74d3}
  {\bibfield  {journal} {\bibinfo  {journal} {\apj}\ }\textbf {\bibinfo
  {volume} {891}},\ \bibinfo {eid} {139} (\bibinfo {year} {2020})},\ \Eprint
  {https://arxiv.org/abs/1911.11480} {arXiv:1911.11480 [astro-ph.GA]}
  \BibitemShut {NoStop}%
\bibitem [{\citenamefont {Miyazaki}\ \emph {et~al.}(2018)\citenamefont
  {Miyazaki} \emph {et~al.}}]{Miyazaki:2018dpu}%
  \BibitemOpen
  \bibfield  {author} {\bibinfo {author} {\bibfnamefont {S.}~\bibnamefont
  {Miyazaki}} \emph {et~al.},\ }\bibfield  {title} {\bibinfo {title} {{A large
  sample of shear selected clusters from the Hyper Suprime-Cam Subaru Strategic
  Program S16A wide field mass maps}},\ }\href
  {https://doi.org/10.1093/pasj/psx120} {\bibfield  {journal} {\bibinfo
  {journal} {Publ. Astron. Soc. Jap.}\ }\textbf {\bibinfo {volume} {70}},\
  \bibinfo {pages} {S27} (\bibinfo {year} {2018})},\ \Eprint
  {https://arxiv.org/abs/1802.10290} {arXiv:1802.10290 [astro-ph.CO]}
  \BibitemShut {NoStop}%
\bibitem [{\citenamefont {Oguri}\ \emph {et~al.}(2021)\citenamefont {Oguri}
  \emph {et~al.}}]{Oguri:2021vro}%
  \BibitemOpen
  \bibfield  {author} {\bibinfo {author} {\bibfnamefont {M.}~\bibnamefont
  {Oguri}} \emph {et~al.},\ }\bibfield  {title} {\bibinfo {title} {{Hundreds of
  weak lensing shear-selected clusters from the Hyper Suprime-Cam Subaru
  Strategic Program S19A data}},\ }\href {https://doi.org/10.1093/pasj/psab047}
  {\bibfield  {journal} {\bibinfo  {journal} {Publ. Astron. Soc. Jap.}\
  }\textbf {\bibinfo {volume} {73}},\ \bibinfo {pages} {817} (\bibinfo {year}
  {2021})},\ \Eprint {https://arxiv.org/abs/2103.15016} {arXiv:2103.15016
  [astro-ph.CO]} \BibitemShut {NoStop}%
\bibitem [{\citenamefont {Giles}\ \emph {et~al.}(2015)\citenamefont {Giles},
  \citenamefont {Maughan}, \citenamefont {Hamana}, \citenamefont {Miyazaki},
  \citenamefont {Birkinshaw}, \citenamefont {Ellis},\ and\ \citenamefont
  {Massey}}]{Giles:2014bea}%
  \BibitemOpen
  \bibfield  {author} {\bibinfo {author} {\bibfnamefont {P.~A.}\ \bibnamefont
  {Giles}}, \bibinfo {author} {\bibfnamefont {B.~J.}\ \bibnamefont {Maughan}},
  \bibinfo {author} {\bibfnamefont {T.}~\bibnamefont {Hamana}}, \bibinfo
  {author} {\bibfnamefont {S.}~\bibnamefont {Miyazaki}}, \bibinfo {author}
  {\bibfnamefont {M.}~\bibnamefont {Birkinshaw}}, \bibinfo {author}
  {\bibfnamefont {R.~S.}\ \bibnamefont {Ellis}},\ and\ \bibinfo {author}
  {\bibfnamefont {R.}~\bibnamefont {Massey}},\ }\bibfield  {title} {\bibinfo
  {title} {{The X-ray Properties of Weak Lensing Selected Galaxy Clusters}},\
  }\href {https://doi.org/10.1093/mnras/stu2679} {\bibfield  {journal}
  {\bibinfo  {journal} {Mon. Not. Roy. Astron. Soc.}\ }\textbf {\bibinfo
  {volume} {447}},\ \bibinfo {pages} {3044} (\bibinfo {year} {2015})},\ \Eprint
  {https://arxiv.org/abs/1402.4484} {arXiv:1402.4484 [astro-ph.CO]}
  \BibitemShut {NoStop}%
\bibitem [{\citenamefont {Cooray}\ and\ \citenamefont
  {Sheth}(2002)}]{Cooray:2002dia}%
  \BibitemOpen
  \bibfield  {author} {\bibinfo {author} {\bibfnamefont {A.}~\bibnamefont
  {Cooray}}\ and\ \bibinfo {author} {\bibfnamefont {R.~K.}\ \bibnamefont
  {Sheth}},\ }\bibfield  {title} {\bibinfo {title} {{Halo Models of Large Scale
  Structure}},\ }\href {https://doi.org/10.1016/S0370-1573(02)00276-4}
  {\bibfield  {journal} {\bibinfo  {journal} {Phys. Rept.}\ }\textbf {\bibinfo
  {volume} {372}},\ \bibinfo {pages} {1} (\bibinfo {year} {2002})},\ \Eprint
  {https://arxiv.org/abs/astro-ph/0206508} {arXiv:astro-ph/0206508}
  \BibitemShut {NoStop}%
\bibitem [{\citenamefont {Hayashi}\ and\ \citenamefont
  {White}(2008)}]{Hayashi:2007uk}%
  \BibitemOpen
  \bibfield  {author} {\bibinfo {author} {\bibfnamefont {E.}~\bibnamefont
  {Hayashi}}\ and\ \bibinfo {author} {\bibfnamefont {S.~D.~M.}\ \bibnamefont
  {White}},\ }\bibfield  {title} {\bibinfo {title} {{Understanding the shape of
  the halo-mass and galaxy-mass cross-correlation functions}},\ }\href
  {https://doi.org/10.1111/j.1365-2966.2008.13371.x} {\bibfield  {journal}
  {\bibinfo  {journal} {Mon. Not. Roy. Astron. Soc.}\ }\textbf {\bibinfo
  {volume} {388}},\ \bibinfo {pages} {2} (\bibinfo {year} {2008})},\ \Eprint
  {https://arxiv.org/abs/0709.3933} {arXiv:0709.3933 [astro-ph]} \BibitemShut
  {NoStop}%
\bibitem [{\citenamefont {Gao}\ and\ \citenamefont {White}(2007)}]{Gao:2006qz}%
  \BibitemOpen
  \bibfield  {author} {\bibinfo {author} {\bibfnamefont {L.}~\bibnamefont
  {Gao}}\ and\ \bibinfo {author} {\bibfnamefont {S.~D.~M.}\ \bibnamefont
  {White}},\ }\bibfield  {title} {\bibinfo {title} {{Assembly bias in the
  clustering of dark matter haloes}},\ }\href
  {https://doi.org/10.1111/j.1745-3933.2007.00292.x} {\bibfield  {journal}
  {\bibinfo  {journal} {Mon. Not. Roy. Astron. Soc.}\ }\textbf {\bibinfo
  {volume} {377}},\ \bibinfo {pages} {L5} (\bibinfo {year} {2007})},\ \Eprint
  {https://arxiv.org/abs/astro-ph/0611921} {arXiv:astro-ph/0611921}
  \BibitemShut {NoStop}%
\bibitem [{\citenamefont {Faltenbacher}\ and\ \citenamefont
  {White}(2010)}]{Faltenbacher:2009fj}%
  \BibitemOpen
  \bibfield  {author} {\bibinfo {author} {\bibfnamefont {A.}~\bibnamefont
  {Faltenbacher}}\ and\ \bibinfo {author} {\bibfnamefont {S.~D.~M.}\
  \bibnamefont {White}},\ }\bibfield  {title} {\bibinfo {title} {{Assembly bias
  and the dynamical structure of dark matter halos}},\ }\href
  {https://doi.org/10.1088/0004-637X/708/1/469} {\bibfield  {journal} {\bibinfo
   {journal} {Astrophys. J.}\ }\textbf {\bibinfo {volume} {708}},\ \bibinfo
  {pages} {469} (\bibinfo {year} {2010})},\ \Eprint
  {https://arxiv.org/abs/0909.4302} {arXiv:0909.4302 [astro-ph.CO]}
  \BibitemShut {NoStop}%
\bibitem [{\citenamefont {Mao}\ \emph {et~al.}(2018)\citenamefont {Mao},
  \citenamefont {Zentner},\ and\ \citenamefont {Wechsler}}]{Mao:2017aym}%
  \BibitemOpen
  \bibfield  {author} {\bibinfo {author} {\bibfnamefont {Y.-Y.}\ \bibnamefont
  {Mao}}, \bibinfo {author} {\bibfnamefont {A.~R.}\ \bibnamefont {Zentner}},\
  and\ \bibinfo {author} {\bibfnamefont {R.~H.}\ \bibnamefont {Wechsler}},\
  }\bibfield  {title} {\bibinfo {title} {{Beyond Assembly Bias: Exploring
  Secondary Halo Biases for Cluster-size Haloes}},\ }\href
  {https://doi.org/10.1093/mnras/stx3111} {\bibfield  {journal} {\bibinfo
  {journal} {Mon. Not. Roy. Astron. Soc.}\ }\textbf {\bibinfo {volume} {474}},\
  \bibinfo {pages} {5143} (\bibinfo {year} {2018})},\ \bibinfo {note}
  {[Erratum: Mon.Not.Roy.Astron.Soc. 481, 3167 (2018)]},\ \Eprint
  {https://arxiv.org/abs/1705.03888} {arXiv:1705.03888 [astro-ph.CO]}
  \BibitemShut {NoStop}%
\bibitem [{\citenamefont {Chue}\ \emph {et~al.}(2018)\citenamefont {Chue},
  \citenamefont {Dalal},\ and\ \citenamefont {White}}]{Chue:2018hxk}%
  \BibitemOpen
  \bibfield  {author} {\bibinfo {author} {\bibfnamefont {C.~Y.~R.}\
  \bibnamefont {Chue}}, \bibinfo {author} {\bibfnamefont {N.}~\bibnamefont
  {Dalal}},\ and\ \bibinfo {author} {\bibfnamefont {M.}~\bibnamefont {White}},\
  }\bibfield  {title} {\bibinfo {title} {{Some assembly required: assembly bias
  in massive dark matter halos}},\ }\href
  {https://doi.org/10.1088/1475-7516/2018/10/012} {\bibfield  {journal}
  {\bibinfo  {journal} {JCAP}\ }\textbf {\bibinfo {volume} {10}},\ \bibinfo
  {pages} {012}},\ \Eprint {https://arxiv.org/abs/1804.04055} {arXiv:1804.04055
  [astro-ph.CO]} \BibitemShut {NoStop}%
\bibitem [{\citenamefont {Shirasaki}\ \emph {et~al.}(2016)\citenamefont
  {Shirasaki}, \citenamefont {Nagai},\ and\ \citenamefont
  {Lau}}]{Shirasaki:2016buk}%
  \BibitemOpen
  \bibfield  {author} {\bibinfo {author} {\bibfnamefont {M.}~\bibnamefont
  {Shirasaki}}, \bibinfo {author} {\bibfnamefont {D.}~\bibnamefont {Nagai}},\
  and\ \bibinfo {author} {\bibfnamefont {E.~T.}\ \bibnamefont {Lau}},\
  }\bibfield  {title} {\bibinfo {title} {{Covariance in the thermal
  SZ\textendash{}weak lensing mass scaling relation of galaxy clusters}},\
  }\href {https://doi.org/10.1093/mnras/stw1263} {\bibfield  {journal}
  {\bibinfo  {journal} {Mon. Not. Roy. Astron. Soc.}\ }\textbf {\bibinfo
  {volume} {460}},\ \bibinfo {pages} {3913} (\bibinfo {year} {2016})},\ \Eprint
  {https://arxiv.org/abs/1603.08609} {arXiv:1603.08609 [astro-ph.CO]}
  \BibitemShut {NoStop}%
\bibitem [{\citenamefont {Sunayama}\ \emph {et~al.}(2020)\citenamefont
  {Sunayama}, \citenamefont {Park}, \citenamefont {Takada}, \citenamefont
  {Kobayashi}, \citenamefont {Nishimichi}, \citenamefont {Kurita},
  \citenamefont {More}, \citenamefont {Oguri},\ and\ \citenamefont
  {Osato}}]{Sunayama:2020wvq}%
  \BibitemOpen
  \bibfield  {author} {\bibinfo {author} {\bibfnamefont {T.}~\bibnamefont
  {Sunayama}}, \bibinfo {author} {\bibfnamefont {Y.}~\bibnamefont {Park}},
  \bibinfo {author} {\bibfnamefont {M.}~\bibnamefont {Takada}}, \bibinfo
  {author} {\bibfnamefont {Y.}~\bibnamefont {Kobayashi}}, \bibinfo {author}
  {\bibfnamefont {T.}~\bibnamefont {Nishimichi}}, \bibinfo {author}
  {\bibfnamefont {T.}~\bibnamefont {Kurita}}, \bibinfo {author} {\bibfnamefont
  {S.}~\bibnamefont {More}}, \bibinfo {author} {\bibfnamefont {M.}~\bibnamefont
  {Oguri}},\ and\ \bibinfo {author} {\bibfnamefont {K.}~\bibnamefont {Osato}},\
  }\bibfield  {title} {\bibinfo {title} {{The impact of projection effects on
  cluster observables: stacked lensing and projected clustering}},\ }\href
  {https://doi.org/10.1093/mnras/staa1646} {\bibfield  {journal} {\bibinfo
  {journal} {Mon. Not. Roy. Astron. Soc.}\ }\textbf {\bibinfo {volume} {496}},\
  \bibinfo {pages} {4468} (\bibinfo {year} {2020})},\ \Eprint
  {https://arxiv.org/abs/2002.03867} {arXiv:2002.03867 [astro-ph.CO]}
  \BibitemShut {NoStop}%
\bibitem [{\citenamefont {Wu}\ \emph {et~al.}(2022)\citenamefont {Wu} \emph
  {et~al.}}]{DES:2022qkn}%
  \BibitemOpen
  \bibfield  {author} {\bibinfo {author} {\bibfnamefont {H.-Y.}\ \bibnamefont
  {Wu}} \emph {et~al.} (\bibinfo {collaboration} {DES}),\ }\bibfield  {title}
  {\bibinfo {title} {{Optical selection bias and projection effects in stacked
  galaxy cluster weak lensing}},\ }\href
  {https://doi.org/10.1093/mnras/stac2048} {\bibfield  {journal} {\bibinfo
  {journal} {Mon. Not. Roy. Astron. Soc.}\ }\textbf {\bibinfo {volume} {515}},\
  \bibinfo {pages} {4471} (\bibinfo {year} {2022})},\ \Eprint
  {https://arxiv.org/abs/2203.05416} {arXiv:2203.05416 [astro-ph.CO]}
  \BibitemShut {NoStop}%
\bibitem [{\citenamefont {Bohringer}\ \emph {et~al.}(2007)\citenamefont
  {Bohringer} \emph {et~al.}}]{Bohringer:2007vs}%
  \BibitemOpen
  \bibfield  {author} {\bibinfo {author} {\bibfnamefont {H.}~\bibnamefont
  {Bohringer}} \emph {et~al.},\ }\bibfield  {title} {\bibinfo {title} {{The
  Representative XMM-Newton Cluster Structure Survey (REXCESS) of an X-ray
  Luminosity Selected Galaxy Cluster Sample}},\ }\href
  {https://doi.org/10.1051/0004-6361:20066740} {\bibfield  {journal} {\bibinfo
  {journal} {Astron. Astrophys.}\ }\textbf {\bibinfo {volume} {469}},\ \bibinfo
  {pages} {363} (\bibinfo {year} {2007})},\ \Eprint
  {https://arxiv.org/abs/astro-ph/0703553} {arXiv:astro-ph/0703553}
  \BibitemShut {NoStop}%
\bibitem [{\citenamefont {Nagai}\ \emph {et~al.}(2007)\citenamefont {Nagai},
  \citenamefont {Kravtsov},\ and\ \citenamefont {Vikhlinin}}]{Nagai:2007mt}%
  \BibitemOpen
  \bibfield  {author} {\bibinfo {author} {\bibfnamefont {D.}~\bibnamefont
  {Nagai}}, \bibinfo {author} {\bibfnamefont {A.~V.}\ \bibnamefont
  {Kravtsov}},\ and\ \bibinfo {author} {\bibfnamefont {A.}~\bibnamefont
  {Vikhlinin}},\ }\bibfield  {title} {\bibinfo {title} {{Effects of Galaxy
  Formation on Thermodynamics of the Intracluster Medium}},\ }\href
  {https://doi.org/10.1086/521328} {\bibfield  {journal} {\bibinfo  {journal}
  {Astrophys. J.}\ }\textbf {\bibinfo {volume} {668}},\ \bibinfo {pages} {1}
  (\bibinfo {year} {2007})},\ \Eprint {https://arxiv.org/abs/astro-ph/0703661}
  {arXiv:astro-ph/0703661} \BibitemShut {NoStop}%
\bibitem [{\citenamefont {Miyatake}\ \emph {et~al.}(2015)\citenamefont
  {Miyatake}, \citenamefont {More}, \citenamefont {Mandelbaum}, \citenamefont
  {Takada}, \citenamefont {Spergel}, \citenamefont {Kneib}, \citenamefont
  {Schneider}, \citenamefont {Brinkmann},\ and\ \citenamefont
  {Brownstein}}]{Miyatake:2013bha}%
  \BibitemOpen
  \bibfield  {author} {\bibinfo {author} {\bibfnamefont {H.}~\bibnamefont
  {Miyatake}}, \bibinfo {author} {\bibfnamefont {S.}~\bibnamefont {More}},
  \bibinfo {author} {\bibfnamefont {R.}~\bibnamefont {Mandelbaum}}, \bibinfo
  {author} {\bibfnamefont {M.}~\bibnamefont {Takada}}, \bibinfo {author}
  {\bibfnamefont {D.~N.}\ \bibnamefont {Spergel}}, \bibinfo {author}
  {\bibfnamefont {J.-P.}\ \bibnamefont {Kneib}}, \bibinfo {author}
  {\bibfnamefont {D.~P.}\ \bibnamefont {Schneider}}, \bibinfo {author}
  {\bibfnamefont {J.}~\bibnamefont {Brinkmann}},\ and\ \bibinfo {author}
  {\bibfnamefont {J.~R.}\ \bibnamefont {Brownstein}},\ }\bibfield  {title}
  {\bibinfo {title} {{The Weak Lensing Signal and the Clustering of BOSS
  Galaxies I: Measurements}},\ }\href
  {https://doi.org/10.1088/0004-637X/806/1/1} {\bibfield  {journal} {\bibinfo
  {journal} {Astrophys. J.}\ }\textbf {\bibinfo {volume} {806}},\ \bibinfo
  {pages} {1} (\bibinfo {year} {2015})},\ \Eprint
  {https://arxiv.org/abs/1311.1480} {arXiv:1311.1480 [astro-ph.CO]}
  \BibitemShut {NoStop}%
\bibitem [{\citenamefont {Hamana}\ and\ \citenamefont
  {Mellier}(2001)}]{Hamana:2001vz}%
  \BibitemOpen
  \bibfield  {author} {\bibinfo {author} {\bibfnamefont {T.}~\bibnamefont
  {Hamana}}\ and\ \bibinfo {author} {\bibfnamefont {Y.}~\bibnamefont
  {Mellier}},\ }\bibfield  {title} {\bibinfo {title} {{Numerical study of
  statistical properties of the lensing excursion angles}},\ }\href
  {https://doi.org/10.1046/j.1365-8711.2001.04685.x} {\bibfield  {journal}
  {\bibinfo  {journal} {Mon. Not. Roy. Astron. Soc.}\ }\textbf {\bibinfo
  {volume} {327}},\ \bibinfo {pages} {169} (\bibinfo {year} {2001})},\ \Eprint
  {https://arxiv.org/abs/astro-ph/0101333} {arXiv:astro-ph/0101333}
  \BibitemShut {NoStop}%
\bibitem [{\citenamefont {{Becker}}(2013)}]{2013MNRAS.435..115B}%
  \BibitemOpen
  \bibfield  {author} {\bibinfo {author} {\bibfnamefont {M.~R.}\ \bibnamefont
  {{Becker}}},\ }\bibfield  {title} {\bibinfo {title} {{CALCLENS: weak lensing
  simulations for large-area sky surveys and second-order effects in cosmic
  shear power spectra}},\ }\href {https://doi.org/10.1093/mnras/stt1352}
  {\bibfield  {journal} {\bibinfo  {journal} {\mnras}\ }\textbf {\bibinfo
  {volume} {435}},\ \bibinfo {pages} {115} (\bibinfo {year}
  {2013})}\BibitemShut {NoStop}%
\bibitem [{\citenamefont {Shirasaki}\ \emph {et~al.}(2015)\citenamefont
  {Shirasaki}, \citenamefont {Hamana},\ and\ \citenamefont
  {Yoshida}}]{Shirasaki:2015dga}%
  \BibitemOpen
  \bibfield  {author} {\bibinfo {author} {\bibfnamefont {M.}~\bibnamefont
  {Shirasaki}}, \bibinfo {author} {\bibfnamefont {T.}~\bibnamefont {Hamana}},\
  and\ \bibinfo {author} {\bibfnamefont {N.}~\bibnamefont {Yoshida}},\
  }\bibfield  {title} {\bibinfo {title} {{Probing cosmology with weak lensing
  selected clusters \textendash{} I. Halo approach and all-sky simulations}},\
  }\href {https://doi.org/10.1093/mnras/stv1854} {\bibfield  {journal}
  {\bibinfo  {journal} {Mon. Not. Roy. Astron. Soc.}\ }\textbf {\bibinfo
  {volume} {453}},\ \bibinfo {pages} {3043} (\bibinfo {year} {2015})},\ \Eprint
  {https://arxiv.org/abs/1504.05672} {arXiv:1504.05672 [astro-ph.CO]}
  \BibitemShut {NoStop}%
\bibitem [{\citenamefont {Behroozi}\ \emph {et~al.}(2013)\citenamefont
  {Behroozi}, \citenamefont {Wechsler},\ and\ \citenamefont
  {Wu}}]{Behroozi:2011ju}%
  \BibitemOpen
  \bibfield  {author} {\bibinfo {author} {\bibfnamefont {P.~S.}\ \bibnamefont
  {Behroozi}}, \bibinfo {author} {\bibfnamefont {R.~H.}\ \bibnamefont
  {Wechsler}},\ and\ \bibinfo {author} {\bibfnamefont {H.-Y.}\ \bibnamefont
  {Wu}},\ }\bibfield  {title} {\bibinfo {title} {{The Rockstar Phase-Space
  Temporal Halo Finder and the Velocity Offsets of Cluster Cores}},\ }\href
  {https://doi.org/10.1088/0004-637X/762/2/109} {\bibfield  {journal} {\bibinfo
   {journal} {Astrophys. J.}\ }\textbf {\bibinfo {volume} {762}},\ \bibinfo
  {pages} {109} (\bibinfo {year} {2013})},\ \Eprint
  {https://arxiv.org/abs/1110.4372} {arXiv:1110.4372 [astro-ph.CO]}
  \BibitemShut {NoStop}%
\bibitem [{\citenamefont {Shirasaki}\ \emph {et~al.}(2019)\citenamefont
  {Shirasaki}, \citenamefont {Hamana}, \citenamefont {Takada}, \citenamefont
  {Takahashi},\ and\ \citenamefont {Miyatake}}]{Shirasaki:2019gya}%
  \BibitemOpen
  \bibfield  {author} {\bibinfo {author} {\bibfnamefont {M.}~\bibnamefont
  {Shirasaki}}, \bibinfo {author} {\bibfnamefont {T.}~\bibnamefont {Hamana}},
  \bibinfo {author} {\bibfnamefont {M.}~\bibnamefont {Takada}}, \bibinfo
  {author} {\bibfnamefont {R.}~\bibnamefont {Takahashi}},\ and\ \bibinfo
  {author} {\bibfnamefont {H.}~\bibnamefont {Miyatake}},\ }\bibfield  {title}
  {\bibinfo {title} {{Mock galaxy shape catalogues in the Subaru Hyper
  Suprime-Cam Survey}},\ }\href {https://doi.org/10.1093/mnras/stz791}
  {\bibfield  {journal} {\bibinfo  {journal} {Mon. Not. Roy. Astron. Soc.}\
  }\textbf {\bibinfo {volume} {486}},\ \bibinfo {pages} {52} (\bibinfo {year}
  {2019})},\ \Eprint {https://arxiv.org/abs/1901.09488} {arXiv:1901.09488
  [astro-ph.CO]} \BibitemShut {NoStop}%
\bibitem [{\citenamefont {{Comparat}}\ \emph {et~al.}(2020)\citenamefont
  {{Comparat}}, \citenamefont {{Eckert}}, \citenamefont {{Finoguenov}},
  \citenamefont {{Schmidt}}, \citenamefont {{Sanders}}, \citenamefont
  {{Nagai}}, \citenamefont {{Lau}}, \citenamefont {{Kaefer}}, \citenamefont
  {{Pacaud}}, \citenamefont {{Clerc}}, \citenamefont {{Reiprich}},
  \citenamefont {{Bulbul}}, \citenamefont {{Chitham}}, \citenamefont
  {{Chiang}}, \citenamefont {{Ghirardini}}, \citenamefont {{Gonzalez-Perez}},
  \citenamefont {{Gozaliasl}}, \citenamefont {{Fitzpatrick}}, \citenamefont
  {{Klypin}}, \citenamefont {{Merloni}}, \citenamefont {{Nandra}},
  \citenamefont {{Liu}}, \citenamefont {{Prada}}, \citenamefont {{Ramos-Ceja}},
  \citenamefont {{Salvato}}, \citenamefont {{Seppi}}, \citenamefont
  {{Tempel}},\ and\ \citenamefont {{Yepes}}}]{comparat_etal20}%
  \BibitemOpen
  \bibfield  {author} {\bibinfo {author} {\bibfnamefont {J.}~\bibnamefont
  {{Comparat}}}, \bibinfo {author} {\bibfnamefont {D.}~\bibnamefont
  {{Eckert}}}, \bibinfo {author} {\bibfnamefont {A.}~\bibnamefont
  {{Finoguenov}}}, \bibinfo {author} {\bibfnamefont {R.}~\bibnamefont
  {{Schmidt}}}, \bibinfo {author} {\bibfnamefont {J.~S.}\ \bibnamefont
  {{Sanders}}}, \bibinfo {author} {\bibfnamefont {D.}~\bibnamefont {{Nagai}}},
  \bibinfo {author} {\bibfnamefont {E.~T.}\ \bibnamefont {{Lau}}}, \bibinfo
  {author} {\bibfnamefont {F.}~\bibnamefont {{Kaefer}}}, \bibinfo {author}
  {\bibfnamefont {F.}~\bibnamefont {{Pacaud}}}, \bibinfo {author}
  {\bibfnamefont {N.}~\bibnamefont {{Clerc}}}, \bibinfo {author} {\bibfnamefont
  {T.~H.}\ \bibnamefont {{Reiprich}}}, \bibinfo {author} {\bibfnamefont
  {E.}~\bibnamefont {{Bulbul}}}, \bibinfo {author} {\bibfnamefont {J.~I.}\
  \bibnamefont {{Chitham}}}, \bibinfo {author} {\bibfnamefont {C.-H.}\
  \bibnamefont {{Chiang}}}, \bibinfo {author} {\bibfnamefont {V.}~\bibnamefont
  {{Ghirardini}}}, \bibinfo {author} {\bibfnamefont {V.}~\bibnamefont
  {{Gonzalez-Perez}}}, \bibinfo {author} {\bibfnamefont {G.}~\bibnamefont
  {{Gozaliasl}}}, \bibinfo {author} {\bibfnamefont {C.~C.}\ \bibnamefont
  {{Fitzpatrick}}}, \bibinfo {author} {\bibfnamefont {A.}~\bibnamefont
  {{Klypin}}}, \bibinfo {author} {\bibfnamefont {A.}~\bibnamefont {{Merloni}}},
  \bibinfo {author} {\bibfnamefont {K.}~\bibnamefont {{Nandra}}}, \bibinfo
  {author} {\bibfnamefont {T.}~\bibnamefont {{Liu}}}, \bibinfo {author}
  {\bibfnamefont {F.}~\bibnamefont {{Prada}}}, \bibinfo {author} {\bibfnamefont
  {M.}~\bibnamefont {{Ramos-Ceja}}}, \bibinfo {author} {\bibfnamefont
  {M.}~\bibnamefont {{Salvato}}}, \bibinfo {author} {\bibfnamefont
  {R.}~\bibnamefont {{Seppi}}}, \bibinfo {author} {\bibfnamefont
  {E.}~\bibnamefont {{Tempel}}},\ and\ \bibinfo {author} {\bibfnamefont
  {G.}~\bibnamefont {{Yepes}}},\ }\bibfield  {title} {\bibinfo {title}
  {{Full-sky photon simulation of clusters and active galactic nuclei in the
  soft X-rays for eROSITA}},\ }\href
  {https://doi.org/10.21105/astro.2008.08404} {\bibfield  {journal} {\bibinfo
  {journal} {The Open Journal of Astrophysics}\ }\textbf {\bibinfo {volume}
  {3}},\ \bibinfo {eid} {13} (\bibinfo {year} {2020})},\ \Eprint
  {https://arxiv.org/abs/2008.08404} {arXiv:2008.08404 [astro-ph.CO]}
  \BibitemShut {NoStop}%
\bibitem [{\citenamefont {{Voit}}(2005)}]{voit05}%
  \BibitemOpen
  \bibfield  {author} {\bibinfo {author} {\bibfnamefont {G.~M.}\ \bibnamefont
  {{Voit}}},\ }\bibfield  {title} {\bibinfo {title} {{Tracing cosmic evolution
  with clusters of galaxies}},\ }\href
  {https://doi.org/10.1103/RevModPhys.77.207} {\bibfield  {journal} {\bibinfo
  {journal} {Reviews of Modern Physics}\ }\textbf {\bibinfo {volume} {77}},\
  \bibinfo {pages} {207} (\bibinfo {year} {2005})},\ \Eprint
  {https://arxiv.org/abs/astro-ph/0410173} {arXiv:astro-ph/0410173 [astro-ph]}
  \BibitemShut {NoStop}%
\bibitem [{\citenamefont {Kravtsov}\ \emph {et~al.}(2018)\citenamefont
  {Kravtsov}, \citenamefont {Vikhlinin},\ and\ \citenamefont
  {Meshscheryakov}}]{Kravtsov:2014sra}%
  \BibitemOpen
  \bibfield  {author} {\bibinfo {author} {\bibfnamefont {A.}~\bibnamefont
  {Kravtsov}}, \bibinfo {author} {\bibfnamefont {A.}~\bibnamefont
  {Vikhlinin}},\ and\ \bibinfo {author} {\bibfnamefont {A.}~\bibnamefont
  {Meshscheryakov}},\ }\bibfield  {title} {\bibinfo {title} {{Stellar mass --
  halo mass relation and star formation efficiency in high-mass halos}},\
  }\href {https://doi.org/10.1134/S1063773717120015} {\bibfield  {journal}
  {\bibinfo  {journal} {Astron. Lett.}\ }\textbf {\bibinfo {volume} {44}},\
  \bibinfo {pages} {8} (\bibinfo {year} {2018})},\ \Eprint
  {https://arxiv.org/abs/1401.7329} {arXiv:1401.7329 [astro-ph.CO]}
  \BibitemShut {NoStop}%
\end{thebibliography}%

\appendix

\section{Universal Pressure Profile in Arnaud et al. (2010)}\label{apdx:UPP}

We here summarize the model of gas pressure profiles for galaxy clusters as obtained in Ref.~\cite{Arnaud:2009tt}. The model has been calibrated with a sample of 33 clusters at $z<0.33$ using X-ray data taken in an XMM-Newton large program, the representative XMM-Newton cluster structure survey (REXCESS) \cite{Bohringer:2007vs}. 

The model assumes that the spherically-averaged electron pressure profile in a cluster can be approximated in the functional form as proposed in Ref~\cite{Nagai:2007mt};
\beqa
P_\mathrm{e}(r) = \frac{P_0}{(c_{500} x)^{\gamma}\left[1+\left(c_{500} x\right)^{\alpha}\right]^{(\beta-\gamma)/\alpha}},
\eeqa
where $r$ is the cluster-centric radius defined in physical coordinates, 
$x=r/r_\mathrm{bd}$ is the dimensionless radius normalized with 
some halo boundary radius of $r_\mathrm{bd}$, 
and there are five free parameters 
($P_0$, $c_{500}$, $\alpha$, $\beta$ and $\gamma$) in the model. 
The parameters have been constrained with the observed pressure profiles in the REXCESS sample as well as their X-ray scaling relation under the assumption of hydrostatic equilibrium.
Taking into account a marginal radial dependence in the hydrostatic mass-pressure scaling relation, Ref.~\cite{Arnaud:2009tt} found that the REXCESS data can be well explained with the parameters below;
\beqa
P_0(M_\mathrm{UPP}, z) &=& 1.34145\times 10^{-2} \left(\frac{H_0}{70\,\mathrm{km/s/Mpc}} \right)^{1/2} \nonumber \\
&&
\qquad
\times \left(\frac{M_\mathrm{UPP}}{2.1\times10^{14}\, h^{-1}M_\odot}\right)^{2/3+\alpha(x)}
\nonumber \\
&&
\qquad
\times
\left(\frac{H(z)}{H_0}\right)^{8/3}\, \mathrm{keV}\,\mathrm{cm}^{-3}, \\
\alpha(x) &=& \frac{0.22}{1+(x/0.5)^{3}}, \\ 
r_\mathrm{bd}(M_\mathrm{UPP}, z) &=& \left[\frac{3 M_\mathrm{UPP}}{4 \pi \, 500\rho_\mathrm{crit}(z)}\right]^{1/3},
\eeqa
where 
$H(z)$ is the Hubble parameter at $z$,
$H_0 = H(z=0)$, and 
they introduce a mass parameter $M_\mathrm{UPP}$ which is equal to 
the spherical over-density mass of $M_\mathrm{500c}$ in hydrostatic equilibrium.
For other parameters, they found the best-fit values to be $c_{500} = 1.177$, $\alpha=1.0510$, $\beta=5.4905$, and $\gamma=0.3081$.
If necessary, we assume a fully ionized gas in a cluster to compute the thermal gas pressure profile as $P_\mathrm{gas}(r) = (5X+3)/(2X+2) P_\mathrm{e}(r)$ where $X=0.76$ is the primordial hydrogen abundance.

\section{Systematic tests of stacked cluster lensing measurements}\label{apdx:test_dSigma_measurement}

\begin{figure*}[!t]
\includegraphics[clip, width=2.0\columnwidth]{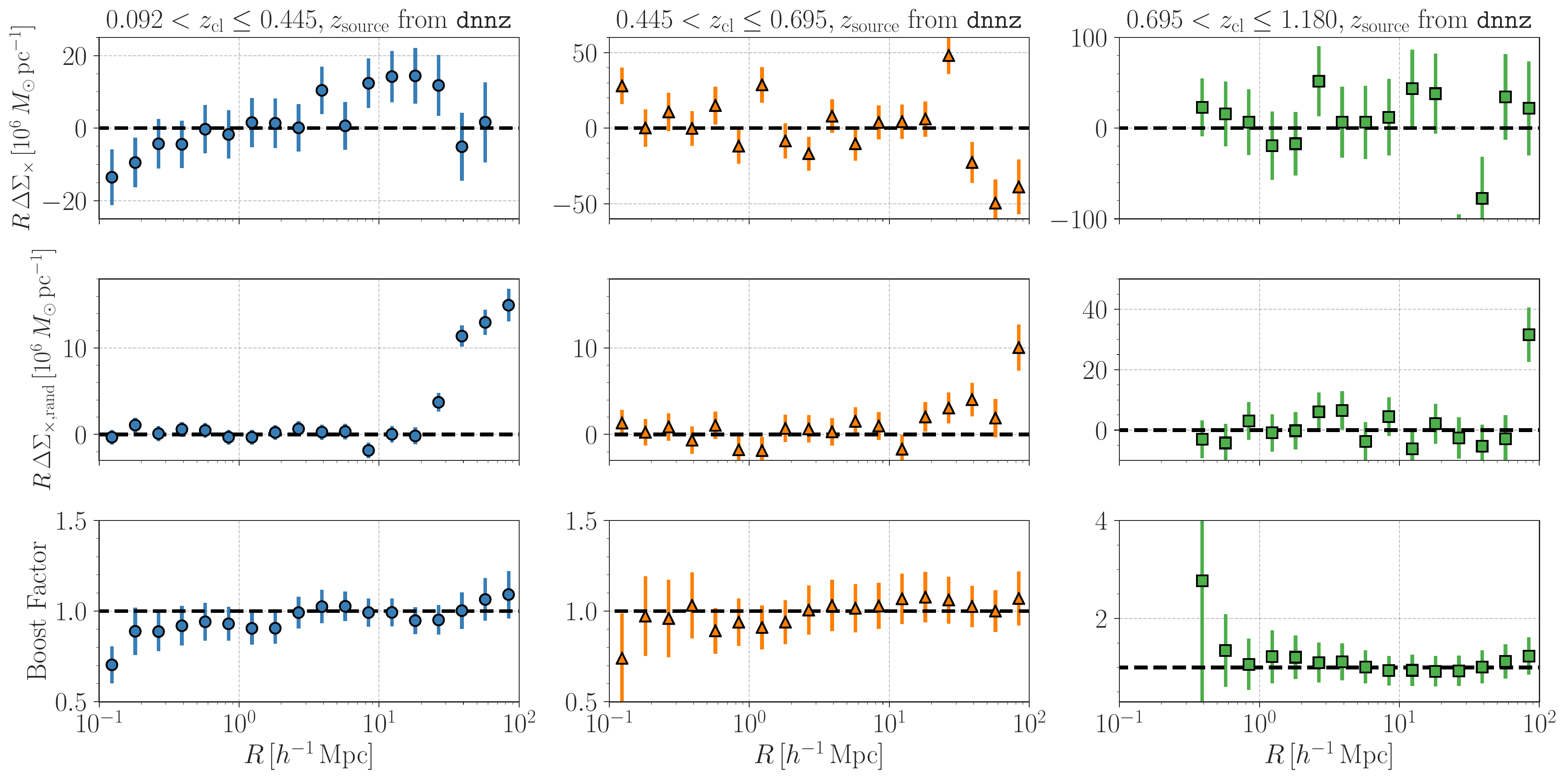}
\caption{\label{fig:Dsigma_null_tests} 
A summary of systematic tests for our stacked lensing measurements assuming the background selection with the photo-z information from ${\tt dnnz}$ (see Section~\ref{subsec:background_selection} for details).
The upper three panels show the B-mode signal (a 45-degree rotated component from tangential shear) around the ACT cluster samples in different redshift bins. 
The middle three panels present the B-mode signal around random points with the same redshift distribution as the actual cluster sample. 
In the lower three panels, we show our measurement of the boost factor 
as a test of our background selection.
In each row from left to right, we show the results at $0.092 < z \le 0.445$, $0.445< z \le 0.695$, and $0.695 < z \le 1.180$, respectively.
}
\end{figure*}

In this appendix, we summarize the results of several systematic tests in our stacked lensing measurements. Those include basic null tests of the stacked lensing analysis with the measurements of a 45-degree rotated component from tangential shear (so-called B-mode) around clusters and random points, and the ``boost-factor" test to quantify an excess or deficiency in the number of pairs of lens and source galaxies compared to that of random point and source galaxy pairs. The results for the measurements of the B mode and boost factor are summarized in Fig.~\ref{fig:Dsigma_null_tests}.

\begin{figure*}[!t]
\includegraphics[clip, width=2.0\columnwidth]{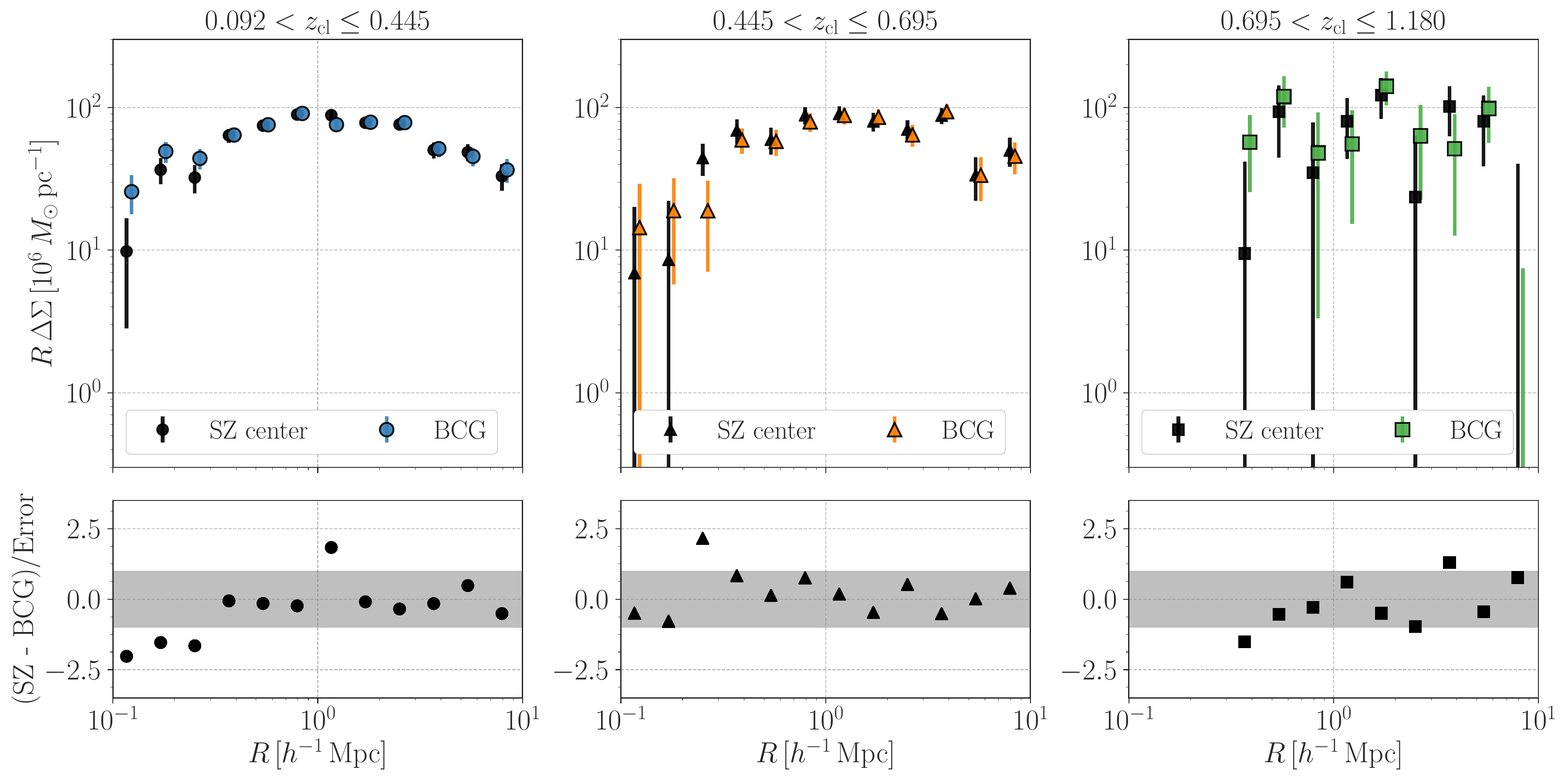}
\caption{\label{fig:Dsigma_varycenter} 
Stacked lensing signals using different definitions of the cluster centers.
The colored symbols in the top three panels present the stacked lensing signal with the SZ centers, while the black points in each top panel stand for our fiducial measurement centered on the BCGs. The data points in the top panels for the SZ centers are shifted along the x-axis for illustrative purposes. At the top, we compute the error bars by the shape noise alone.
At the bottom, we show the residual between the two lensing measurements 
normalized by the $1\sigma$ shape-noise errors. 
The gray region in the bottom panels highlights a $\pm1\sigma$ level.
We see a marginal difference between using the BCG and the SZ centers at the scales of $R \simlt 0.3 \, h^{-1}\mathrm{Mpc}$ in the cluster redshift range of $0.092 < z \le 0.445$. We use the BCG center for our fiducial measurements, whereas we include possible off-centering effects in our modeling of the stacked lensing signals.
}
\end{figure*}

In addition, we look into off-centering effects in our stacked lensing measurement by setting  the center of each cluster to the SZ peak in the ACT observation (the SZ center), instead of using the BCG position. The measured signals with the SZ center are shown in Fig.~\ref{fig:Dsigma_varycenter}.
Furthermore, we examine possible systematic effects arising from misestimation of source photo-z in our default code of ${\tt dnnz}$ by using a different photo-z method of 
${\tt mizuki}$. 
Fig.~\ref{fig:Dsigma_varypz} shows the stacked lensing signals with the ${\tt mizuki}$ photo-z information.

\subsection{B-mode signal around clusters and random points}

If a scalar potential governed by underlying density fields 
is responsible for the weak lensing, 
the average B-mode signal around clusters should be 
statistically consistent with zero. 
The top panels of Fig.~\ref{fig:Dsigma_null_tests} show the stacked B-mode signals 
around our ACT cluster subsamples at different redshift bins. 
To define the detection significance, we compute the chi-square quantity $\chi^2_\mathrm{B} = \sum_i [\Delta \Sigma_\times(R_i)/\sigma_\times(R_i)]^2$, 
where 
$\Delta \Sigma_\times(R_i)$ represents the B-mode signal at the i-th radius bin,
$\sigma_\times(R_i)$ is the root-mean-square (RMS) error due to the shape noise,
and the index $i$ runs over the radius range of $0.1<R \, [h^{-1}\mathrm{Mpc}]<10$.
Note that we exclude the B-mode signals at large radii of $R\ge 10\, h^{-1}\mathrm{Mpc}$
because the azimuthal average can be incomplete at such large radii in the presence of the survey boundary.
We find $\chi^2_\mathrm{B}/\mathrm{dof} = 11.9/12$, $18.6/12$ and $3.18/9$ at the cluster redshift bin of $0.092<z\le0.445$, $0.445<z\le0.695$, and
$0.695<z\le1.180$, respectively. 
We thus conclude that our measurement of the B-model signal is consistent with a null detection.

Imperfect PSF correction in galaxy shape measurements can introduce 
a significant detection of the B-mode signal around random points.
The middle panels of Fig.~\ref{fig:Dsigma_null_tests} show 
the B-mode signal around random points. 
For each of three redshift bins, we find $\chi^2_\mathrm{B}/\mathrm{dof} = 7.73/12$, $5.95/12$ and $3.76/9$ (from lower to higher redshifts), after 
removing points at $R\ge 10 \, h^{-1}\mathrm{Mpc}$.
These results show that the B-mode signal around random points is also consistent with zero, highlighting the imperfect PSF correction is not important in our analysis.

\subsection{Boost factor}

The boost factor is defined as the ratio of the number of source-cluster pairs to the source-random counterpart as a function of $R$. 
When the boost factor differs from unity, this indicates that some source galaxies 
are actually physically associated with lens clusters, calling into question the robustness of the background selection in the stacked lensing analysis.
We compute the boost factor as in Ref.~\cite{Miyatake:2013bha} and 
set a typical statistical uncertainty with the standard deviation over 100 realizations of random points. 
The bottom panels in Fig.~\ref{fig:Dsigma_null_tests} show 
the boost factor at the three cluster redshift bins. 
We confirm that our measurement of the boost factor does not show 
a significant deviation from unity over a wide range of $R$, 
ensuring our background selection is not biased.

\subsection{Off centering}

The cluster center used in the lensing measurement could be offset from the true gravitational
potential minimum, leading to some dilution in the lensing signals at inner radii around
and below the scale of off-centering.
We here examine the impact of off-centering by comparing the
lensing signal calculated with the BCG centers to the counterpart using the
SZ centers instead. The comparison is summarized in Fig.~\ref{fig:Dsigma_varycenter}. 
We find a marginal off-centering trend for the clusters in  
the lowest redshift bin in the range of $R\simlt 0.3\, h^{-1}\mathrm{Mpc}$, but the deviation is still at a $1\sigma$ level. 
For other redshift bins, we do not see any prominent off-centering effects in our stacked lensing measurements.
Nevertheless, we take into account possible off-centering effects in our model when comparing it with the data.

\subsection{Lensing signals with different photo-z methods}

We adopt ${\tt dnnz}$ photo-z estimates for our fiducial measurement
as described in Section~\ref{subsec:dSigma_estimator}. 
In this section, we describe the details of another photo-z method in the HSC catalog 
and check the consistency between ACT cluster lensing signals
based on two different photo-z methods.
For this comparison, we consider the photo-z method of ${\tt mizuki}$ as a representative example
of template fitting approaches. 
The ${\tt mizuki}$ code relies on a Bayesian template fitting
method which allows for simultaneously constraining physical
properties of galaxies such as star formation and photo-z \cite{2015ApJ...801...20T}.
The model template of galaxy spectra in ${\tt mizuki}$ is constructed from a stellar population synthesis code with a redshift evolution motivated by previous galaxy observations.
To break a severe parameter degeneracy in the model, ${\tt mizuki}$ introduces informative prior on physical quantities of galaxies such as stellar mass and star formation rate.
Hence, the ${\tt mizuki}$ photo-z estimate can be affected by the assumptions of galaxy physics, while our fiducial method of ${\tt dnnz}$ is purely a data-driven approach and is free from the choice of physical prior by construction.
In this sense, any change in the stacked lensing measurement with the ${\tt mizuki}$ photo-z information allows us to get some sense of how our results are sensitive to photo-z estimation.

\begin{figure*}[!t]
\includegraphics[clip, width=2.0\columnwidth]{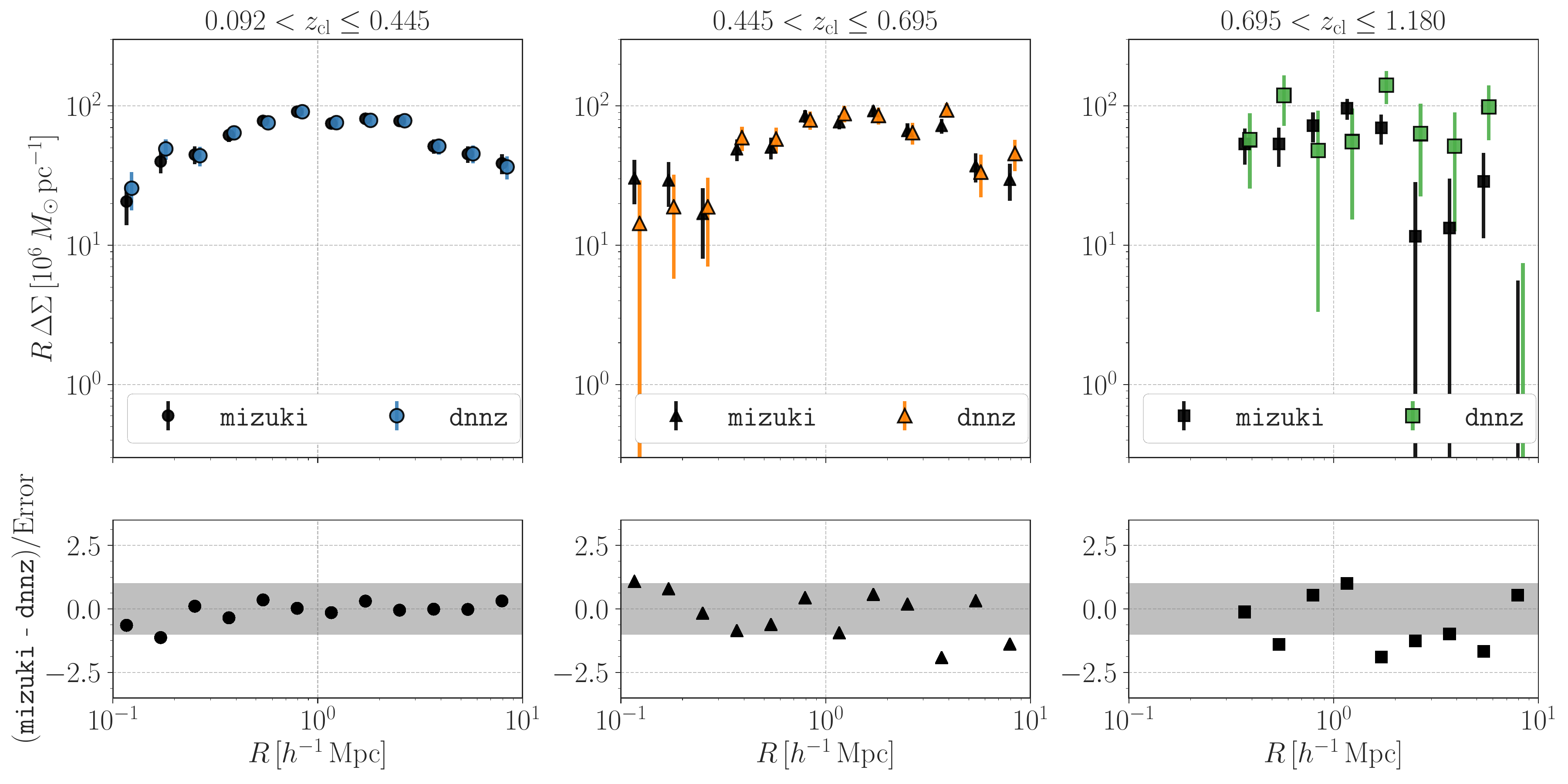}
\caption{\label{fig:Dsigma_varypz} 
Similar to Fig.~\ref{fig:Dsigma_varycenter}, but we compare the stacked lensing signals with two different photo-z estimates, ${\tt dnnz}$ and ${\tt mizuki}$. 
}
\end{figure*}

The colored symbols in the top panels of Fig.~\ref{fig:Dsigma_varypz} show lensing signals measured with the ${\tt mizuki}$ photo-z information.
We also show the difference between ${\tt dnnz}$- and ${\tt mizuki}$-based measurements in the bottom panels.
Note that we normalize the difference with the RMS error due to the shape noise in our fiducial measurement.
According to the figure, we conclude that the typical differences
between photo-z methods are within the statistical uncertainties.

\section{Mock catalogs of ACT clusters and HSC Y3}\label{apdx:mock}

In this appendix, we describe the production of mock catalogs of ACT clusters as well as the HSC Y3 source galaxies. 

Our mock production relies on the full-sky ray-tracing simulations and 
the halo catalogs in the light-cone simulation realization in Ref.~\cite{Takahashi:2017hjr}. 
The full-sky simulations have been constructed from 
a set of N-body simulations with $2048^3$ particles in various cosmological volumes. 
Note that Ref.~\cite{Takahashi:2017hjr} adopted the same $\Lambda$CDM cosmology 
as our fiducial model.
Weak gravitational lensing fields on a full sky have
been computed by the standard multiple lens-plane algorithms \cite{Hamana:2001vz, 2013MNRAS.435..115B, Shirasaki:2015dga} with an onion-like configuration of 38 projected mass density shells.
Each of the mass density shells was obtained by projecting N-body particles over a radial width of $150\, h^{-1}\mathrm{Mpc}$, allowing individual full-sky simulations to cover a cosmological volume up to $z=5.3$.
In the end, the lensing simulations consist of the shear field at each of 38 different source redshifts with an angular resolution of $0.43\, \mathrm{arcmin}$, which are suitable for the simulation of the current generation of galaxy imaging surveys. We caution that the effective resolution in  these mock, stacked lensing analyses can be worse than the original angular resolution in the full-sky simulation, because of the non-local nature of tangential shear (see Ref.~\cite{Takahashi:2017hjr} for details).
In each output of the N-body simulation, Ref.~\cite{Takahashi:2017hjr} identified dark matter halos using the Rockstar algorithm \cite{Behroozi:2011ju}. The N-body simulations allow us to resolve dark matter halos with masses greater than $8.9\times 10^{12}\, h^{-1}M_\odot$ 
with more than 50 N-body particles at redshifts $z < 1.17$.
Hence, the mass resolution in the light-cone halo catalogs is expected to be well below the minimum halo mass in the ACT clusters. 

\begin{figure*}[!t]
\includegraphics[clip, width=2.1\columnwidth]{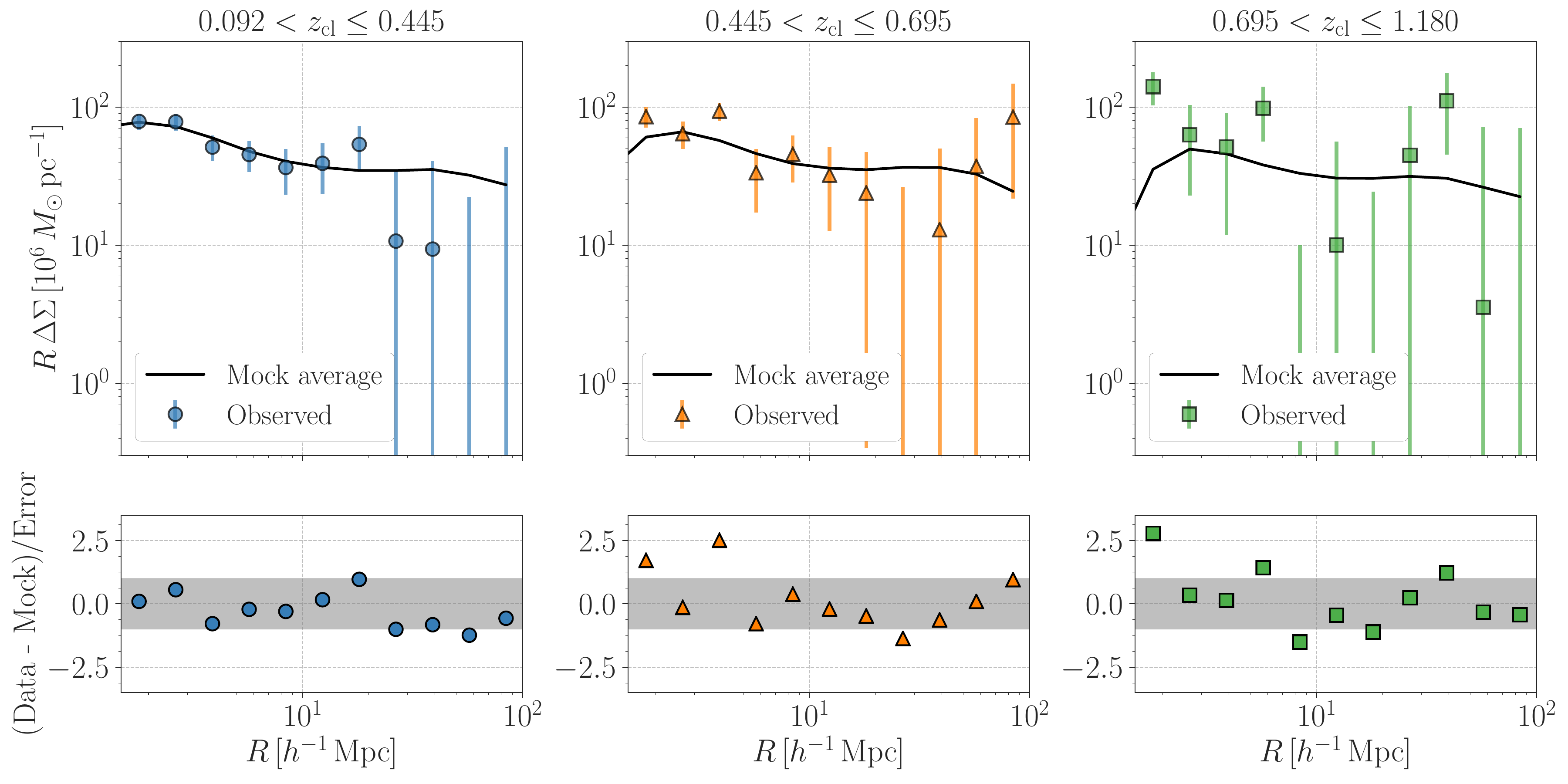}
\caption{\label{fig:Dsigma_mockavg} 
Comparison of the stacked lensing signal with our mock and real datasets.
The colored symbols in the top three panels show the stacked lensing signal with 
the real data at different cluster redshift bins,
while the solid line stands for the average signal over our 1404 mock realizations.
At the bottom, we show the difference between the two signals normalized to the $1\sigma$ statistical error. The gray region in the bottom represents a $\pm1\sigma$ level.
Note that we include the sample variance for the error bars in this figure because it is a dominant source of the scatter at $R\simgt 3\, h^{-1}\mathrm{Mpc}$.
}
\end{figure*}

For mock catalogs of the ACT clusters, we first define an effective selection function of dark matter halos in the ACT cluster search by using our halo-based model. The selection function is computed as;
\beqa
S_h(M) &=& \int \mathrm{d}\log M_\mathrm{UPP}\, S(\log M_\mathrm{UPP}) \nonumber \\
&&
\qquad \qquad
\times
{\cal P}(\log M_\mathrm{UPP} | M), \label{eq:selec_func_halo}
\eeqa
where $S(\log M_\mathrm{UPP})$ represents the completeness of the ACT clusters as given in Fig.~\ref{fig:ACT_selecfunc}, ${\cal P}(\log M_\mathrm{UPP} | M)$ is the probability distribution function of $\log M_\mathrm{UPP}$ at the halo mass $M$ as in Eq.~(\ref{eq:prob_MUPP_given_M}), and we omit the redshift $z$ 
in Eq.~(\ref{eq:selec_func_halo}) for simplicity.
To have a reasonable estimate of $S_h(M)$, we perform the MCMC analysis with the log-likelihood of Eq.~(\ref{eq:logL}), but ignore the sample covariance in the stacked lensing measurements and assume no baryonic effects in our model of stacked lensing signals.
In the MCMC analysis, we vary six model parameters ($a_0$, $a_1$, $C_0$, $f_0$, $f_1$ and $\alpha_\mathrm{off}$) by using ${\tt emcee}$ with 21 walkers, 20000 steps for burn-in, and 20000 steps for sampling the likelihood.
To mitigate possible systematic effects due to the sample variance, we set the radius range of 
$0.1<R\, [h^{-1}\mathrm{Mpc}]<10$ in this MCMC analysis. 
We then find the best-fit parameters of 
$(a_0, a_1, C_0) =(0.283, 0.197, 0.241)$,
$(0.290, 0.0725, 0.217)$, and 
$(0.358, 0.149, 0.210)$ at the cluster redshift bins of
$0.092 < z \le 0.445$,
$0.445 < z \le 0.695$,
and $0.695 < z \le 1.180$,
respectively.
Using these best-fit parameters, we infer the selection function of $S_h(M)$ as Eq.~(\ref{eq:selec_func_halo}) at a given redshift.
Once having the selection function, we populate the mock ACT clusters into halos in the full-sky light-cone catalogs by following $S_h(M)$ and then extract the objects in the relevant survey footprint (see Fig.~\ref{fig:survey_overlap}) 13 times from a single full-sky realization.
At the same time, we generate random points for each mock realization of the ACT clusters so that the redshift distribution of random points is the same as that of the mock clusters. We set the number of random points to be 100 times as large as the number of mock ACT clusters. In the end, we have 1404 mock realizations of the ACT clusters
from 108 full-sky halo catalogs.

For mock source galaxies of the HSC Y3, we follow the method given in Ref.~\cite{Shirasaki:2019gya}.
The production of mock source catalogs is carefully designed to account for
various observational effects such as the survey footprint, 
inhomogeneous angular distribution of source galaxies, statistical uncertainties in
photometric redshift estimate, variations in the lensing
weight, and the statistical noise in galaxy shape measurements 
including both intrinsic shapes and measurement errors. 
We produce 1404 mock catalogs from
108 full-sky ray-tracing simulations of gravitational lensing. Note that our HSC Y3 mock catalogs share exactly the same information on angular positions, redshifts, lensing weights, and shear responsivity with the real catalog.
Hence, we can use the same pipeline for mock lensing analyses as done in the real data,
allowing a direct estimate of the covariance matrix with 1404 realizations of mock stacked lensing measurements.

To validate our mock production, we show the average lensing signal around the mock ACT clusters over 1404 realizations in Fig.~\ref{fig:Dsigma_mockavg}.
In the figure, we limit the radius range to be $R>3\, h^{-1}\mathrm{Mpc}$ because the numerical resolution in our parent full-sky simulations is not sufficient to resolve the stacked lensing signal at $R \simlt 3-5\, h^{-1}\mathrm{Mpc}$. 
The figure demonstrates that the average signal of our mock lensing analyses is in good agreement with the real measurement. 

\section{Calibration of scatter in the gas pressure profile with Illustris TNG}\label{apdx:scatter_Pgas}

Under the assumption of self-similarity, the pressure profile for a halo with a given mass and redshift should be identical to that of a halo with the same mass and redshift. However, galaxy clusters are not exactly self-similar. There are intrinsic variations in halo profiles due to the different halo formation histories. Halos that are formed earlier usually have more concentrated dark matter halo profiles, leading to a steeper potential well and thus denser gas in the halo core. Radiative cooling, star formation, and feedback effects will also result in different gas profiles between halos, owing to the stochastic nature of how accretion onto and feedback from supermassive black holes in cluster cores are triggered. Outside the core, we also expect the gas in recently merged clusters to be more turbulent and thus less thermalized. 

To compute the intrinsic scatter of the pressure profile, we use a non-parametric model to generate samples of thermal pressure profiles. Specifically, we use the thermal pressure covariance matrix measured from simulations to model intrinsic variations in thermal pressure profile, following a similar method Ref.~\cite{comparat_etal20} for X-ray surface brightness profiles. 

Before computing the covariance matrices of the thermal pressure profiles in groups and clusters in the TNG300 simulation, we first account for the dependence of the pressure profile on mass and redshift given self-similarity. To do so we scale the radius by $R_\mathrm{200c,p}$ and normalize the thermal pressure profiles with the self-similar quantity (see, e.g. Ref.~\cite{voit05}),
\begin{eqnarray}
    P_\mathrm{200c}(M_\mathrm{200c},z) &=& \frac{GM_\mathrm{200c}}{2R_\mathrm{200c,p}} 200 \rho_{\rm crit}(z) f_{\rm baryon},
\end{eqnarray} 
where 
$M_\mathrm{200c}$ is the spherical overdensity mass defined as $M_\mathrm{200c} = 4 \pi \cdot 200\, \rho_\mathrm{crit}(z) R^3_\mathrm{200c,p}/3$, 
$f_{\rm baryon} = \Omega_b / \Omega_M$ is the cosmic baryon fraction. 

We then account for additional halo mass dependence in the pressure profiles due to baryonic physics, which effects are stronger in lower mass halos, by applying the Kernel Localized Linear Regression (KLLR) method \cite{kllr1, kllr2} to estimate the average halo mass trend of the pressure profile in each scaled radial bin in $r/R_\mathrm{200c,p}$. Here we use KLLR to perform linear regression on pressure profile value at each scaled radial bin for every halo with a Gaussian kernel in $M_\mathrm{200c}$ to get the average pressure as a function of $M_\mathrm{200c}$, with a bootstrap re-sampling to estimate the uncertainties. 

Next, we compute the covariance matrix $\mathcal{C}$ for the normalized thermal pressure profiles over the sample of TNG clusters as
\begin{equation}
    \mathcal{C}(x, x') = 
    \langle (\log f(x)-\overline{\log f}(x))(\log f(x')-\overline{\log f}(x')) \rangle
    \label{eqn:cov_matrix}
\end{equation}
where $f(x) = P(x)/P_{200c}$, $x = r/R_\mathrm{200c,p}$
and $\overline{\log f}(x)$ is the  logarithmic mean of the profile, estimated with KLLR, and $r$ and $r'$ are the two radii in consideration. 

Finally, we generate the realizations of variations of the thermal profile by sampling the covariance matrix, treating the covariance matrix as a multivariate Gaussian distribution. For $N$ radial bins, the multivariate Gaussian distribution is 
\begin{eqnarray}
    &&\mathcal{N}(\delta \log f_1, \delta \log f_2, \dots, \delta 
    \log f_N) \nonumber \\ &&=\frac{1}{\sqrt{2\pi|\mathcal{C}|^N}}\exp\left(-\frac{1}{2}(\boldmath{\delta \log f})^T\mathcal{C}^{-1}(\boldmath{\delta \log f} )\right),
    \label{eqn:log_normal}
\end{eqnarray}
where $\boldmath{\delta \log f} = (\delta \log f_1, \delta \log f_2, \dots, \delta \log f_N)$ is the deviation from the mean log profile. Once we specify the covariance matrix $\mathcal{C}$, we can draw a realization of the variation in the profile $\delta \log f$ from the multivariate Gaussian distribution. The resulting realization of the profile will then be the sum of the mean profile and the variation: $ \log f = \overline{\log f} + \delta \log f$. 


\section{Details in Baryonification model}\label{apdx:baryonification}

We here describe details of our implementation of the baryonification model introduced in Section~\ref{subsec:baryonic_effects}. In this appendix, we omit the halo mass $M$ and redshift $z$ except when necessary.

\subsection{Truncation of the Dark-Emulator output}

The baryonification model requires a converged enclosed halo mass for an input dark-matter-only (dmo) profile in the limit of infinite boundary radius, but the prediction by the Dark Emulator includes the halo-matter cross-correlation caused by neighboring halos, referred to as the two-halo term in the literature. 
To meet the requirement in the baryonification, 
we truncate the Dark-Emulator output 
to
reduce effects associated with two-halo terms in the inherent mapping process of the baryonification.
In this paper, we first perform a least-square fitting to the dmo density profile by the Dark Emulator with the functional form below \cite{Diemer:2014xya};
\beqa
\rho_\mathrm{DK14}(r) &=& \rho_\mathrm{inner}(r) f_\mathrm{trans}(r) + \rho_\mathrm{outer}(r), \\
\rho_\mathrm{inner}(r) &=& \rho_s \exp\left\{-\frac{2}{\alpha}\left[\left(\frac{r}{r_s}\right)^\alpha-1\right]\right\}, \\
f_\mathrm{trans}(r) &=& \left[1+\left(\frac{r}{r_t}\right)^{\beta}\right]^{-\gamma/\beta}, \\
\rho_\mathrm{outer}(r) &=& \bar{\rho}_\mathrm{m0}\left[b_e \left(\frac{r}{5 R_\mathrm{200b}}\right)^{-s_e}+1\right],
\eeqa
where 
we have eight fitting parameters in total ($\rho_s, r_s, \alpha, \beta, \gamma, r_t, b_e$, and $s_e$). Among those parameters, we vary only five parameters of $\rho_s, r_s, r_t, b_e$ and $s_e$ but fix the other parameters as $\alpha = 0.155 + 0.0095 \nu^2$, $\beta=4.0$, and $\gamma=8.0$ where $\nu$ is the peak height parameter given by the ratio of 1.686 with the standard deviation of the linear density fluctuation smoothed by a top-hat filter at a scale of $R = [3M/(4\pi \bar{\rho}_\mathrm{m0})]^{1/3}$.
Finding the best-fit parameters to the Dark-Emulator prediction of $\rho_\mathrm{hm}$, we then infer the boundary radius of $r_\mathrm{out}$ at which the logarithmic slope of $\mathrm{d}\ln\rho_\mathrm{DK14}/\mathrm{d}\ln r$ becomes the smallest.
Once the estimate of $r_\mathrm{out}$ is obtained, we truncate the density profile $\rho_\mathrm{hm}$ as in Eq.~(\ref{eq:Emulator_truncation}).

\subsection{Dark Matter}

We simply set the dark matter component in the dark-matter-baryon (dmb) profile as
\beqa
\rho_\mathrm{DM}(r) = \left(1-\frac{\Omega_\mathrm{b}}{\Omega_\mathrm{m}}\right) \rho_\mathrm{hm+t}(r),
\eeqa
where the factor of $1-\Omega_\mathrm{b}/\Omega_\mathrm{m}$ represents the global mass fraction of cosmic dark matter, and $\rho_\mathrm{hm+t}$ is given by Eq.~(\ref{eq:Emulator_truncation}).

\subsection{Gas}

For the gas density profile, we adopt the functional form as proposed in Ref.~\cite{Giri:2021qin};
\beqa
\rho_\mathrm{gas}(r) &\propto& \left[1+\left(\frac{r}{0.1R_\mathrm{200c}}\right)\right]^{-\beta_g} \nonumber \\
&&
\qquad
\times \left[1+\left(\frac{r}{\theta_\mathrm{ej}R_\mathrm{200c}}\right)^{\gamma_g}\right]^{-(\delta_g - \beta_g)/\gamma_g}, \label{eq:gas_density}
\eeqa
where $\beta_g, \gamma_g, \delta_g$, and $\theta_\mathrm{ej}$ are free parameters in the model, and we define $R_\mathrm{200c}$ with the Dark-Emulator output $\rho_\mathrm{hm}$ as in Eqs.~(\ref{eq:def_M500c_base}) and (\ref{eq:def_M500c}) but replace the factor of 500 with 200.
Throughout this paper, we fix $\gamma_g = 2$ and $\delta_g = 7$ as motivated in the comparison with hydrodynamical simulations \cite{Giri:2021qin} but allow for the mass dependence of $\beta_g$ as
\beqa
\beta_g(M) = \frac{3(M/M_c)^{\mu}}{1+(M/M_c)^{\mu}},
\eeqa
where we introduce two free parameters of $\mu$ and $M_c$.
The normalization of Eq.~(\ref{eq:gas_density}) is set by
\beqa
\int_0^{\infty} 4\pi r^2 \mathrm{d}r\, \rho_\mathrm{gas}(r) = \left[\frac{\Omega_\mathrm{b}}{\Omega_\mathrm{m}}-f_\mathrm{star}(M)\right] M,
\eeqa
where $f_\mathrm{star}(M)$ is the stellar-to-halo mass relation defined below.

\subsection{Stars}

For the stellar density profile, we decompose the stars in a halo into two components.
One contribution is from the central galaxy’s mass density profile, which is parameterized as follows:
\beqa
\rho_\mathrm{cga}(r) = \frac{f_\mathrm{cga} M}{4\pi^{3/2} R_{*} r^2}\exp \left[-\left(\frac{r}{2R_{*}}\right)^2\right],
\eeqa
where $f_\mathrm{cga}$ describes the fraction of stars in the central galaxy,
and $R_*$ is the stellar half-light radius which is set to $0.015$ times the spherical over-density radius of $R_{200c}$ \cite{Kravtsov:2014sra}.
Another component consists of the satellite stars in a halo and we assume that their distribution follows the dark matter density profile as
\beqa
\rho_\mathrm{sga}(r) = f_\mathrm{sga} \rho_\mathrm{hm+t}(r),
\eeqa
where $f_\mathrm{sga}$ represents the fraction of satellite stars in the halo and average mass density profile at halo mass $M$ predicted by the Dark Emulator, 
$\rho_{hm+t}(r)$, is specified in Eq~(\ref{eq:Emulator_truncation}).
The total stellar density profile in a halo is given by 
$\rho_\mathrm{star}(r)=\rho_\mathrm{cga}(r)+\rho_\mathrm{sga}(r)$.
We parameterize the stellar-to-halo mass relation as
\beqa
f_\mathrm{star}(M) &=& 0.09\left(\frac{M}{M_s}\right)^{-\eta_\mathrm{star}}, \\
f_\mathrm{cga}(M)  &=& 0.09 \left(\frac{M}{M_s}\right)^{-\eta_\mathrm{cga}}, \\
f_\mathrm{sga}(M)  &=& f_\mathrm{star}(M) - f_\mathrm{cga}(M),
\eeqa
where $M_s = 2.5\times 10^{11}\, h^{-1}M_\odot$,
$\eta_\mathrm{star}$ and $\eta_\mathrm{cga}$ are free parameters in the model.
In this paper, we fix $\eta_\mathrm{star}=0.32$ and $\eta_\mathrm{cga}=0.60$ as constrained in Ref.~\cite{Schneider:2018pfw}. 
Note that Ref~\cite{Giri:2021qin} demonstrated that varying those parameters does not introduce significant impacts on matter clustering at Mpc scales.
The impact of the stellar density profile on the stacked lensing can be significant at 
$R\simlt0.1\, h^{-1}\mathrm{Mpc}$, but off-centering effects also affect the stacked lensing signal at similar scales.   
When comparing our model with the real data, 
we fix the stellar density profile but vary off-centering effects 
to marginalize over the modeling uncertainty at the small scale of $R\sim 0.1\, h^{-1}\mathrm{Mpc}$.

\section{Full posterior distributions in parameter inference}\label{apdx:details_mcmc}

\begin{figure*}[!t]
\includegraphics[clip, width=2.0\columnwidth]{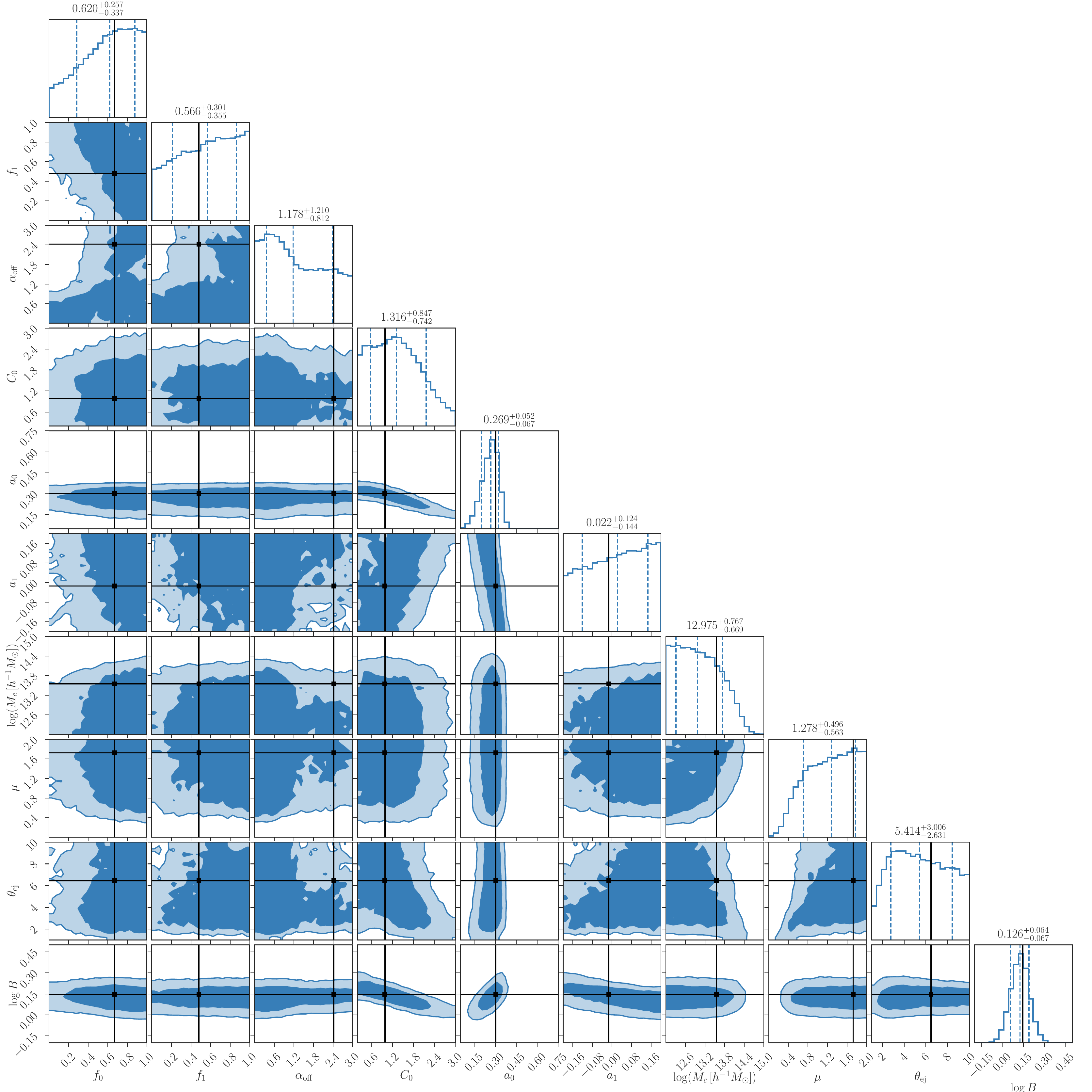}
\caption{\label{fig:full_posterior_z1} 
The posterior distribution of our nine model parameters at the cluster redshift of 
$0.092 < z \le 0.445$. The darker and lighter regions represent the 68\% and 95\% credible intervals, respectively. The black point with connected lines in each panel highlights the set of the best-fit parameters. The one-dimensional histograms in some panels shows the posterior distribution of a single model parameter. For referece, we also include the posterior distribution of the mass bias $\log B$ at the average $M_\mathrm{UPP}$ at the bottom. Note that $\log B$ is a derived parameter in our model. 
}
\end{figure*}

\begin{figure*}[!t]
\includegraphics[clip, width=2.0\columnwidth]{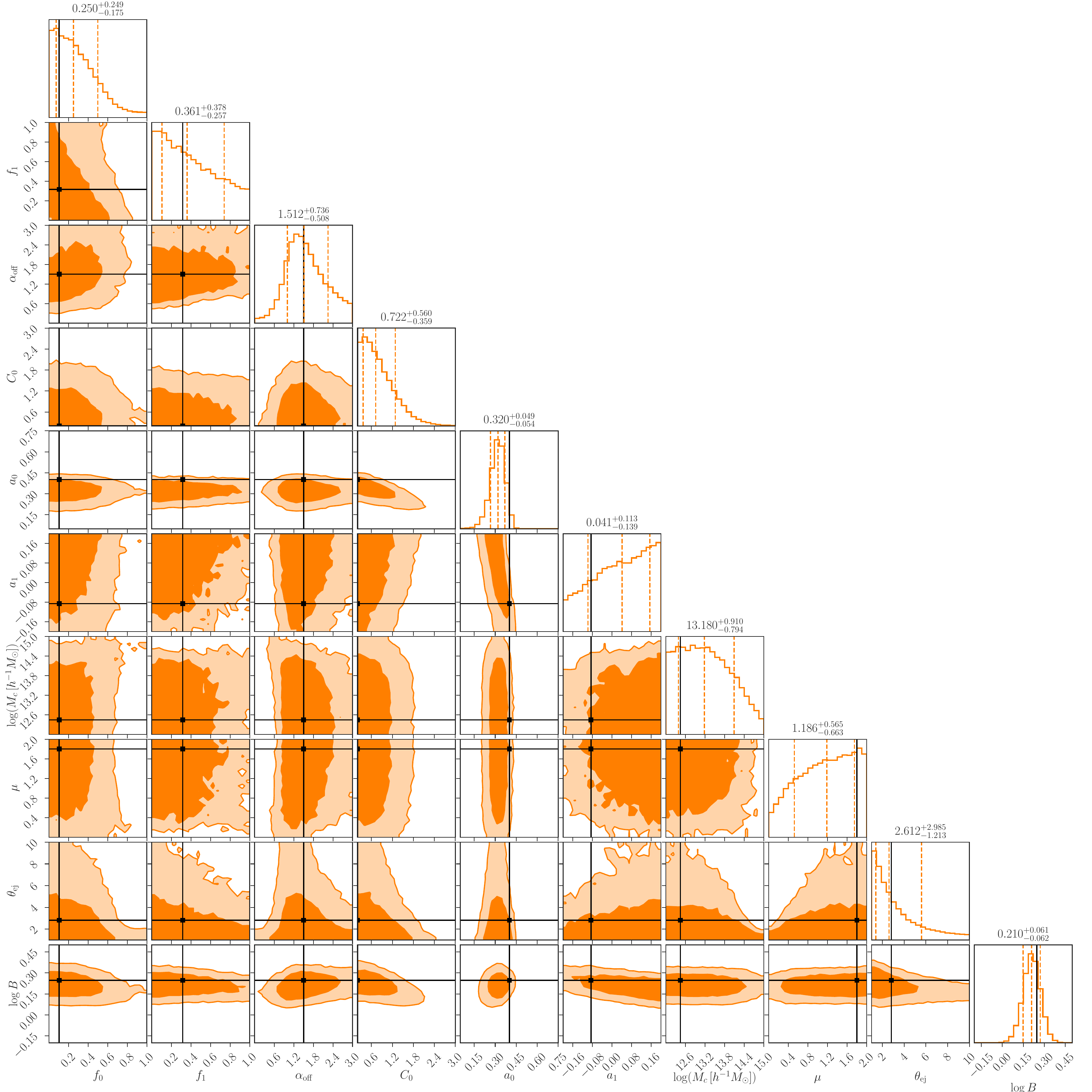}
\caption{\label{fig:full_posterior_z2} 
Similar to Fig.~\ref{fig:full_posterior_z1}, but showing the case at the cluster redshift of
$0.445 < z \le 0.695$. 
}
\end{figure*}

\begin{figure*}[!t]
\includegraphics[clip, width=2.0\columnwidth]{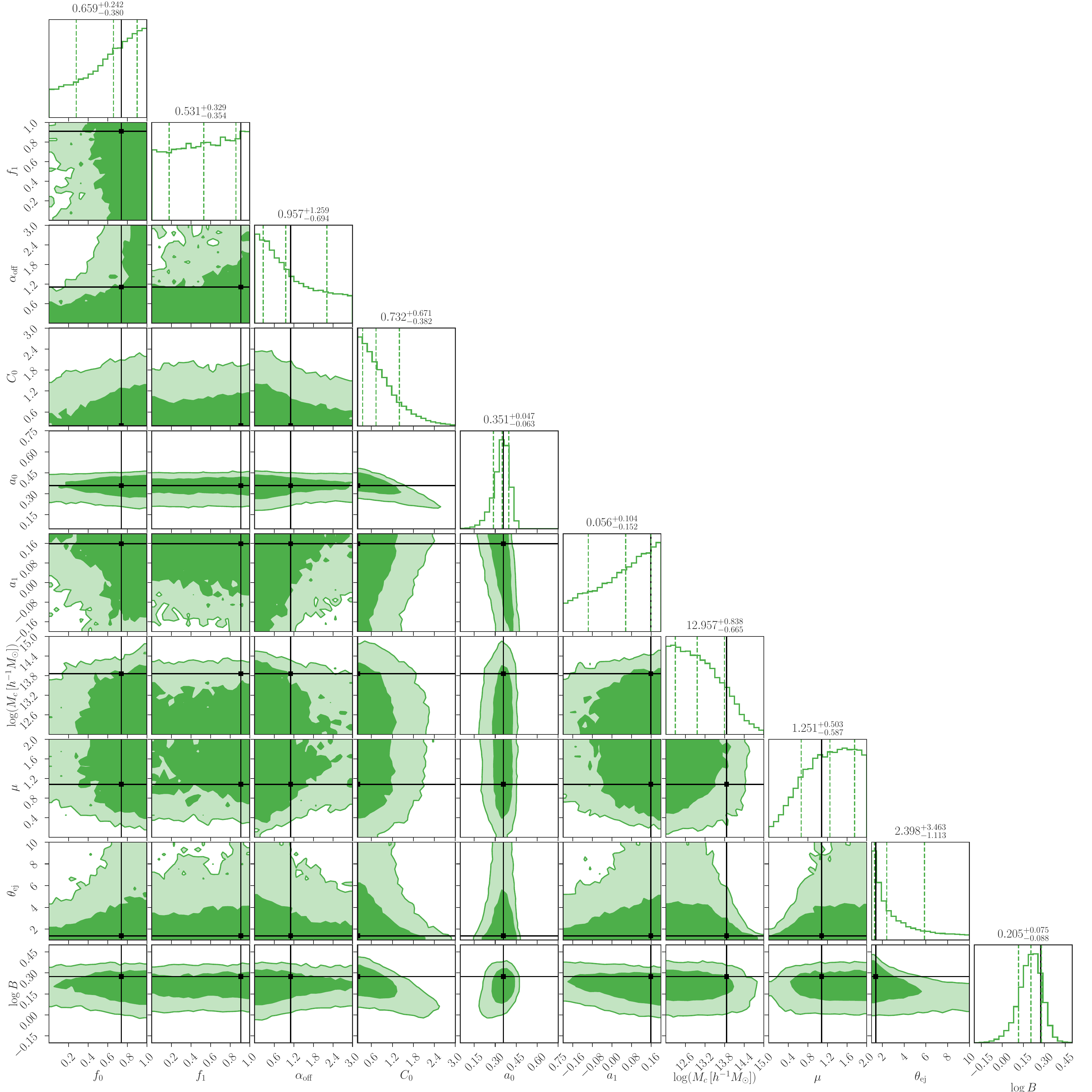}
\caption{\label{fig:full_posterior_z3}
Similar to Fig.~\ref{fig:full_posterior_z1}, but showing the case at the cluster redshift of
$0.695 < z \le 1.180$.
}
\end{figure*}

In this appendix, we provide the posterior distribution of our nine model parameters obtained through the MCMC analysis. Here, we adopt our fiducial cosmology, include the baryonic effects as in Section~\ref{subsec:baryonic_effects}, and adopt the log-likelihood function of Eq.~(\ref{eq:logL}). The results are summarized in Figs.~\ref{fig:full_posterior_z1}-\ref{fig:full_posterior_z3}.

For the sake of completeness, we also note that the best-fit parameters at each cluster redshift bin; $(f_0, f_1, \alpha_\mathrm{eff}, a_0, a_1, C_0, \log (M_c [h^{-1}M_\odot]), \mu, \theta_\mathrm{ej}) = 
    (0.669, 0.481, 2.422, 0.304, -0.0137, 0.986, 13.55, 1.723, 6.462)$,
$(0.104, 0.316, 1.502, 0.401, -0.0859, 0.205, 12.44, 1.802, 2.823)$, 
and
$(0.738, 0.909, 1.106, 0.358, 0.158, 0.206, 13.85, 1.081, 1.389)$ from lower to higher cluster redshifts.

\end{document}